\documentclass[aps,prd,twocolumn,floatfix,10pt,amssymb,amsfont,amsmath,superscriptaddress,float,nofootinbib,longbibliography]{revtex4-2}

\usepackage{preamb}

\begin{document}

\title{Dualities in one-dimensional quantum lattice models:\\topological sectors}

\author{\textsc{Laurens Lootens}}
\email{laurens.lootens@ugent.be}
\affiliation{Department of Physics and Astronomy, Ghent University, Krijgslaan 281, 9000 Gent, Belgium}
\affiliation{Department of Applied Mathematics and Theoretical Physics, University of Cambridge,\\ Wilberforce Road, Cambridge, CB3 0WA, United Kingdom}
\author{\textsc{Clement Delcamp}}
\email{clement.delcamp@ugent.be}
\affiliation{Department of Physics and Astronomy, Ghent University, Krijgslaan 281, 9000 Gent, Belgium}
\author{\textsc{Frank Verstraete}}
\affiliation{Department of Physics and Astronomy, Ghent University, Krijgslaan 281, 9000 Gent, Belgium}
\affiliation{Department of Applied Mathematics and Theoretical Physics, University of Cambridge,\\ Wilberforce Road, Cambridge, CB3 0WA, United Kingdom}

\begin{abstract}
    \noindent
    It has been a long-standing open problem to construct a general framework for relating the spectra of dual theories to each other. Here, we solve this problem for the case of one-dimensional quantum lattice models with symmetry-twisted boundary conditions. In ref.~\href{https://doi.org/10.1103/PRXQuantum.4.020357}{[PRX Quantum 4, 020357]}, dualities are defined between (categorically) symmetric models that only differ in a choice of module category. Using matrix product operators, we construct from the data of module functors explicit symmetry operators preserving boundary conditions as well as intertwiners mapping topological sectors of dual models onto one another. We illustrate our construction with a family of examples that are in the duality class of the spin-$\frac{1}{2}$ Heisenberg XXZ model. One model has symmetry operators forming the fusion category $\Rep(\mc S_3)$ of representations of the group $\mc S_3$. We find that the mapping between its topological sectors and those of the XXZ model is associated with the non-trivial braided auto-equivalence of the Drinfel'd center of $\Rep(\mc S_3)$.

\end{abstract}

\maketitle

\section{Introduction}

\noindent
Over the past few years tremendous progress has been achieved in our understanding of quantum theories by interpreting \emph{symmetries} in terms of \emph{topological operators}. More specifically, correlation functions of the theories including symmetry operators are insensitive to topology-preserving deformations of the submanifolds supporting the operators, unless they pass through charged operators \cite{Gaiotto:2014kfa}. In this context, ordinary global symmetries are generated by codimension-one invertible operators, which together with the requirements that symmetry operators can be fused, implies that these furnish a representation of a group. This new approach has led to generalizations of the notion of symmetry, whereby operators are not necessarily supported on one-codimensional submanifolds and/or are not necessarily invertible. This manuscript is concerned with such generalized symmetries in the context of translation invariant one-dimensional quantum lattice models. 

Relaxing the invertibility condition leads to symmetry operators whose properties are encoded into abstract higher mathematical structures known as \emph{spherical fusion categories} \cite{ENO,etingof2016tensor}. These so-called \emph{categorical symmetries} have been under intense scrutiny in recent years \cite{Thorngren:2019iar,Thorngren:2021yso,Albert:2021vts,PhysRevResearch.2.033417,PhysRevResearch.2.043086,Chatterjee:2022kxb,McGreevy:2022oyu,Kaidi:2021gbs,Inamura:2021szw,Vanhove:2021zop,Huang:2021nvb,Huang:2021ytb}. Crucially, the corresponding operators are typically non-local, in the sense that they cannot be written as tensor products of local operators, and are realized instead by \emph{matrix product operators} (MPOs) \cite{PhysRevB.79.085119,Buerschaper:2013nga,bultinckAnyons,bultinckMPOs,Bultinck_2017,hauruDefects,Lootens:2020mso}.
Although exotic, such categorical symmetries are not uncommon in one-dimensional quantum models and are typically related to rational Conformal Field Theories (CFTs) \cite{PhysRevLett.84.1659,Frohlich:2006ch,Fuchs:2007tx,hauruDefects,aasenDefects,Aasen:2020jwb,PhysRevLett.121.177203}. For instance, the Kramers-Wannier duality defect of the Ising CFT provides such a categorical symmetry operator for the critical transverse-field Ising model \cite{PhysRev.60.252}. Moreover, a large family of lattice models known as \emph{anyonic chains} that commute with symmetry operators organized into fusion categories can be readily constructed \cite{PhysRevLett.98.160409,PhysRevB.87.235120,PhysRevLett.101.050401,PhysRevLett.103.070401,ardonneAnyonChain,Buican:2017rxc,Finch_2013,PhysRevB.94.085138}, including spin systems with quantum group symmetries \cite{pasquier1990common,couvreur2022matrix}. 

In virtue of their topological nature, any categorically symmetric model in (1+1)d can be lifted to a (gapped) boundary condition of the \emph{Turaev-Viro-Barrett-Westbury} topological quantum field theory (TQFT) \cite{Turaev:1992hq,Barrett:1993ab} with input datum the corresponding spherical fusion category \cite{Thorngren:2019iar,Thorngren:2021yso,Lin:2022dhv}. Mathematically, gapped boundary conditions admit a classification in terms of \emph{module categories} over the input category \cite{Fuchs:2012dt,kongBdries,KONG2014436,KONG201762,Freed:2020qfy}, the case of pure gauge theories having received special attention \cite{Fuchs:2013gha,PhysRevB.102.045139,beigi2011quantum,PhysRevLett.119.170504,Cong:2017hcl,PhysRevB.96.165138,Hu:2017faw,Bullivant:2020xhy}. This \emph{holographic} viewpoint on symmetries has also garnered a lot of interest \cite{PhysRevResearch.2.033417,Albert:2021vts,PhysRevResearch.2.043086,PhysRevB.104.075141,Moradi:2022lqp}. Crucially, the bulk TQFT can be reconstructed from any choice of gapped boundary condition, so that bulk topological lines are encoded into the \emph{Drinfel'd center} of the corresponding spherical fusion category of (boundary) topological lines \cite{Freed:2020qfy}. This suggests a notion of duality between models canonically associated with distinct boundary conditions of the same bounding TQFT.

\bigskip \noindent
Inspired by these developments, we initiated in ref.~\cite{Lootens:2021tet} a systematic study of dualities in one-dimensional quantum lattice models from the viewpoint of their (categorical) symmetries. One merit of our approach is to make very concrete the concepts and results alluded to above---which are often formal and abstract otherwise---as well as demonstrate that this approach to dualities agrees and extends traditional ones \cite{bondAlgebra}.
Within our framework, an equivalence class of dual models is given by a choice of input (spherical) fusion category together with an algebra of local operators. A representative of such a class then corresponds to a specific lattice realization of the underlying theory. Choosing a lattice realization loosely boils down to picking a collection of degrees of freedom, which happen to be encoded into a choice of module category over the input fusion category. This means that models that only differ in a choice of module category are dual to one another. Importantly, dualities thus defined are such that  any symmetric local operator is mapped to a dual symmetric local operator, whereas non-symmetric local operators in one theory are mapped to non-local non-symmetric operators in the dual theory. Generally speaking, we can interpret such dualities as arising from some generalized \emph{gauging} procedure of the categorical symmetry \cite{Freed:2018cec,CDMod2,weyl1929elektron,PhysRevD.11.395,RevModPhys.51.659, Fisher2004, RevModPhys.52.453,Buerschaper:2010yf,doi:10.1063/1.1665530,Tachikawa:2017gyf,Bhardwaj:2017xup}. It follows from our construction that duality transformations are naturally associated with maps between module categories.

Practically, given a known one-dimensional quantum lattice model, a suitable choice of input fusion category can be extracted from a detailed understanding of its symmetries---which are typically generated by non-local operators. The Hamiltonian itself is then built from linear combinations of symmetric local operators obtained from the data of a module category over the input fusion category. The algebra entering the characterization of the equivalence class of dual models is that generated by the set of local operators. Keeping the same linear combination of symmetric operators, but choosing a different realization via a choice of module category, yields a dual model \cite{Lootens:2021tet}.

A key technical novelty of our approach is our ability to explicitly write down, in the form of matrix product operators \cite{pirvu2010matrix,haegeman2017diagonalizing,cirac2020matrix}, the non-local lattice operators generating the symmetries of a given family of dual models. Similarly, we are able to implement a duality relation via MPOs that transmute the local operators of a given Hamiltonian into those of one of its duals \cite{Lootens:2021tet}. The main teaching of the present manuscript is that these MPOs can be further exploited so as to construct \emph{isometries} relating the full spectra of dual models. A crucial aspect of such a mapping is the delicate interplay between duality relations and \emph{sectors} of the models that come into play.

Indeed, the study of sectors cannot be dissociated from the constructions of duality mappings. Consider for instance the Kramers-Wannier duality of the transverse field Ising model \cite{PhysRev.60.252}. Given a closed chain, one formulation  of this duality identifies the simultaneous action of two Pauli $Z$ operators on qubits located at neighbouring vertices with that of a single Pauli $Z$ operator on a qubit located along the edge bounding these vertices, and vice versa for Pauli $X$ operators. It turns out that such a duality mapping imposes \emph{kinematical constraints} for both the original Hamiltonian and its dual. Indeed, it follows from the definition that acting simultaneously on all qubits with Pauli $X$ or Pauli $Z$ operators, respectively, must leave every state invariant. In other words, these kinematical constraints force both models into the even charge sectors of their respective $\mathbb Z_2$-symmetry. This means in particular that the symmetry cannot be spontaneously broken on either side of the duality \cite{BenTov:2014eea,Radicevic:2018okd}. Accessing the odd charge sector of the original model requires locally altering the duality mapping, which in turn modifies the boundary condition of the dual model from periodic to antiperiodic. As such, it is not possible to define a mapping of local operators without addressing the mapping of sectors.

The purpose of the present manuscript is to completely address the fate of sectors upon dualizing for the case of \emph{closed boundary conditions}. Here, closed boundary conditions include the familiar case of periodic boundary conditions, but more generally also contains \emph{symmetry-twisted} boundary conditions \cite{PETKOVA2001157,FUCHS2002353}. These boundary conditions are special in the sense that while they do break invariance with respect to the original translation operator, they do so in a way allowing us to define a \emph{twisted} translation operator, together with corresponding twisted momentum, with respect to which invariance is preserved. For these kinds of boundary conditions, sectors are labelled by combinations of \emph{fluxes} given by symmetry twists and symmetry \emph{charges} that decompose the remaining twisted symmetry. Borrowing terminology from the study of topological order, we refer to these super-selection sectors labelled by symmetry-twisted boundary conditions and twisted symmetry sectors as \emph{topological sectors} \cite{Aasen:2017ubm,PhysRevLett.121.177203,Aasen:2020jwb,Lin:2022dhv}.

Recently there has been renewed interest in the physical realization of duality transformations and their applications in quantum technologies. In the context of quantum simulation, the fact that dualities typically change the phase of the states on which they act can be exploited to efficiently prepare states in a given phase \cite{Verresen:2021wdv,Tantivasadakarn:2021vel,Ashkenazi:2021ieg,Tantivasadakarn:2022hgp}. A prototypical example is the case of the Kramers-Wannier duality, where the duality MPO generates a long-range entangled GHZ state from a trivially entangled product state. The same duality transformations can also be used to generate quantum circuits that permute the anyons of a topologically ordered state \cite{PhysRevB.105.085130}, which has applications in the construction of topological quantum memories \cite{PhysRevB.106.085122}. Both of these applications ultimately rely on an understanding of the non-trivial action of a duality on the topological sectors of a model, as well as the explicit realization of operators that implement these transformations. The framework we present in this manuscript provides both of these features, and we expect the duality operators we construct here to guide the physical realization of more general duality transformations.

\bigskip \noindent
We will make use of several categorical concepts, which for convenience are summarized in tab.~\ref{tab:dictionary} together with their physical meaning in this work. Concretely, our construction goes as follows: We begin by choosing an input fusion category $\mc D$, from which we can construct an (abstract) algebra of local operators referred to as the \emph{bond algebra} \cite{bondAlgebra,cobanera2010unified}. At this stage, these local operators do not yet admit an explicit matrix representation on a Hilbert space. Instead, they are written in terms of \emph{string diagrams}, which allows us to compute their operator products using the diagrammatic manipulations of the input category. A particular equivalence class of dual models is then built by taking certain linear combinations of such local operators. The choice of these operators completely determines the spectrum of these models, and in this sense captures the physical properties that can be directly inferred from the spectrum. The next step is to find explicit matrix representations of the local operators that build up these models, which then specify the Hilbert space and the Hamiltonian. These are classified by different choices of module categories $\mc M$ over the input category $\mc D$. The choice of module category therefore provides a particular physical realization of the physical properties captured in the spectrum associated with an equivalence class of dual models.

\begin{table}
    \begin{align*}
        \setlength{\tabcolsep}{0.6em}
        \begin{tabularx}{0.95\columnwidth}{l|l}
             Fusion category $\mc D$ & Abstract algebra of operators\\
             \midrule
             $\mc D$-module category $\mc M$ & Degrees of freedom\\
             \midrule
             $\mc D$-module functors & Duality operators\\
             \midrule
             $\mc D$-module endofunctors & Symmetry operators\\
             \midrule
             Drinfel'd center & Topological sectors
        \end{tabularx}
    \end{align*}
    \caption{Dictionary between categorical concepts and their role in this work.}
    \label{tab:dictionary}
\end{table}

Ignoring boundary conditions by considering infinite chains, one can construct operators that intertwine between dual representations of the local operators determined by different choices of (indecomposable) module categories $\mc M$ and $\mc N$. Explicitly, such an operator can be written in terms of an MPO intertwiner that acts as a map between module categories; the consistency conditions on this MPO intertwiner are equivalent to those of a so-called $\mc D$-module functor $\mc M \to \mc N$ in $\Fun_{\mc D}(\mc M,\mc N)$. In the special case where the module categories are the same on both sides, we interpret the MPO as a symmetry, which is labeled by a $\mc D$-module endofunctor in $\Fun_{\mc D}(\mc M,\mc M)$. Importantly, the composition of $\mc D$-module endofunctors endows the category of endofunctors with the structure of a fusion category. This fusion category is denoted as $\mc D_{\mc M}^\star$ and referred to as the \emph{Morita dual} of $\mc D$ with respect to $\mc M$; it describes the symmetries of the Hamiltonian associated with the specific choice of module category $\mc M$.

As alluded to above, in order to realize these dualities as explicit isometries one needs to carefully consider the boundary conditions. In our setup, the symmetry-twisted boundary conditions discussed above are given by an endofunctor of the module category, which tells us the way in which the degrees of freedom at either end of the chain have to be glued together. The additional condition that translation invariance is preserved up to a local unitary transformation is satisfied by requiring this endofunctor to possess a $\mc D$-module structure. As expected , these are organized into the same fusion category $\mc D_{\mc M}^\star$ that describes the symmetries, hence the name. The symmetries of a model interact with these boundary conditions and typically one is left with less symmetry in the presence of a non-trivial symmetry twist. In general, this can be understood using `tubes', which are modifications of the usual periodic MPOs to include the action on the symmetry twist. The topological sectors of the model are then characterized as irreducible representations of the category of tubes, which are well known to be in correspondence with simple objects in the Drinfel'd center $\mc Z(\mc D^\star_{\mc M})$ \cite{Aasen:2020jwb,Albert:2021vts,Xu:2022rtj,Lin:2022dhv,Kong:2022hjj}. We confirm this directly on the lattice and explicitly compute the projectors onto the various topological sectors.

Importantly, the symmetry fusion categories $\mc D_{\mc M}^\star$ for different choices of $\mc M$ are all \emph{Morita equivalent} \cite{MUGER200381,MUGER2003159}. It is well known that this guarantees that the centers $\mc Z(\mc D^\star_{\mc M})$ are equivalent as categories, which guarantees that the topological sectors of dual models can be mapped to one another. By generalizing the MPO intertwiners constructed for the infinite case to accommodate the twisted boundary conditions, we can explicitly compute these maps and construct the isometries that relate the full set of eigenvectors of one model to those of any of its dual models.

\bigskip \noindent
We illustrate our construction with a family of dual models whose fusion categories of symmetry operators are in the Morita class of the fusion category $\Rep(\mc S_3)$ of representations of the symmetric group $\mc S_3$. We distinguish four models associated with the four indecomposable module categories over $\Rep(\mc S_3)$. One of them is the spin-$\frac{1}{2}$ XXZ Heisenberg model. In the infinite case, the Hamiltonian commutes in particular with non-local operators labelled by group elements in $\mc S_3$. Similarly, equivalence classes of boundary conditions that preserves the translation invariance of the model are shown to be labelled by conjugacy classes of $\mc S_3$. Given a conjugacy class and one of its representatives, the corresponding boundary condition is such that the resulting Hamiltonian only commutes with operators labelled by group elements in the centralizer of the representative. It follows that this Hamiltonian decomposes into twisted symmetry sectors indexed by irreducible representations of the centralizer. Putting everything together, we find eight topological sectors, which are in one-to-one correspondence with simple objects in the Drinfel'd center $\mc Z(\Vect_{\mc S_3})$ of the fusion category $\Vect_{\mc S_3}$ of $\mc S_3$-graded vector spaces, or equivalently irreducible representations of the quantum double $\mc D(\mc S_3)$. These simple objects also encode elementary anyonic excitations in Hamiltonian realizations of the $\mc S_3$ Dijkgraaf-Witten theory. As evoked above, this is no mere coincidence and confirms the fact that the spin-$\frac{1}{2}$ XXZ Heisenberg model can indeed arise as effective theory on the boundary of a (2+1)d topological model with $\mc Z(\Vect_{\mc S_3})$ topological order.   

By construction, Hamiltonian models that only differ from the spin-$\frac{1}{2}$ XXZ Heisenberg model in the choice of module category over $\Rep(\mc S_3)$ are dual to it, and also dual to one another. We explore these dual models within our framework. We highlight in particular a model whose Hamiltonian commutes in the infinite case with non-local symmetry operators labelled by irreducible representation of $\mc S_3$, as opposed to group elements. Correspondingly, boundary conditions are also labelled by irreducible representations. We explicitly construct the boundary terms as well as the lattice symmetry operators that preserve the boundary conditions. The symmetry charge sectors decomposing the corresponding Hamiltonians are presented in detail. As for the XXZ Heisenberg model, we find a total of eight topological sectors, which are now in one-to-one correspondence with simple objects of the Drinfel'd center $\mc Z(\Rep(\mc S_3))$. Duality is then guaranteed by the Morita equivalence between $\Vect_{\mc S_3}$ and $\Rep(\mc S_3)$. Furthermore, we study in detail how topological sectors of this model are mapped to those of the XXZ Heisenberg model upon duality. We show in particular how the two-dimensional charge sector of the XXZ Heisenberg model with periodic boundary conditions is mapped onto the trivial charge sector of the dual model with a non-abelian boundary condition. This is a concrete physical realization of the non-trivial permutation of anyons taking place at the only non-trivial invertible domain wall between topological orders described by $\mc Z(\Vect_{\mc S_3})$ and $\mc Z(\Rep(\mc S_3))$, respectively.

\bigskip \noindent
The manuscript is organized as follows: We begin by reviewing in sec.~\ref{sec:infinite} the case of infinite chains with an emphasis on the lattice duality operators transmuting local symmetric operators into one another. This section is also the opportunity to introduce the relevant technical preliminaries together with the graphical calculus that is used throughout our work. The characterization of topological sectors is presented in sec.~\ref{sec:topoSectors} together with the operators mapping sectors of dual models onto one another, as well as the isometries realizing the dualities between the Hamiltonians. Finally, we present sec.~\ref{sec:Examples} several examples that illustrate the various results obtained in this work. Our manuscript is complemented by app.~\ref{sec:app_Morita} and app.~\ref{sec:app_double} that compile numerous results regarding categorical Morita equivalence and the quantum double of a finite group, respectively.

\section{Infinite chains\label{sec:infinite}}

\noindent
\emph{After introducing some technical definitions, we review in this section our systematic and constructive approach to symmetry and duality operators for infinite one-dimensional lattice models.}

\subsection{Technical preliminaries\label{sec:prelim}}

\noindent
We consider in this manuscript families of one-dimensional quantum lattice models that are \emph{dual} to one another. These dual models are characterized by distinct lattice realizations of a (categorical) symmetry generated by operators that are organized into a (spherical) fusion category. We like to think of choosing a fusion category as picking a \emph{backbone} that supports the various theories. Lattice realizations are then obtained by choosing collections of physical degrees of freedom that are compatible with the backbone. Mathematically, a choice of lattice realization is associated with a module category over the input fusion category. In this context, a map between distinct module categories amounts to a duality operator, whereas a map from a module category to itself amounts to a symmetry operator. This is the program that was presented in ref.~\cite{Lootens:2021tet} in the case of infinite one-dimensional lattices. The purpose of the present manuscript is to continue this systematic study and explain the subtle interplay between duality relations, boundary conditions and topological sectors. Let us first briefly review the infinite case. 

\medskip \noindent
We set the stage by introducing the technical ingredients alluded to above. We encourage the reader to consult ref.~\cite{etingof2016tensor} for details. Succinctly, a fusion category $\mc D$ encodes a collection of objects interpreted as (possibly non-elementary) topological charges that can be fused to one another. Throughout this manuscript, we notate via $Y_1,Y_2,\ldots \in \mc I_\mc D$ representatives of isomorphism classes of \emph{simple} objects in $\mc D$ and their respective quantum dimensions via $d_{Y_1},d_{Y_2},\ldots \in \mathbb C$. The fusion of objects is encoded into the \emph{monoidal} structure $(\otimes,\mathbb 1,F)$ of $\mc D$ consisting a product rule $\otimes : \mc D \times \mc D \to \mc D$, a distinguished object $\mathbb 1$ referred to as the trivial charge or \emph{vacuum}, and a (natural) isomorphism $F : (- \otimes -) \otimes - \xrightarrow{\sim} - \otimes (- \otimes -)$ satisfying a coherence relation known as a `pentagon axiom'. The isomorphism $F$ shall be referred to as the \emph{monoidal associator} and its components written as $F^{Y_1Y_2Y_3}: ( Y_1 \otimes Y_2) \otimes Y_3 \xrightarrow{\sim} Y_1 \otimes (Y_2 \otimes Y_3)$ for all $Y_1,Y_2,Y_3 \in \mc I_\mc D$. Introducing the notation $\mc H^{Y_3}_{Y_1Y_2} := \Hom_{\mc D}(Y_1 \otimes Y_2,Y_3) \ni |Y_1Y_2Y_3,i \ra$ for the vector space of maps from the (typically not simple) object $Y_1 \otimes Y_2$ to $Y_3$, we have $Y_1 \otimes Y_2 \simeq \bigoplus_{Y_3}N^{Y_3}_{Y_1Y_2} \, Y_3$, where $N^{Y_3}_{Y_1Y_2} := {\rm dim}_{\mathbb C} \, \mc H^{Y_3}_{Y_1Y_2} \in \mathbb N$. It follows that the monoidal associator boils down to a collection of complex matrices
\begin{equation}
    \F{}^{Y_1Y_2Y_3}_{Y_4}: 
    \bigoplus_{Y_5}\mc H_{Y_1Y_2}^{Y_5} \otimes \mc H_{Y_5Y_3}^{Y_4}
    \xrightarrow{\sim} 
    \bigoplus_{Y_6}\mc H_{Y_2Y_3}^{Y_6} \otimes \mc H_{Y_1Y_6}^{Y_4} 
    \label{eq:Fdef}
\end{equation}
whose actions can be conveniently depicted in terms of \emph{string diagrams} as 
\begin{equation}
    \label{eq:monoAssociativity}
    \includeTikz{0}{monoidalAssociator-1}{\monoidalAssociator{1}} \!\!\!\!\!\! 
    = \sum_{Y_6} \sum_{j,l}
    \big( \F{}^{Y_1Y_2Y_3}_{Y_4}\big)^{Y_6,lj}_{Y_5,ik}
    \!\! \includeTikz{0}{monoidalAssociator-2}{\monoidalAssociator{2}} ,
\end{equation}
where the indices $i,j,k,l$ label basis vectors in the hom-spaces $\mc H^{Y_5}_{Y_1Y_2}$, $\mc H^{Y_4}_{Y_1Y_6}$, $\mc H^{Y_4}_{Y_5Y_3}$ and $\mc H^{Y_6}_{Y_2Y_3}$, respectively. Henceforth, we refer to the entries of these complex matrices as the \emph{$F$-symbols}.

Given a fusion category $\mc D$, a  module category $\mc M$ over it is roughly speaking a collection of representatives of simple objects $M_1,M_2,\ldots \in \mc I_\mc M$ that are acted upon by the objects in $\mc D$. More concretely, a \emph{right} module category $\mc M$ over $\mc D$ is a triple $(\mc M,\cat,\F{\cat})$ that consists of a category $\mc M$, an action $\cat: \mc M \times \mc D \to \mc M$ and a (natural) isomorphism $\F{\cat}: (- \cat -) \cat - \xrightarrow{\sim} - \cat (- \otimes -)$ satisfying a `pentagon axiom' involving the monoidal associator $F$ of $\mc D$.
The isomorphism $\F{\cat}$ shall be referred to as the (right) \emph{module associator} and its components written as $\F{\cat}^{M Y_1Y_2}:  (M \cat Y_1) \cat Y_2 \xrightarrow{\sim} M \cat (Y_1 \otimes Y_2)$ for all $Y_1,Y_2 \in \mc I_\mc D$ and $M \in \mc I_\mc M$. For instance, every fusion category $\mc D$ has the structure of a module category over itself via its monoidal structure known as the \emph{regular} module category, whereby the module associator $\F{\cat}$ is provided by the monoidal associator $F$. Introducing the notation $\mc V^{M_2}_{M_1Y} := \Hom_{\mc M}(M_1 \cat Y,M_2) \ni |M_1YM_2,i \ra$ for the vector space of maps from the (typically not simple) object $M_1 \cat Y$ to $M_2$, we have $M_1 \cat  Y \simeq \bigoplus_{M_2}N^{M_2}_{M_1 Y} \, M_2$, where $N^{M_2}_{M_1 Y} := {\rm dim}_{\mathbb C} \, \mc V^{M_2}_{M_1 Y} \in \mathbb N$. It follows that the module associator boils down to a collection of complex matrices
\begin{equation}
    \F{\cat}^{M_1 Y_1Y_2}_{M_2}: 
    \bigoplus_{M_3}\mc V_{M_1 Y_1}^{M_3} \otimes \mc V_{M_3 Y_2}^{M_2}
    \xrightarrow{\sim} 
    \bigoplus_{Y_3}\mc H_{Y_1Y_2}^{Y_3} \otimes \mc V_{M_1 Y_3}^{M_2}
\end{equation}
whose actions can be depicted in terms of string diagrams as
\begin{equation}
    \label{eq:McatMove}
    \includeTikz{0}{moduleAssociator-1}{\moduleAssociator{1}} \!\!\!\!\!\! 
    = \sum_{Y_3} \sum_{j,l}
    \big( \F{\cat}^{M_1Y_1Y_2}_{M_2}\big)^{Y_3,lj}_{M_3,ik}
    \;\; 
    \includeTikz{0}{moduleAssociator-2}{\moduleAssociator{2}} ,
\end{equation}
where the indices $i,j,k,l$ label basis vectors in the hom-spaces $\mc V_{M_1Y_1}^{M_3}$, $\mc V_{M_1 Y_3}^{M_2}$, $\mc V_{M_3 Y_2}^{M_2}$ and $\mc H_{Y_1 Y_2}^{Y_3}$, respectively. Henceforth, we refer to the entries of these complex matrices as $\F{\cat}$-symbols.

Mirroring the concepts above, we can define a notion of \emph{left} module category $(\mc M, \act, \F{\act})$. Combining the notions of left and right module category then yields the concept of \emph{bimodule category}. Concretely, a $(\mc C,\mc D)$-bimodule category is a sextuple $(\mc M,\act,\cat, \F{\act},\F{\cat},\F{\caact})$ such that the triples $(\mc M,\act,\F{\act})$ and $(\mc M,\cat,\F{\cat})$ define left $\mc C$- and right $\mc D$-module categories, respectively, and where $\F{\caact} :  (- \act -) \cat - \xrightarrow{\sim} - \act (- \cat -)$ is a (natural) isomorphism satisfying two `pentagon axioms' involving either $\F{\cat}$ or $\F{\act}$. The isomorphism $\F{\caact}$ shall be referred to as the \emph{bimodule associator} and its components written as $\F{\caact}^{XM_1Y} : (X \act M_1) \cat Y \xrightarrow{\sim} X \act (M_1 \cat Y)$ for all $X \in \mc I_\mc C$, $Y \in \mc I_\mc D$ and $M_1 \in \mc I_\mc M$. As before, the bimodule associator boils down to a collection of complex matrices
\begin{equation}
    \F{\caact}^{XM_1Y}_{M_2}:
    \bigoplus_{M_3}\mc V_{XM_1}^{M_3} \otimes \mc V_{M_3Y}^{M_2}
    \xrightarrow{\sim} 
    \bigoplus_{M_4}\mc V_{M_1Y}^{M_4} \otimes \mc V_{XM_4}^{M_2}
\end{equation}
whose actions can be depicted in terms of string diagrams as
\begin{equation}
    \includeTikz{0}{BimoduleAssociator-2}{\BimoduleAssociator{2}} \!\!\!\!
    = \sum_{M_4} \sum_{j,l}
    \big( \F{\caact}^{XM_1Y}_{M_2}\big)^{M_4,lj}_{M_3,ik} 
    \includeTikz{0}{BimoduleAssociator-1}{\BimoduleAssociator{1}} ,
\end{equation}
where the indices $i,j,k,l$ label basis vectors in the hom-spaces $\mc V_{XM_1}^{M_3}$, $\mc V_{XM_4}^{M_2}$, $\mc V_{M_3 Y}^{M_2}$ and $\mc V_{M_1Y}^{M_4}$, respectively. Henceforth, we refer to the entries of these complex matrices as $\F{\caact}$-symbols.

The final ingredient we require is a notion of structure-preserving map between module categories over the same fusion category. 
Given two (right) module $\mc D$-categories $(\mc M,\cat,\F{\cat})$ and $(\mc N, \catb, \F{\catb})$, we define a $\mc D$\emph{-module functor} between them as a pair $(\fr F,\omega)$ that consists of a functor $\fr F: \mc M \to \mc N$ and a (natural) isomorphism $\omega : \fr F(- \cat -) \xrightarrow{\sim} \fr F(-) \catb -$ satisfying a `pentagon axiom' involving both $\F{\cat}$ and $\F{\catb}$. Components of the isomorphism $\omega$ shall be written as $\omega^{M_1Y} :  \fr F(M_1 \cat Y) \xrightarrow{\sim} \fr F(M_1) \catb Y$ for all $Y \in \mc I_\mc D$ and $M_1 \in \mc I_{\mc M}$, and boil to a collection of complex matrices
\begin{equation}
    \label{eq:ModFunMat}
    \omega^{M_1Y}_{N_1}  : 
    \bigoplus_{N_1} \mc V^{N_1}_{M_1} \otimes \mc V^{N_1}_{N_2Y} 
    \xrightarrow{\sim}
    \bigoplus_{M_2} \mc V^{M_2}_{M_1Y} \otimes \mc V^{N_2}_{M_2}\, , 
\end{equation}
where $\mc V_{M}^N := \Hom_{\mc N}(\fr F(M),N)$.
Such module functors form a category denoted by $\Fun_{\mc D}(\mc M,\mc N)$. Throughout this manuscript, we notate via $X_1,X_2,\ldots$ representatives of isomorphism classes of simple objects in such categories of module functors. Moreover, we shall typically refer to a given module functor as an object $X$ in the corresponding category, in which case the actual functor shall be denoted by $\Func{X}$ and the module structure by $\omF{X}$. In other words, we employ the shorthand notation $X \equiv (\Func{X},\omF{X})$.
Keeping this convention in mind, the action of the matrices \eqref{eq:ModFunMat} can also be depicted in terms of string diagrams as
\begin{equation}
    \label{eq:FcatMove}
    \includeTikz{0}{moduleFunctor-2}{\moduleFunctor{2}} \!\!\!\!\!\! 
    = \sum_{M_2} \sum_{j,l}
    \big( \omF{X}^{M_1Y}_{N_2}\big)^{M_2,lj}_{N_1,ik}
    \includeTikz{0}{moduleFunctor-1}{\moduleFunctor{1}}  ,
\end{equation}
where the indices $i,j,k,l$ label basis vectors in the hom-spaces $\mc V_{M_1}^{N_1}$, $\mc V_{M_2}^{N_2}$, $\mc V_{N_1 Y}^{N_2}$,  and $\mc V_{M_1Y}^{M_2}$, respectively. Henceforth, we refer to the entries of these complex matrices as $\omF{X}$-symbols. Notice that we are performing a slight abuse of notation so as to have string diagrams akin to those associated with bimodule associators. The reason is the following: We shall often consider $\mc D$-module endofunctors in categories $\mc D^\star_\mc M := \Fun_{\mc D}(\mc M,\mc M)$. But every category $\mc D^\star_{\mc M}$ can be equipped with a monoidal structure via  the \emph{composition} of module functors. As a matter of fact, since we shall focus on cases where $\mc M$ is \emph{indecomposable}, $\mc D^\star_\mc M$ even has the structure of a fusion category \cite{etingof2016tensor}. 
Moreover, the category $\mc M$ is naturally endowed with the structure of a \emph{left} module category over $\mc D^\star_\mc M$. Indeed, notice that every $\mc D$-module functor in $\Fun_{\mc D}(\mc D, \mc M)$ is of the form $(- \cat M,\F{\cat}^{M--})$ with $M$ any object in $\mc M$, establishing the equivalence $\mc M \cong \Fun_\mc D(\mc D,\mc M)$. But $\Fun_{\mc D}(\mc D,\mc M)$ has the structure of a left $\mc D^\star_{\mc M}$-module category via the composition $\mc D^\star_\mc M  \times \Fun_{\mc D}(\mc D,\mc M) \to \Fun_{\mc D}(\mc D,\mc M)$ of $\mc D$-module functors. It follows that $\mc M$ has the structure of a $(\mc D^\star_\mc M,\mc D)$-bimodule category. In this context, the module functor structure $\omF{X}^{--}$ of an object $X$ in $\mc D^\star_{\mc M}$ coincides with the bimodule associator $\F{\caact}^{X--}$ of $\mc M$ as an \emph{invertible} $(\mc D^\star_{\mc M},\mc D)$-bimodule category. 

Given three (right) module $\mc D$-categories $\mc M$, $\mc N$ and $\mc O$,  $\mc D$-module functors in $\Fun_{\mc D}(\mc N,\mc O)$ and $\Fun_{\mc D}(\mc M,\mc N)$ can be composed so as to yield module functors in $\Fun_{\mc D}(\mc M,\mc O)$. Given representatives $X_1$ and $X_2$ of isomorphism classes of simple objects in $\Fun_{\mc D}(\mc N,\mc O)$ and $\Fun_{\mc D}(\mc M,\mc N)$, respectively, composition yields a new functor $\Func{X_1}(\Func{X_2}(-))$ in $\Fun_{\mc D}(\mc M,\mc O)$ with the obvious module structure. Crucially, even though $X_1$ and $X_2$ are simple, the composite of the corresponding module functors is typically not a simple object in $\Fun_{\mc D}(\mc M,\mc O)$. Therefore, there exist a collection of complex matrices whose entries can be defined graphically following the convention of eq.~\eqref{eq:FcatMove} as
\begin{equation}
    \label{eq:compFunctor}
    \includeTikz{0}{moduleFunctorComp-2}{\moduleFunctorComp{2}}  
    = \sum_{N} \sum_{j,l}
    \big( \F{\fr F}^{X_1X_2M}_{O}\big)^{N,lj}_{X_3,ik}
    \!\!\!\!\!\!
    \includeTikz{0}{moduleFunctorComp-1}{\moduleFunctorComp{1}} ,
\end{equation}
where the indices $i,j,k,l$ label basis vectors in the hom-spaces $\mc V_{X_1 X_2}^{X_3} := \Hom_{\Fun_{\mc D}(\mc M,\mc O)}(\Func{X_1}(\Func{X_2}(-)),\Func{X_3}(-))$, $\mc V^O_N$, $\mc V^O_M$ and $\mc V^N_M$, respectively. Henceforth, we refer to the entries of these complex matrices as $\F{\fr F}$-symbols. All the symbols introduced so far fulfill consistency conditions descending from the pentagon axioms satisfied by the corresponding isomorphisms. Writing an analogous consistency condition for the composition of module functors in terms of these $\F{\fr F}$-symbols would require the introduction of yet another set of symbols referred to as $\F{\circ}$-symbols, thereby defining these new symbols in terms of $\F{\fr F}$-symbols. With our notations, we would have
\begin{equation}
    \includeTikz{0}{moduleFunctorAssoc-2}{\moduleFunctorAssoc{2}}  
    \! = \sum_{X_6} \sum_{j,l}
    \big( \F{\circ}^{X_1X_2X_3}_{X_4}\big)^{X_6,lj}_{X_5,ik}
    \hspace{-3em}
    \includeTikz{0}{moduleFunctorAssoc-1}{\moduleFunctorAssoc{1}} ,
\end{equation}
so that $\F{\circ}$-symbols encode the associativity condition for the composition of module functors. But such a consistency condition may not exist in general. In other words, there may be an obstruction for the composition of module functors to be associative up to an isomorphism \cite{etingof2010fusion}. That being said, such obstructions do not occur within our construction due to the specific compositions of module functors considered. In addition to this pentagon axiom defining the associativity of the the composition of module functors, note that the $\F{\fr F}$-symbols are involved in a coherence relation involving $\omF{-}$-symbols (see eq.~\eqref{eq:fusionMPO} below). 

We defined in this section seven sets of symbols, namely $\F{}$-, $\F{\act}$-, $\F{\cat}$-, $\F{\caact}$-, $\omF{-}$-, $\F{\fr F}$- and $\F{\circ}$-symbols. Note that, in general, there are neither explicit formulas nor direct ways to compute these various symbols. Rather, they are obtained by solving the consistency conditions they are required to satisfy. This implies in particular that these symbols are typically defined up to basis choices for the various hom-spaces. The results presented in this manuscript hold regardless of these choices, and as such we shall implicitly choose them so as to simplify the values of the symbols considered.\footnote{For instance, we work in a basis where $F^{Y_1Y_2Y_3}_{Y_4}$ evaluates to the identity matrix whenever $Y_1,Y_2$ or $Y_3$ is the unit object; we make similar choices for the other symbols in this manuscript.} Besides, it is not necessary to solve for all these symbols individually. Indeed, it is possible to deduce the $\F{\caact}$-, $\F{\act}$-, $\omF{-}$-, $\F{\fr F}$- and $\F{\circ}$-symbols from the knowledge of the $\F{}$- and $\F{\cat}$-symbols for every possible choice of indecomposable (right) module category over $\mc D$. We include this data for the example considered below where $\mc D = \Rep(\mc S_3)$ as supplementary material.

\subsection{Tensor networks}

\noindent
A key aspect of our construction is the use of \emph{tensor networks} \cite{cirac2020matrix}, as a way to parameterize lattice models as well as their symmetry and duality operators. As we shall emphasize, this language is not only very natural but quickly becomes necessary as we consider non-elementary models, and translating the tensor networks into more explicit or familiar objects often turns out be a tedious exercise. We hope that the new results presented in sec.~\ref{sec:Examples} will convince the most skeptical readers of the benefits of this approach.

The types of tensor networks we consider build upon the graphical calculus of string diagrams, which was briefly employed above. First of all, the $\F{\cat}$-symbols associated with a (right) $\mc D$-module category $\mc M$ can be depicted as
\begin{align}
    \label{eq:PEPS}
    \includeTikz{0}{PEPS-mod-i-l-j-k-M_1-M_2-M_3-Y_1-Y_2-Y_3-1}{\PEPS{mod}{i}{l}{j}{k}{M_1}{M_2}{M_3}{Y_1}{Y_2}{Y_3}{1}} :=
    &\big( \F{\cat}^{M_1 Y_1 Y_2}_{M_2}\big)^{Y_3,lj}_{M_3,ik} \, ,
    \\[-3em] \nn
    &\big( \Fbar{\cat}^{M_1 Y_1 Y_2}_{M_2}\big)^{Y_3,lj}_{M_3,ik} =:
    \includeTikz{0}{PEPS-mod-i-l-j-k-M_1-M_2-M_3-Y_1-Y_2-Y_3-2}{\PEPS{mod}{i}{l}{j}{k}{M_1}{M_2}{M_3}{Y_1}{Y_2}{Y_3}{2}} 
    \, ,
\end{align}
where we should think of the first diagram as the pasting of the string diagrams appearing on the l.h.s. and r.h.s. of eq.~\eqref{eq:FcatMove} into a tetrahedron. By convention, $\F{\cat}$-symbols for which the fusion rules are not everywhere satisfied vanish. We can now construct tensors whose non-vanishing entries are provided by these $\F{\cat}$-symbols. We shall do so graphically. Let us first introduce a couple of graphical conventions. Unlabelled purple strings denote the following morphism:
\begin{equation}
    \label{eq:convModStrand}
    \includeTikz{0}{object-mod}{\object{mod}{}} \equiv \sum_{M \in \mc I_{\mc M}} \includeTikz{0}{object-mod-M}{\object{mod}{M}} \in \End_{\mc M} \big(\!\! \bigoplus_{M \in \mc I_{\mc M}}\!\!\! M \big) \, ,
\end{equation}
where the relevant module category will always be clear from the context.
In the same spirit, unlabelled gray patches denote the following formal vectors:
\begin{equation}
    \begin{split}
        \label{eq:unlabelledPatch}
        \includeTikz{0}{splitSpace-M_1-Y-M_2-1}{\splitSpace{M_1}{Y}{M_2}{}{1}} &\equiv
        \sum_{i}
        \includeTikz{0}{splitSpace-M_1-Y-M_2-i-1}{\splitSpace{M_1}{Y}{M_2}{i}{1}}
        |M_1YM_2, i \ra \, , 
        \\
        \includeTikz{0}{splitSpace-M_1-Y-M_2-2}{\splitSpace{M_1}{Y}{M_2}{}{2}} &\equiv
        \sum_{i}
        \includeTikz{0}{splitSpace-M_1-Y-M_2-i-2}{\splitSpace{M_1}{Y}{M_2}{i}{2}} \la M_1YM_2, i | \, . 
    \end{split}
\end{equation}
Putting these graphical conventions together, we define the following collection of tensors labelled by simple objects in $\mc D$ and a choice of basis vector:
\begin{align}
    \label{eq:PEPSFcat}
    \includeTikz{0}{PEPS-mod-l-Y_1-Y_2-Y_3-1}{\PEPS{mod}{}{l}{}{}{}{}{}{Y_1}{Y_2}{Y_3}{1}}
    \!\!\! \equiv 
    \sum_{\{M\}} \sum_{i,j,k}
    & \includeTikz{0}{PEPS-mod-i-l-j-k-M_1-M_2-M_3-Y_1-Y_2-Y_3-1}{\PEPS{mod}{i}{l}{j}{k}{M_1}{M_2}{M_3}{Y_1}{Y_2}{Y_3}{1}}
    \\ \nn
    \times \, & |M_1 Y_1 M_3,i \ra 
    \otimes |M_3 Y_2 M_2,k \ra
    \\ \nn
    \otimes \, & \la M_1 Y_3 M_2,j | \, ,
\end{align}
whose entries are provided by $\F{\cat}$-symbols. As we shall recall below, these tensors play a crucial role within our framework as they can be exploited to generate algebras of (categorically) symmetric operators. For this reason, we refer to them as \emph{symmetric tensors} for the remainder of this manuscript.

Let us now introduce another type of tensors that evaluate to the module structure of functors between $\mc D$-module categories. Let us first provide a graphical representation for the entries of the matrices specifying the module structure $\omF{X}$ of a $\mc D$-module functor $\Func{X}$ associated with an object $X$ in $\Fun_\mc D(\mc M,\mc N)$: 
\begin{align}
    \includeTikz{0}{MPO-mod-mod-i-j-k-l-N_1-N_2-M_2-M_1-Y-X-1}{\MPO{mod}{mod}{i}{j}{k}{l}{N_1}{N_2}{M_2}{M_1}{Y}{X}{1}} :=
    &\big( \omF{X}^{M_1Y}_{N_2}\big)^{M_2,lj}_{N_1,ik} \, ,
    \\[-3em] \nn
    &\big( \omFbar{X}^{N_1Y}_{M_2}\big)^{N_2,lj}_{M_1,ik}
    =:\includeTikz{0}{MPO-mod-mod-i-j-l-k-N_1-N_2-M_2-M_1-Y-X-6}{\MPO{mod}{mod}{i}{j}{l}{k}{N_1}{N_2}{M_2}{M_1}{Y}{X}{6}} \, ,
\end{align}
where we should think of the diagram as the pasting of the string diagrams appearing on the l.h.s. and r.h.s. of eq.~\eqref{eq:FcatMove}. Adapting the conventions introduced above, we define the following collection of tensors labelled by simple objects in $\mc D$ and $\Fun_{\mc D}(\mc M,\mc N)$: 
\begin{align}
    \label{eq:MPO}
    \includeTikz{0}{MPO-mod-mod-Y-X-1}{\MPO{mod}{mod}{}{}{}{}{}{}{}{}{Y}{X}{1}} 
    \equiv
    \sum_{\substack{\{M\} \\ \{N\} }} \sum_{i,j,k,l} 
    \;\; &  \includeTikz{0}{MPO-mod-mod-i-j-k-l-N_1-N_2-M_2-M_1-Y-X-1}{\MPO{mod}{mod}{i}{j}{k}{l}{N_1}{N_2}{M_2}{M_1}{Y}{X}{1}}
    \\
    \nn
    \times \, &  |M_1XN_1,i \ra \la M_2XN_2,j |
    \\
    \nn
    \otimes \, & |N_1YN_2,k \ra\la M_1YM_2,l |\, .
\end{align}
As we mentioned above, in the case of module endofunctors in $\mc D^\star_\mc M$, we can think of the corresponding tensors as evaluating to the $\F{\caact}$-symbols of $\mc M$ as a $(\mc D^\star_{\mc M},\mc D)$-bimodule category. Contracting two such tensors whose labelling objects match is accomplished by concatenating them, horizontally or vertically, identifying the objects labelling the module strings, and tracing over the basis vectors along which the contraction takes place. Given our diagrammatic conventions, we have for instance
\begin{equation*}
    \includeTikz{0}{contraction-1}{\contraction{1}}
    \hspace{-1pt} \equiv \hspace{-1pt}
    \includeTikz{0}{contraction-2}{\contraction{2}} \, .
\end{equation*}
Tensor networks of this kind are referred to as \emph{Matrix Product Operators} (MPOs).  

Finally, we require another family of tensors whose non-vanishing entries evaluate to the $\F{\fr F}$-symbols associated with a triple $(\mc M,\mc N,\mc O)$ or right $\mc D$-module categories. Specifically, we define
\begin{align}
    \label{eq:PEPSComp}
    \includeTikz{0}{PEPS-mod-j-i-k-l-O-M-N-X_1-X_2-X_3-3}{\PEPS{mod}{j}{i}{k}{l}{O}{M}{N}{X_1}{X_2}{X_3}{3}} :=
    &\big( \F{\fr F}^{X_1 X_2 M}_{O}\big)^{N,lj}_{X_3,ik} \, ,
    \\[-3em] \nn
    &\big( \Fbar{\fr F}^{X_1 X_2 M}_{O}\big)^{N,lj}_{X_3,ik} =:
    \includeTikz{0}{PEPS-mod-j-i-k-l-O-M-N-X_1-X_2-X_3-4}{\PEPS{mod}{j}{i}{k}{l}{O}{M}{N}{X_1}{X_2}{X_3}{4}}
    \, ,
\end{align}
and the corresponding tensor is constructed following the same steps as for symmetric tensors. Tensors of this kind shall be referred to as \emph{fusion tensors}, as they locally implement the fusion of MPOs.

\bigskip \noindent
We mentioned above that the various isomorphisms entering the definitions of module categories and module functors must satisfy some `pentagon axioms' ensuring the self-consistency of the constructions. In terms of tensors defined in this section, these axioms translate into `pulling-through conditions', whereby MPOs built out of tensors of the form \eqref{eq:MPO} are pulled through symmetric tensors of the form \eqref{eq:PEPSFcat}. Concretely, the pentagon axiom fulfilled by the module structure of a $\mc D$-module functor $X$ in $\Fun_{\mc D}(\mc M,\mc N)$ translates into
\begin{equation}
    \label{eq:pulling}
    \includeTikz{0}{pulling-1}{\pulling{1}} \, = \!\!\!\! \raisebox{-9pt}{\includeTikz{0}{pulling-2}{\pulling{2}}} \!\!\! ,
\end{equation}
which is true for any labelling of the various strings and basis vectors. Notice that as we pull the MPO encoding the $\mc D$-module functor $X$ through the symmetric tensor that evaluates to the $\F{\cat}$-symbols of $\mc M$, it transforms into a new symmetric tensor that evaluates to the $\F{\catb}$-symbols of $\mc N$. Making specific choices for the $\mc D$-module categories $\mc M$ and $\mc N$ yields the pulling-through conditions associated with all the other pentagon axioms mentioned so far. Specializing to the case $\mc M = \mc N$, we find the pentagon axiom satisfied by the bimodule associator $\F{\caact}$ and the (right) module associator $\F{\cat}$. Choosing $\mc M = \mc D$ and $\mc N$ arbitrary yields the pentagon axiom fulfilled by the module associator $\F{\cat}$ that involves the monoidal associator $F$. Finally, when $\mc M = \mc N = \mc D$, the pulling-through condition amounts to the pentagon axiom satisfied by the monoidal associator $\F{}$. As we shall review below, these various pulling-through conditions encode the action of symmetry and duality operators.

Another family of tensor network relations will play a crucial role in the following. These encode the composition rule of module functors. We mentioned earlier that given two representatives $X_1$ and $X_2$ of isomorphism classes of simple objects in categories $\Fun_{\mc D}(\mc M,\mc N)$ and $\Fun_{\mc D}(\mc N,\mc O)$, respectively, these could be composed so as to yield a $\mc D$-module functor between $\mc M$ and $\mc O$ that decomposes into simple objects of $\Fun_{\mc D}(\mc M,\mc O)$. Graphically, this translates into the composition of the corresponding MPO tensors by means the fusion tensors that evaluate to the $\F{\fr F}$-symbols introduced previously:
\begin{equation}
    \label{eq:fusionMPO}
    \includeTikz{0}{fusion-1}{\fusion{1}}
    =
    \sum_{X_3}\sum_o\includeTikz{0}{fusion-2}{\fusion{2}} ,
\end{equation}
which is true for any labeling of the various strings and basis vectors.
Specializing to the case $\mc M=\mc N$, the composition of $\mc D$-module functors provides the monoidal structure of $\mc D^\star_\mc M$ so that the fusion tensor evaluates to the $\F{\act}$-symbols of the left $\mc D^\star_\mc M$-module category $\mc M$. The diagrammatic relation above then provides the fusion of the corresponding MPOs symmetries.

\subsection{Symmetric Hamiltonians and dualities\label{sec:symOp}}

\noindent
Let us now put together the ingredients presented earlier into a recipe for constructing symmetric Hamiltonians on infinite one-dimensional lattices, and duality relations between them \cite{Lootens:2021tet}. First we need to pick a microscopic Hilbert space. Let $\mc D$ be a fusion category and $\mc M$ a (right) indecomposable module category over it. We consider the following $\mathbb C$-linear span
\begin{align}
    \label{eq:infHilbert}
    \mathbb C \bigg[\!\!
    \includeTikz{0}{infChain}{\infChain{i_{\msf i-\frac{1}{2}}}{i_{\msf i+\frac{1}{2}}}{i_{\msf i+\frac{3}{2}}}{M_{\msf i-2}}{M_{\msf i-1}}{M_{\msf i}}{M_{\msf i+1}}{Y_{\msf i - \frac{1}{2}}}{Y_{\msf i+\frac{1}{2}}}{Y_{\msf i + \frac{3}{2}}}} \!\!\! \bigg] 
\end{align}
over $\{M \in \mc I_\mc M\}$, $\{Y \in \mc I_\mc D\}$ and basis vectors $\{i\}$ in the hom-spaces defined following the convention of eq.~\eqref{eq:unlabelledPatch}. Notice that this Hilbert space is typically not a tensor product of local Hilbert spaces. We then define \emph{local} operators acting on this microscopic Hilbert space via matrix multiplication of the form\footnote{Note that local operators in ref.~\cite{Lootens:2021tet} are defined in a equivalent but different way. The alternative definition we employ in this manuscript will make accounting for boundary conditions more convenient.}
\begin{equation}
    \label{eq:localOp}
    \mathbb b^\mc M_{\msf i,n}
    \equiv 
    \sum_{\{Y\}}\sum_{j,k} \,
    b_{\msf i,n} \big(\{Y\},j,k \big) \, 
    \includeTikz{0}{ladder-j-k-Y_-msfi-frac-1-2-tildeY_-msfi-frac-1-2-Y_-msfi+frac-1-2-tildeY_-msfi+frac-1-2-Y_msfi-1}{\ladder{j}{k}{Y_{\msf i-\frac{1}{2}}}{\tilde Y_{\msf i-\frac{1}{2}}}{Y_{\msf i + \frac{1}{2}}}{\tilde Y_{\msf i+\frac{1}{2}}}{Y_\msf i}{1}} \, ,
\end{equation}
obtained by taking linear combinations of contractions of two symmetric tensors as defined in eq.~\eqref{eq:PEPS}. Any combinations of such local operators can then be organized into a local Hamiltonian $\mathbb H^\mc M = \sum_{\msf i}\sum_n \mathbb b^{\mc M}_{\msf i,n}$. Notice that the definition of the local operators fixes certain combinations of objects and morphisms in $\mc D$, thereby imposing \emph{kinematical constraints} on the genuine physical degrees of freedom of the model, which are provided by object and morphisms in $\mc M$. This implies in particular that we can often consider a subspace of the microscopic Hilbert defined above when dealing with a specific Hamiltonian. Crucially, any such Hamiltonian $\mathbb H^\mc M$ remains invariant under the action of symmetry operators. Indeed, it follows immediately from the pulling-through condition eq.~\eqref{eq:pulling} that the MPO symmetry
\begin{equation}
    \label{eq:symMPO}
    \sum_{\{Y \in \mc I_\mc D \}}
    \includeTikz{0}{intertwiner-1}{\intertwiner{1}}
\end{equation}
labelled by a simple object $X$ in $\mc D^\star_\mc M = \Fun_\mc D(\mc M,\mc M)$ commutes with the Hamiltonian $\mathbb H^\mc M$. Remark that this symmetry condition relies solely on the local operators being constructed out of symmetric tensors evaluating to the $\F{\cat}$-symbols of $\mc M$, and is oblivious to the specific definitions of these local operators, i.e. choices of objects and morphisms in $\mc D$ as well as complex coefficients $b_{n,\msf i}$. Importantly, for any $\mc D$-module category $\mc M$ that we consider in this manuscript, the fusion category $\mc D^\star_\mc M$ is found to be \emph{Morita equivalent} to $\mc D$ (see app.~\ref{sec:app_Morita}). The physical implications of this mathematical result are discussed in the next section. 

If symmetry operators are labelled by objects in $\mc D^\star_\mc M$, duality operators are labelled by objects in $\Fun_{\mc D}(\mc M,\mc N)$ with $\mc N$ a $\mc D$-module category distinct from $\mc M$. These duality operators have exactly the same form as the symmetry operators \eqref{eq:symMPO} with the difference that the top purple strings are now labelled by objects in $\mc N$ and the tensor evaluates to $\omF{X}$-symbols, i.e. entries of the matrices specifying the module structure $\omF{X}$ of the module functor $\Func{X}$ corresponding to the simple object $X$ in $\Fun_\mc D(\mc M, \mc N)$. Denoting by $\fr T_{\mc M | \mc N}^X$ such an intertwining MPO, from the pulling-through conditions \eqref{eq:pulling} follows the commutation relation
\begin{equation}
    \fr T_{\mc M|\mc N}^{X} \circ \mathbb H^\mc M = \mathbb H^\mc N \circ \fr T_{\mc M|\mc N}^X \, ,
\end{equation}
where $\mathbb H^\mc N := \sum_\msf i \sum_n \mathbb b_{\msf i,n}^\mc N$ and $\mathbb b_{\msf i,n}^\mc N$ defined exactly as in eq.~\eqref{eq:localOp} but with respect to $\mc N$. It follows immediately from our construction that the Hamiltonian $\mathbb H^\mc N$ thus constructed remains invariant under the action of symmetry operators labelled by objects in the Morita dual $\mc D^\star_\mc N$ of $\mc D$ with respect to $\mc N$. 

The duality operator performs the transmutations of the local operators defining the Hamiltonians $\mathbb H^\mc M$ and $\mathbb H^\mc N$. However the knowledge of this operator is not quite sufficient to write down a set of isometric transformations mapping models to one another. Obtaining such transformations indeed requires an analysis of the topological sectors of the corresponding models. This is the main teaching of this manuscript and the purpose of the following sections. 

\subsection{Illustration\label{sec:illustration}}

\noindent
Before concluding this section, let us consider an illustrative example, namely the transverse-field Ising model. We encourage the reader to consult ref.~\cite{Lootens:2021tet} for additional examples. Let $\mc D$ be the fusion category $\Vect_{\mathbb Z_2}$ of $\mathbb Z_2$-graded vector spaces. This fusion category has two simple objects, which we denote by $\mathbb 1$ and $m$. The fusion rules read $\mathbb 1 \otimes \mathbb 1 \simeq \mathbb 1 \simeq m \otimes m$ and $\mathbb 1 \otimes m \simeq m \simeq m \otimes \mathbb 1$. Given any $\Vect_{\mathbb Z_2}$-module category $\mc M$, we consider the Hamiltonian
\begin{equation}
    \label{eq:infIsing}
    \mathbb H^\mc M = -J \sum_{\msf i} \mathbb b_{\msf i,1}^\mc M -Jg \sum_{\msf i} \mathbb b_{\msf i,2}^\mc M
\end{equation}
with local operators given by
\begin{align}
    \nn
    \mathbb b^\mc M_{\msf i,1} = \;
    &\includeTikz{0}{ladder-1-1-mathbb1-m-mathbb1-m-m-2}{\ladder{1}{1}{\mathbb 1}{m}{\mathbb 1}{m}{m}{2}}
    +
    \includeTikz{0}{ladder-1-1-mathbb1-m-m-mathbb1-m-2}{\ladder{1}{1}{\mathbb 1}{m}{m}{\mathbb 1}{m}{2}}  
    \\
    +\, &\includeTikz{0}{ladder-1-1-m-mathbb1-m-mathbb1-m-2}{\ladder{1}{1}{m}{\mathbb 1}{m}{\mathbb 1}{m}{2}}
    +
    \includeTikz{0}{ladder-1-1-m-mathbb1-mathbb1-m-m-2}{\ladder{1}{1}{m}{\mathbb 1}{\mathbb 1}{m}{m}{2}}  
\end{align}
and 
\begin{align}
    \nn
    \mathbb b^\mc M_{\msf i,2} = 
    &\includeTikz{0}{ladder-1-1-mathbb1-mathbb1-mathbb1-mathbb1-mathbb1-2}{\ladder{1}{1}{\mathbb 1}{\mathbb 1}{\mathbb 1}{\mathbb 1}{\mathbb 1}{2}} - 
    \includeTikz{0}{ladder-1-1-mathbb1-mathbb1-m-m-mathbb1-2}{\ladder{1}{1}{\mathbb 1}{\mathbb 1}{m}{m}{\mathbb 1}{2}}\\
    +\, &\includeTikz{0}{ladder-1-1-m-m-mathbb1-mathbb1-mathbb1-2}{\ladder{1}{1}{m}{m}{\mathbb 1}{\mathbb 1}{\mathbb 1}{2}} - 
    \includeTikz{0}{ladder-1-1-m-m-m-m-mathbb1-2}{\ladder{1}{1}{m}{m}{m}{m}{\mathbb 1}{2}}\, .
\end{align}
By definition, only hom-spaces for which the corresponding objects satisfy the fusion rules are non-vanishing, which is the case of all the hom-spaces appearing above. Since the outcome of the fusion of two objects is uniquely determined, hom-spaces are necessarily one-dimensional and we labelled by $1$ the corresponding unique basis vectors. Let us now choose specific $\Vect_{\mathbb Z_2}$-module categories. 

Let $\mc M=\Vect_{\mathbb Z_2}$ be the regular module category. Recall that in this case the module associator $\F{\cat}$ boils down to the monoidal associator $F$ of $\Vect_{\mathbb Z_2}$, which happens to be trivial. It means that the $\F{\cat}$-symbols equal $1$ whenever all the fusion rules are satisfied, and $0$ otherwise. Consider the microscopic Hilbert space \eqref{eq:infHilbert}. It follows from the fusion rules that objects in $\mc I_{\mc D}$ are fully determined by a choice of objects in $\mc I_{\mc M}$. This means that the physical degrees of freedom are labelled by objects in $\mc I_\mc M$ and located in the `middles' of the corresponding strings so that the effective microscopic Hilbert spaces is isomorphic to $\bigotimes_{\msf i}\mathbb C^2$ with $\mathbb C^2 \simeq \mathbb C[\includeTikz{0}{object-mod}{\object{mod}{}}]$. The operator $\mathbb b_{\msf i,1}^\mc M$ acts on this Hilbert space as $|\mathbb 1 / m \ra \mapsto | m / \mathbb 1 \ra$ on the site $\msf i$, where we identify $| \mathbb 1 \ra$ and $|m \ra$ with the $+1$ and $-1$ eigenvectors of the Pauli $S^z$ operator, respectively. The operator $\mathbb b_{\msf i,2}^\mc M$ acts as the identity operator whenever the degrees of freedom at sites $\msf i$ and $\msf i+1$ agree, and minus the identity operator otherwise. Putting everything together, we find that the Hamiltonian \eqref{eq:infIsing} boils down to
\begin{equation}
    \label{eq:infIsingVecZ2}
    \mathbb H^{\Vect_{\mathbb Z_2}} = -J \sum_{\msf i} (S^x_\msf i +g S^z_\msf i S^z_{\msf i+1}) \, ,
\end{equation}
which we recognize as the transverse-field \emph{Ising} model. This model has a (global) $\mathbb Z_2$ symmetry generated by $\prod_{\msf i}S^x_\msf i$.

The fusion category $\Vect_{\mathbb Z_2}$ admits another (indecomposable) module category over it, namely the category $\mc M = \Vect$ of vector spaces. This category has a unique object, which we denote by $\mathbb 1$, and the module structure is provided by $\mathbb 1 \cat \mathbb 1 \simeq \mathbb 1 \simeq \mathbb 1 \cat m$. As for the previous case, the module associator is trivial. For this example, the physical degrees of freedom are identified with the unique basis vectors of the hom-spaces $\Hom_{\mc M}(\mathbb 1 \cat Y_{\msf i-\frac{1}{2}}) \simeq \mathbb C$, with $Y_{\msf i-\frac{1}{2}} \in \{\mathbb 1,m\}$, and thus the effective Hilbert space is still isomorphic to $\bigotimes_\msf i \mathbb C^2$. It readily follows from the definition of the local operators that the Hamiltonian \eqref{eq:infIsing} now boils down to
\begin{equation}
    \label{eq:infIsingVec}
    \mathbb H^{\Vect} = -J \sum_{\msf i} (S^x_{\msf i-\frac{1}{2}} S^x_{\msf i+\frac{1}{2}} + g S^z_{\msf i+\frac{1}{2}}) \, ,
\end{equation}
which we recognize as the \emph{Kramers-Wannier} dual of the transverse-field Ising model. This model also has a $\mathbb Z_2$ symmetry, which is now generated by $\prod_\msf i S^z_{\msf i+\frac{1}{2}}$.

Within our framework, obtaining two models that are dual to one another is guaranteed by the fact that they only differ by a choice of $\Vect_{\mathbb Z_2}$-module category. This implies that altering the definitions of the local operators $\mathbb b_{\msf i,1}^{\mc M}$ and $\mathbb b_{\msf i,2}^{\mc M}$ would still yield two dual models, and the resulting models would always be invariant under the action of operators labelled by objects in the fusion categories $(\Vect_{\mathbb Z_2})^\star_{\Vect_{\mathbb Z_2}} \cong \Vect_{\mathbb Z_2}$ and $(\Vect_{\mathbb Z_2})^\star_{\Vect} \cong \Rep(\mathbb Z_2)$, respectively, which are Morita equivalent \cite{Lootens:2021tet}. 

Let us now explicitly construct the duality operator transmuting the Hamiltonians $\mathbb H^{\Vect_{\mathbb Z_2}}$ and $\mathbb H^{\Vect}$ into one another. Applying the recipe presented above, this duality operator can be written as an MPO labelled by the unique simple object in $\Fun_{\Vect_{\mathbb Z_2}}(\Vect_{\mathbb Z_2},\Vect) \cong \Vect$ such that the individual tensors evaluate to the module structure of the corresponding functor $\Vect_{\mathbb Z_2} \to \Vect$, which happens to be trivial. Graphically, it reads
\begin{equation}
    \sum_{\{Y \in \mc I_{\mc D} \}}\includeTikz{0}{intertwiner-2}{\intertwiner{2}}
\end{equation}
where dotted strings are labelled by the unique simple object $\Vect$ that we have been notating via $\mathbb 1$. This duality operator should be interpreted as a map from any symmetric operator associated with $\Vect_{\mathbb Z_2}$ to a symmetric operator associated with $\Vect$. Concretely, it follows from the fusion rules in $\Vect_{\mathbb Z_2}$ that this operator maps states $| Y_{\msf i} \ra \otimes | Y_{\msf i+1} \ra$ at the bottom to states $| Y_\msf i + Y_{\msf i+1} \ra$ associated with site $\msf i+\frac{1}{2}$ at the top, hence $S^z_{\msf i}S^z_{\msf i+1} \mapsto S^z_{\msf i+\frac{1}{2}}$ and $S^x_\msf i \mapsto S^x_{\msf i-\frac{1}{2}} S^x_{\msf i+\frac{1}{2}}$, as expected.

\section{Topological sectors and dualities\label{sec:topoSectors}}

\noindent
\emph{Building upon the constructions above, we introduce in this section the notion of twisted boundary condition and present a characterization of topological sectors. We subsequently explain a method to compute the mapping of topological sectors under a duality relation.}

\subsection{Boundary conditions and tube category}
\noindent

\noindent
Given an input fusion category $\mc D$, we reviewed in the previous section a recipe to construct dual local operators associated with choices of module categories over $\mc D$. These operators are invariant under the action of symmetry operators labelled by objects in fusion categories that are Morita equivalent. Duality operators can then be constructed from the data of module functors between module categories \cite{Lootens:2021tet}.
However, this is not enough in order to fully establish a duality relation between Hamiltonian models. Indeed, it is further required to establish how the mappings of symmetric operators interact with the topological sectors of the models.

Let us consider a (finite) spin chain of length $L+1$ with total Hilbert space $\mc H^\mc M$ given by 
\begin{equation*}
        \mathbb C \bigg[ \!\!
        \includeTikz{0}{openChain}{\openChain} \!\!\! \bigg] ,
\end{equation*}
over $\{M \in \mc I_\mc M\}$, $\{Y \in \mc I_\mc D\}$ and basis vectors $\{i\}$ in the hom-spaces defined following the convention of eq.~\eqref{eq:unlabelledPatch}. Notice that we have left the boundary condition unspecified. Loosely speaking, choosing a boundary condition amounts to picking a relation, or a map, between degrees of freedom at sites $L+1$ and $1$ so as to close the chain. For instance, \emph{periodic} boundary conditions would be obtained by enforcing $M_{L+1} = M_1$. Within our formalism, such a map between degrees of freedom is provided by an endofunctor of $\mc M$ in $\Fun(\mc M,\mc M)$. But the corresponding (possibly twisted) boundary conditions should not break translation invariance of a Hamiltonian acting on this Hilbert space. In other words, given a choice of boundary condition, there should still be an isomorphism between vector spaces related by a translation by one site. This requires the action of the endofunctor to commute with the module action of $\mc D$ on $\mc M$ up to a natural isomorphism, i.e. it must be equipped with a $\mc D$-module structure. Therefore, we consider boundary conditions classified by $\mc D^\star_\mc M = \Fun_{\mc D}(\mc M,\mc M)$. This is also the fusion category that describes the symmetry of the model, and as such these boundary conditions are referred to as \emph{symmetry-twisted} boundary conditions.

As mentioned in sec.~\ref{sec:prelim}, the category $\mc D^\star_\mc M$ is equipped with a monoidal structure provided by the composition of $\mc D$-module endofunctors. We further commented that $\mc M$ is naturally endowed with the structure of a left module category over $\mc D^\star_\mc M$, and as such we can employ the same graphical calculus for objects in $\mc D^\star_\mc M$ as that for objects in $\mc D$. Specifically, we consider microscopic Hilbert spaces $\mc H^{\mc M}$ of the form
\begin{align}
    \label{eq:Hilbert}
    \mathbb C \bigg[\!\!\!
    \includeTikz{0}{chain-i_-L+frac-1-2-i_frac-1-2-i_frac-3-2-M_L-M_-L+1-M_1-M_2-Y_-L+frac-1-2-A-Y_frac-3-2}{\chain{\, i_{\! L+\frac{1}{2}}}{i_\frac{1}{2}}{i_\frac{3}{2}}{M_L}{M_{L+1}}{M_1}{M_2}{Y_{L+\frac{1}{2}}}{A}{Y_\frac{3}{2}}} \!\!\! \bigg] ,
\end{align}
over $\{M \in \mc I_\mc M\}$, $\{Y \in \mc I_\mc D\}$ and basis vectors $\{i\}$ in the hom-spaces defined following the convention of eq.~\eqref{eq:unlabelledPatch}, and where $\{A \in \mc I_{\mc D^\star_\mc M}\}$ are representatives of isomorphism classes of simple objects in $\mc D^\star_\mc M$ encoding choices of boundary conditions. A couple of important remarks: Firstly, boundary conditions are promoted within our approach to genuine degrees of freedom of the model, which implies in particular that they can be acted upon. Secondly, given an arbitrary boundary condition, the effective number of sites may not be $L$---as would be the case for periodic boundary conditions for instance---but rather $L+1$. This may seem somewhat paradoxical, but as we shall see this is a characteristic feature of \emph{non-abelian} boundary conditions.

Given a choice of $\mc D$-module category $\mc M$, let us study the decomposition of the Hilbert space $\mc H^\mc M$ into super-selection sectors of the symmetry.
In order to perform such a decomposition, we consider tensor networks $\fr T_{\mc M|\mc M}^{A,A',X,X',k,k'}$ that describe the action of the symmetry in the presence of twisted boundary conditions. These are of the form
\begin{align}
    \label{eq:tubes}
    \sum_{\substack{\{Y \in \mc I_\mc D\} \\ \{M,\tilde M \in \mc I_\mc M\} }} \!
    \includeTikz{0}{tube-k-k-tildeM_L-tildeM_-L+1-tildeM_1-tildeM_2-M_L-M_-L+1-M_1-M_2-X-X-X-A-A-Y_-L+frac-1-2-Y_frac-3-2}{\tube{k}{k'}{\tilde M_L}{\tilde M_{L+1}}{\tilde M_1}{\tilde M_2}{M_L}{M_{L+1}}{M_1}{M_2}{X}{X'}{X}{A'}{A}{Y_{\! L +\frac{1}{2}}}{Y_\frac{3}{2}}}  ,
\end{align}
where $A,A',X,X' \in \mc I_{\mc D^\star_\mc M}$, while $k,k'$ label basis vectors in the hom-spaces $\Hom_{\mc D^\star_\mc M}(X',A' \otimes X)$ and $\Hom_{\mc D^\star_\mc M}(X \otimes A,X')$, respectively. We distinguish two types of tensors in this expression. On the one hand, we have the same MPO tensors as in eq.~\eqref{eq:symMPO}, which evaluates to the $\F{\caact}$-symbols of $\mc M$ as a $(\mc D^\star_\mc M,\mc D)$-bimodule category. On the other hand we have two fusion tensors of the form \eqref{eq:PEPS} that evaluates to the $\F{\act}$-symbols of $\mc M$ as \emph{left} module category over $\mc D^\star_\mc M$.

Given the geometry of such tensor networks, we will henceforth refer to them as `tubes'. It follows from our graphical calculus that these symmetry tubes can be interpreted as linear maps $\mc H^{\mc M} \to \mc H^{\mc M}$. Let us now demonstrate that the subspace of linear maps spanned by the symmetry tubes is closed under multiplication. Graphically, multiplication of symmetry tubes is obtained by stacking them on top of one another and contracting the indices along which the stacking takes place, i.e. 
\begin{equation*}
    \sum_{\{Y \in \mc I_\mc D\}} \,
    \includeTikz{0}{tubeAlgebra}{\tubeAlgebra} , 
\end{equation*}
where we are using the convention defined in eq.~\eqref{eq:convModStrand}. Notice that this contraction is accompanied with the identifications of objects in $\mc D$ as well as in $\mc D^\star_\mc M$. At this point, we can use the fusion of MPO tensors defined in eq.~\eqref{eq:fusionMPO} together with the \emph{recoupling theory} of fusion tensors to express this stacking as a complex linear combination of tubes. Recoupling fusion tensors amounts to changing the contraction patterns of a given collection of tensors, which is rendered possible due to the coherence relations satisfied by the $\F{\cat}$-symbols these tensors evaluate to \cite{Lootens:2020mso}. Concretely, the fusion tensors satisfy graphical identities of the form of eq.~\eqref{eq:monoAssociativity}, where the $\F{}$-symbols would be that of $\mc D^\star_\mc M$. These recoupling moves can be explicitly found in ref.~\cite{Lootens:2021tet}. Putting everything together, we obtain the following multiplication rule: 
\begin{align}
    \label{eq:tubeAlgebra}
    &\fr T^{A,A',X_1,X_1',k_1,k_1'}_{\mc M|\mc M} \cdot 
    \fr T^{A',A'',X_2,X_2',k_2,k_2'}_{\mc M|\mc M}
    \\ \nn
    & \q =
    \sum_{\substack{X_3,X_3' \\ k_3,k_3' \\ \{j\}}}
    \big(\Fbar{}^{A'' X_2 X_1}_{X_3'}\big)_{X_2',k_2 j_2}^{X_3,j_1 k_3} \, 
    \big(\F{}^{X_2 A' X_1}_{X_3'}\big)_{X_2',k_2' j_2}^{X_1',k_1 j_3}
    \\[-2.3em] \nn
    &\hspace{5em} \times \big(\Fbar{}^{X_2 X_1 A}_{X_3'}\big)_{X_3,j_1 k_3'}^{X_1',k_1' j_3} \; 
    \fr T^{A,A'',X_3,X_3',k_3,k_3'}_{\mc M|\mc M} \, ,
\end{align}
where it follows from the definition of the $\F{}$-symbols that in particular the first sum is over simple objects $X_3$ appearing in the decomposition of the monoidal product $X_2 \otimes X_1$. 

We exploit the results obtained above to introduce the \emph{tube category} $\Tube(\mc D^\star_\mc M)$,  whose objects are objects $A,A'$ in $\mc D^\star_\mc M$ and hom-spaces $\Hom_{\Tube(\mc D^\star_\mc M)}(A,A')$ are vector spaces spanned by tubes $\fr T^{A,A',-,-,-,-}_{\mc M | \mc M}$ as defined previously \cite{ocneanu1994chirality,ocneanu2001operator}. The composition rule is then provided by the multiplication rule \eqref{eq:tubeAlgebra}. It follows that the Hilbert space $\mc H^\mc M$ defines a \emph{representation} in $\Fun(\Tube(\mc D^\star_\mc M),\Vect)$, and thus admits a decomposition 
\begin{equation}
    \mc H^{\mc M} = \bigoplus_\mc V (\mc H^{\mc M})_\mc V
\end{equation}
into superselection sectors labelled by irreducible representations $\mc V$ of $\Tube(\mc D^\star_\mc M)$. Henceforth, we refer to these superselection sectors as \emph{topological sectors}.
Crucially, there is a well-known equivalence between the category of representations of the tube category and the Drinfel'd center $\mc Z(\mc D^\star_\mc M)$ \cite{doi:10.1142/S0129055X95000232,Izumi:2000qa,MUGER200381,MUGER2003159,2015arXiv151107329P} (see app.~\ref{sec:app_defMorita}),\footnote{We review in app.~\ref{sec:app_double} this correspondence in the case where $\mc D$ is the category $\Vect_G$ of $G$-graded vector spaces.} so that topological sectors of $\mc H^{\mc M}$ can be labelled by simple objects $Z$ in $\mc Z(\mc D^\star_{\mc M})$. As a matter of fact simple objects in $\mc Z(\mc D^\star_\mc M)$ are often obtained by computing the minimal idempotent tubes w.r.t. the multiplication defined in eq.~\eqref{eq:tubeAlgebra}. Such a simple object encodes a (possibly not simple) twisted boundary condition as well as a symmetry charge that decomposes the action of the tubes leaving the boundary condition invariant.

By considering the space of all tubes and introduce the convention that tubes with mismatching objects multiply to zero, we can consider the \emph{tube algebra} of all tubes. In addition to being closed under multiplication, the tube algebra is closed under Hermitian conjugation according to
\begin{align}
    \nn
    \big(\fr T_{\mc M|\mc M}^{A,A',X,X',k,k'}\big)^\dag \! = \!\!\! \sum_{\substack{i,i'\! ,j,j'\\ X''}} 
    &\big(\F{}^{X \bar X X}_{X}\big)_{\mathbb 1,11}^{\mathbb 1,11}
    \big(\F{}^{A' X \bar X}_{A'}\big)_{X',k i}^{\mathbb 1,11}
    \\[-1.3em]
    \nn
    &\times
    \big(\Fbar{}^{X A \bar X}_{A'}\big)_{X',k' i}^{X'',j i'}
    \big(\F{}^{\bar X X X''}_{X}\big)_{\mathbb 1,11}^{A',i' j'}
    \\[.2em]
    \label{eq:tubeDagger}
    & \times
    \fr T_{\mc M|\mc M}^{A',A,\bar X,X'',j,j'} \, .
\end{align}
This closure under Hermitian conjugation equips the tube algebra with the structure of a ${}^*$-algebra. In virtue of its finiteness, it follows that the tube algebra is block diagonal in the topological sectors so we can define a new basis 
\begin{align}
    \nn
    e^{Z,A_i,A'_j}_{\mc M | \mc M} 
    &\equiv \#_Z \!\!\! \sum_{X,X',k,k'} \!\!\!\! d_X \, \Omega^{Z,A_i,A_j'}_{X,X',k,k'} \, \fr T_{\mc M|\mc M}^{A,A',X,X',k,k'} ,
    \\
    \big(e^{Z,A_i,A'_j}_{\mc M | \mc M}\big)^\dag 
    &= e^{Z,A'_j,A_i}_{\mc M | \mc M} \, ,
\end{align}
where $Z$ labels a simple object of $\mc Z(\mc D^\star_\mc M)$, and $A_i \equiv (A,i)$ runs over all simple objects $A$ of $\mc D^\star_\mc M$ decomposing $Z$ as an object of $\mc D^\star_\mc M$ as well as the corresponding degeneracy labels $i$. The normalization factor is chosen to be $\#_Z := {\ell (Z)}/{{\rm FPdim}(\mc D)}$, where $\ell (Z)$ corresponds to the number of simple objects appearing in the decomposition of $Z$ and ${\rm FPdim}(\mc D)$ is the \emph{Frobenius-Perron dimension} of $\mc D$ \cite{etingof2016tensor}. These new basis elements behave like \emph{matrix units} under the tube multiplication, i.e. they diagonalize the multiplication:
\begin{equation}
    \begin{split}
        e^{Z,A_i,A'_j}_{\mc M | \mc M} \cdot e^{Z',A'_k,A''_l}_{\mc M | \mc M} &= \delta_{Z,Z'} \delta_{j,k} \,  e^{Z,A_i,A''_l}_{\mc M | \mc M} \, ,
        \\
        \sum_{Z,A_i} e^{Z,A_i,A_i}_{\mc M | \mc M} &= \mathbb 1^{\mc M} \, ,
    \end{split}
\end{equation}
where $\mathbb 1^{\mc M}$ denotes the identity operator on $\mc H^{\mc M}$. This means that for $A_i = A'_j$ they define a complete set of idempotents that project onto states within a certain topological sector $Z$ with boundary condition $A$ and degeneracy $i$. Consequently, for $A_i \neq A'_j$ they define isometries, which, within a topological sector $Z$, map between states with boundary condition $A$ with degeneracy $i$ and states with boundary condition $A'$ and degeneracy $j$.

\subsection{Symmetric operators}

\noindent
Given a $\mc D$-module category $\mc M$ and a choice of boundary condition $A \in \mc I_{\mc D^\star_\mc M}$, let us now construct local operators that are invariant under the action of symmetry operators. We already explained in sec.~\ref{sec:symOp} how to construct $\mc D^\star_\mc M$-symmetric operators away from the boundary in terms of tensors evaluating to the $\F{\cat}$-symbols of $\mc M$. Local operators involving the boundary conditions are defined as 
\begin{equation}
    \label{eq:bdryOp}
    \mathbb b^{\mc M,A}_n \! \equiv\!\!  \sum_{\substack{\{Y\} \\ j,k}} \! 
    b_{n}(\{Y\},j,k)
    \includeTikz{0}{ladderBdry-j-k-Y_-L+frac-1-2-tildeY_-L+frac-1-2-Y_-frac-3-2-tildeY_-frac-3-2-Y-A-1}{\ladderBdry{j}{k}{\, Y_{L+\frac{1}{2}}}{\, \tilde Y_{L+\frac{1}{2}}}{Y_{\frac{3}{2}}}{\tilde Y_{\frac{3}{2}}}{Y}{A}{1}}  ,
\end{equation}
where in addition to symmetric tensors evaluating to the $\F{\cat}$-symbols of $\mc M$, we now require an MPO tensor evaluating to the $\F{\caact}$-symbols of $\mc M$ as a $(\mc D^\star_\mc M,\mc D)$-bimodule category. Notice that such a definition allows for very general types of boundary conditions. It immediately follows from the pulling-through conditions translating the pentagon axioms of the bimodule associator $\F{\caact}$ of $\mc M$ involving the right module associator $\F{\cat}$ of $\mc M$ on the one hand, and that involving the left module associator $\F{\act}$ of $\mc M$ as a module category over $\mc D^\star_\mc M$ on the other hand, that these local operators can be pulled through the tubes \eqref{eq:tubes} defined previously. By considering arbitrary linear combinations of local operators \eqref{eq:localOp} and \eqref{eq:bdryOp}, we construct families of Hamiltonians associated with boundary condition $A$:
\begin{equation}
    \mathbb H^{\mc M,A} = \sum_n \mathbb b_n^{\mc M,A} + \sum_{\msf i=2}^L \sum_n   \mathbb b_{\msf i,n}^{\mc M}  \, .
\end{equation}
We have already established that away from the boundary, the local operators commute with the tubes. An arbitrary tube would however modify the boundary condition provided by $A$:
\begin{equation}
    \fr T_{\mc M|\mc M}^{A,A',X,X', k, k'} \circ \mathbb H^{\mc M,A} = \mathbb H^{\mc M,A'} \circ \fr T_{\mc M|\mc N}^{A,A',X,X', k, k'} \, .
\end{equation} 
This means that only a subset of tubes of the form $\fr T_{\mc M|\mc M}^{A,A,-,-,-,-}$ would leave the boundary condition invariant and as such commute with the Hamiltonian $\mathbb H^{\mc M,A}$. Practically, this means that in general the symmetry operators leaving $\mathbb H^{\mc M,A}$ invariant are not organized into $\mc D^\star_\mc M$, as is the case in the infinite chain scenario. The symmetry charge sectors decomposing the action of these tubes commuting with $\mathbb H^{\mc M,A}$ then provide topological sectors that are in one-to-one correspondence with simple objects of $\mc Z(\mc D^\star_\mc M)$. Note that in general a given simple object of $\mc Z(\mc D^\star_\mc M)$ is not necessarily associated with a simple boundary condition, and conversely, the same topological sector can be found in the decomposition of Hamiltonians with boundary conditions provided by different simple objects in $\mc D^\star_\mc M$. As alluded to above, this is the statement that, as an object in $\mc D^\star_\mc M$, a simple object $Z$ in $\mc Z(\mc D^\star_\mc M)$ decomposes over simple objects in $\mc D^\star_\mc M$. We shall provide concrete examples of these statements in sec.~\ref{sec:Examples}.

Let us refine the statements above using the matrix unit basis considered above. A given Hamiltonian $\mathbb H^{\mc M,A}$ can be decomposed into topological sectors as
\begin{equation}
    \begin{split}
    \mathbb H^{\mc M,A} &= \sum_{Z,i} \, \mathbb H^{\mc M,A}_{Z,i} \, ,
    \\
    \mathbb H^{\mc M,A}_{Z,i} &\equiv e^{Z,A_i,A_i}_{\mc M | \mc M} \, \mathbb H^{\mc M,A} \, e^{Z,A_i,A_i}_{\mc M | \mc M} \, ,
    \end{split}
\end{equation}
where the sum is over all topological sectors $Z$, such that $A$ appears in the decomposition of $Z$ as an object of $\mc D^\star_\mc M$, and degeneracy labels $i$. The Hamiltonians $\mathbb H^{\mc M,A}_{Z,i}$ can be thought of as the elementary building blocks of a Hamiltonian with a given boundary condition. Importantly, all these elementary Hamiltonians within a given topological sector $Z$ have the same spectrum since they can be related by the isometry
\begin{equation}
    \mathbb H^{\mc M,A'}_{Z,j} = \Big(e^{Z,A_i,A'_j}_{\mc M | \mc M}\Big)^\dagger \mathbb H^{\mc M,A}_{Z,i} e^{Z,A_i,A'_j}_{\mc M | \mc M} \, .
\end{equation}
As pointed out above, this implies that a given topological sector $Z$ can be found in Hamiltonians $\mathbb H^{\mc M,A}$ for different choices of $A$, as long as the boundary condition $A$ appears in the decomposition of $Z$ into simple objects of $\mc D^\star_\mc M$.

\subsection{Intertwining tubes and dualities\label{sec:tubeInt}}

\noindent
Given an input fusion category $\mc D$ and a pair $(\mc M,\mc N)$ of (right) $\mc D$-module categories, we explained in sec.~\ref{sec:symOp} how MPOs \eqref{eq:symMPO} labelled by simple objects in $\Fun_{\mc D}(\mc M,\mc N)$ perform the transmutations of local operators $\mathbb b^\mc M_{\msf i,n}$ into $\mathbb b^{\mc N}_{\msf i,n}$, which only differ by the choice of $\mc D$-module categories. Let us now explain how these mappings of local operators interact with the topological sectors characterized above.

In order to account for boundary conditions, the intertwining MPOs \eqref{eq:symMPO} need to be promoted to intertwining tubes $\fr T_{\mc M|\mc N}^{A,B,X,X', k, k'}$ of the form
\begin{align}
    \label{eq:tubeIntertwiner}
    \bigoplus_{\substack{ \{Y \in \mc I_\mc D\} \\ \{M \in \mc I_\mc M\} \\ \{N \in \mc I_\mc N\}}} 
    \includeTikz{0}{tube-k-k-N_L-N_-L+1-N_1-N_2-M_L-M_-L+1-M_1-M_2-X-X-X-B-A-Y_-L+frac-1-2-Y_frac-3-2}{\tube{k}{k'}{N_L}{N_{L+1}}{N_1}{N_2}{M_L}{M_{L+1}}{M_1}{M_2}{X}{X'}{X}{B}{A}{Y_{L +\frac{1}{2}}}{Y_\frac{3}{2}}}  ,
\end{align}
where $A \in \mc I_{\mc D^\star_\mc M}$, $B \in \mc I_{\mc D^\star_\mc N}$ and $X,X'$ represent isomorphism classes of simple objects in $\Fun_\mc D(\mc M, \mc N)$, while $k,k'$ label basis vectors in the hom-spaces $\Hom_{\Fun_{\mc D}(\mc M,\mc N)}(X',B \act X)$ and $\Hom_{\Fun_{\mc D}(\mc M,\mc N)}(X \cat A,X')$. As before, we distinguish two types of tensors entering the definition of these intertwining tubes. One the one hand, we have MPO tensors evaluating to $\omF{X}$-symbols. On the other hand, we have two fusion tensors of the form \eqref{eq:PEPSComp} that evaluates to the $\F{\fr F}$-symbols associated with the triples $(\mc M,\mc M,\mc N)$ and $(\mc M,\mc N,\mc N)$ of right $\mc D$-module categories, respectively. Note that in virtue of the composition of $\mc D$-module functors, $\Fun_{\mc D}(\mc M,\mc N)$ is equipped with a $(\mc D^\star_\mc N,\mc D^\star_\mc M)$-bimodule structure, hence the definition of the hom-spaces above.

It follows from the various pulling-through conditions that descend from the coherence axioms involving module associators, bimodule associators, module functors and composition of bimodule functors that these intertwining tubes can be pulled through local operators of the form \eqref{eq:bdryOp}. This operation gives rise to commutation relations of the form
\begin{equation}
    \fr T_{\mc M|\mc N}^{A,B,X,X', k, k'} \circ \mathbb H^{\mc M,A} = \mathbb H^{\mc N,B} \circ \fr T_{\mc M|\mc N}^{A,B,X,X', k, k'} \, .
\end{equation} 
Note however that there is no guarantee that there will exist non-vanishing intertwining tubes associated with any pair $(A,B)$ of boundary conditions. Indeed, not every boundary condition of a model associated with the $\mc D$-module category $\mc N$ is compatible with a given boundary condition of a dual model associated with the $\mc D$-module category $\mc M$. In order to obtain the mapping of topological sectors associated with the duality provided by $\mc D$-module functors in $\Fun_\mc D(\mc M, \mc N)$ further require to project $\mathbb H^{\mc M,A}$ and $\mathbb  H^{\mc N,B}$ onto specific symmetry charges. This is done via the symmetry tube idempotents with respect to the multiplication rule \eqref{eq:tubeAlgebra}. Given an intertwining tube associated with a pair $(A,B)$ of boundary conditions, if acting on both side with such charge projectors yields a non-trivial tensor, then the corresponding topological sectors are mapped onto one another by the duality. Repeating this process for every combination of boundary conditions and the corresponding symmetry charges provides the mapping of all the topological sectors realizing an equivalence $\mc Z(\mc D^\star_\mc M) \xrightarrow{\sim} \mc Z(\mc D^\star_\mc N)$. Mathematically, the existence of such an equivalence is guaranteed by the fact that $\mc D^\star_\mc M$ and $\mc D^\star_\mc N$ are Morita equivalent (see app.~\ref{sec:app_Morita}).

\bigskip \noindent
Similar to the symmetry tubes, one can define the multiplication of intertwining tubes via stacking, and use the recoupling theory of the fusion tensors given by the $\F{\circ}$-symbols to express the result as a linear combination of new intertwining tubes.\footnote{Note that one of the $X_i$ in the resulting $\F{\circ}^{X_1 X_2 X_3}$ symbols is always a simple object in some Morita dual of the  input fusion category, which implies that the potential obstructions to defining these symbols vanish.} The computation parallels that of the multiplication of symmetry tubes, yielding
\begin{align}
    \label{eq:tubeAlgebraIntertwiner}
    &\fr T^{A,B,X_1,X_1',k_1,k_1'}_{\mc M|\mc N} \cdot 
    \fr T^{B,C,X_2,X_2',k_2,k_2'}_{\mc N|\mc O}
    \\ \nn
    & \q =
    \sum_{\substack{X_3,X_3' \\ k_3,k_3' \\ \{j\}}}
    \big(\Fbar{\circ}^{C X_2 X_1}_{X_3'}\big)_{X_2',k_2 j_2}^{X_3,j_1 k_3} \, 
    \big(\F{\circ}^{X_2 B X_1}_{X_3'}\big)_{X_2',k_2' j_2}^{X_1',k_1 j_3}
    \\[-2.3em] \nn
    &\hspace{5em} \times \big(\Fbar{\circ}^{X_2 X_1 A}_{X_3'}\big)_{X_3,j_1 k_3'}^{X_1',k_1' j_3} \; 
    \fr T^{A,C,X_3,X_3',k_3,k_3'}_{\mc M|\mc O} \, .
\end{align}
Additionally, the Hermitian conjugate of an intertwining tube $\fr T_{\mc M|\mc N}$ can be expressed as a linear combination of the opposite intertwining tubes $\fr T_{\mc N|\mc M}$:
\begin{align}
    \nn
    \big(\fr T_{\mc M|\mc N}^{A,B,X,X',k,k'}\big)^\dag \! = \!\!\! 
    \sum_{\substack{i,i'\! ,j,j'\\ X''}} 
    &\big(\F{}^{X \bar X X}_{X}\big)_{\mathbb 1,11}^{\mathbb 1,11}
    \big(\F{}^{B X \bar X}_{B}\big)_{X',k i}^{\mathbb 1,11}
    \\[-1.3em]
    \nn
    &\times \big(\Fbar{}^{X A \bar X}_{B}\big)_{X',k' i}^{X'',j i'}
    \big(\F{}^{\bar X X X''}_{X}\big)_{\mathbb 1,11}^{B,i' j'}
    \\[.2em]
    \label{eq:tubeDaggerIntertwiner}
    &\times \fr T_{\mc N|\mc M}^{B,A,X,X',j,j'}.
\end{align}
Putting together symmetry tubes and intertwining tubes, we can construct a finite  ${}^*$-algebra spanned by tubes $\{\fr T_{\mc M|\mc M},\fr T_{\mc M|\mc N},\fr T_{\mc N|\mc M},\fr T_{\mc N|\mc N}\}$ where incompatible tube multiplications such as $\fr T_{\mc M|\mc M} \cdot \fr T_{\mc N|\mc N}$ are defined to be zero \cite{Neshveyev_2018,PhysRevB.105.085130}. As before, this ${}^*$-algebra is block diagonal in the topological sectors and can be decomposed into matrix units. The decomposition of the ${}^*$-subalgebras spanned by the symmetry tubes $\fr T_{\mc M|\mc M}$ and $\fr T_{\mc N|\mc N}$ is unchanged, while we also have
\begin{align}
    \nn
    e^{Z,A_i,\tilde Z,B_j}_{\mc M | \mc N} 
    &\equiv \#_{Z,\tilde Z} \!\!\! \sum_{X,X',k,k'} \!\!\!\!  d_X \, \Omega^{Z,A_i,\tilde Z,B_j}_{X,X',k,k'} \, \fr T_{\mc M|\mc N}^{A,B,X,X',k,k'} ,
    \\
    \big(e^{Z,A_i,\tilde Z,B_j}_{\mc M | \mc N}\big)^\dagger &= e^{\tilde Z,B'_j,Z,A_i}_{\mc N | \mc M} \, ,
\end{align}
where $Z,A_i$ are defined as before, $\tilde Z$ denotes a simple object of $\mc Z(\mc D^\star_\mc N)$ in the image of the topological sector $Z$ under the duality, and $B_j \equiv (B,j)$ refers to the simple object $B$ of $\mc D^\star_\mc N$ appearing in the decomposition into simple objects of $\tilde Z$ as an object of $\mc D^\star_\mc N$ with degeneracy label $i$. The normalization factor is now chosen to be $\#_{Z,\tilde Z} := (\ell(Z) \ell(\tilde Z))^\frac{1}{2} / {\rm FPdim}(\mc D)$. These new matrix units satisfy
\begin{equation}
    \begin{split}
        e^{Z,A_i,A'_j}_{\mc M | \mc M} \cdot e^{Z',A'_k,\tilde Z',B'_l}_{\mc M | \mc N} 
        &= \delta_{Z,Z'} \,  \delta_{j,k} \, e^{Z',A_i,\tilde Z',B'_l}_{\mc M | \mc N} \, ,
        \\
        e^{Z,A_i,\tilde Z,B_j}_{\mc M | \mc N} \cdot e^{\tilde Z',B_k,B'_l}_{\mc N | \mc N} 
        &= \delta_{Z,Z'} \, \delta_{j,k} \, e^{Z,A_i,\tilde Z,B'_l}_{\mc M | \mc N} \, ,
        \\
        e^{Z,A_i,\tilde Z,B_j}_{\mc M | \mc N} \cdot e^{\tilde Z',B_k,Z',A'_l}_{\mc N | \mc M} 
        &= \delta_{Z,Z'} \, \delta_{j,k} \,  e^{Z,A_i,A'_l}_{\mc M | \mc M} \, ,
        \\
        e^{\tilde Z,B_i,Z,A_j}_{\mc N | \mc M} \cdot e^{Z',A_k,\tilde Z',B'_l}_{\mc M | \mc N}  
        &= \delta_{Z,Z'} \, \delta_{j,k} \, e^{\tilde Z,B_i,B'_l}_{\mc N | \mc N} \, .
    \end{split}
\end{equation}
Using these matrix units, one can construct explicit isometries that relate a given Hamiltonian $\mathbb H^{\mc M,A}_{Z,i}$ in the topological sector $Z$ with degeneracy $i$ to a dual Hamiltonian $\mathbb H^{\mc N,B}_{\tilde Z,j}$ in the dual topological sector $\tilde Z$ with degeneracy $j$:
\begin{equation}
    \mathbb H^{\mc N,B}_{\tilde Z,j} 
    = 
    \big(e^{Z,A_i,\tilde Z,B_j}_{\mc M | \mc N}\big)^\dagger 
    \, \mathbb H^{\mc M,A}_{Z,i} \, 
    e^{Z,A_i,\tilde Z,B_j}_{\mc M | \mc N} \, ,
\end{equation}
thereby demonstrating that duality transformations do preserve the spectrum. 

\subsection{Illustration}

\noindent
Let us illustrate the concepts presented in this section with the case of the transverse-field Ising model. Starting from the fusion category $\mc D = \Vect_{\mathbb Z_2}$, we constructed in sec.~\ref{sec:illustration} local symmetric operators associated with a choice of module category $\mc M$ over $\Vect_{\mathbb Z_2}$. We identified the model associated with $\mc M = \Vect_{\mathbb Z_2}$ as the transverse-field Ising model and that associated with $\mc M = \Vect$ as its Kramers-Wannier dual. Moreover, we provided an explicit lattice operator that transforms symmetric operators of one model into symmetric operators of the other. Let us now examine the topological sectors of these two models and their mappings under the duality. 

Let us first consider the case $\mc M = \Vect_{\mathbb Z_2}$. By construction, boundary conditions are labelled by simple objects in $\Fun_{\Vect_{\mathbb Z_2}}(\Vect_{\mathbb Z_2},\Vect_{\mathbb Z_2}) \cong \Vect_{\mathbb Z_2}$ and by convention the site $L+2$ is defined to be the site 1. In order to construct the boundary operators as per eq.~\eqref{eq:bdryOp}, we require the MPO tensors that evaluate to the $\F{}$-symbols of $\Vect_{\mathbb Z_2}$:
\begin{equation}
    \includeTikz{0}{MPO-obj-obj-1-1-1-1-M_2-M_2-M_1-M_1-mathbb1-mathbb1m-3}{\MPO{obj}{obj}{1}{1}{1}{1}{M_2}{M_2}{M_1}{M_1}{\mathbb 1}{\mathbb 1/m}{3}} = 1
    = \includeTikz{0}{MPO-obj-obj-1-1-1-1-M_2-qmotimesM_2-qmotimesM_1-M_1-m-mathbb1m-3}{\MPO{obj}{obj}{1}{1}{1}{1}{M_2}{\q\;\, m \otimes M_2}{\q\;\, m \otimes M_1}{M_1}{m}{\mathbb 1/m}{3}} \, ,
\end{equation}
for any $M_1,M_2 \in \mc I_{\Vect_{\mathbb Z_2}}$ such that the fusion rules are everywhere satisfied. Let us focus for now on the Hamiltonian twisted by the identity object $\mathbb 1$. By definition, the Hamiltonian $\mathbb H^{\Vect_{\mathbb Z_2},\mathbb 1}$ acts on a Hilbert space consisting  of $L+1$ degrees of freedom. But the boundary condition imposes the constraint $M_{L+1} = M_1$. Consider the controlled NOT gate ${\rm c}X_{L+1,1}: \mathbb C^2 \otimes \mathbb C^2 \to \mathbb C^2 \otimes \mathbb C^2$ whose target and controlled qubits are those at sites $L+1$ and $1$, respectively. Concretely, we have
\begin{equation}
    \label{eq:cX}
    \begin{split}
        {\rm c}X_{L+1,1} &: | \mathbb 1 \ra \otimes | \mathbb 1 \ra \mapsto | \mathbb 1 \ra \otimes | \mathbb 1 \ra
        \\
        &: | \mathbb 1 \ra \otimes | m \ra \mapsto | m \ra \otimes | m \ra
        \\
        &: | m \ra \otimes | \mathbb 1 \ra \mapsto | m \ra \otimes | m \ra
        \\
        &: | m \ra \otimes | m \ra \mapsto | \mathbb 1 \ra \otimes | m \ra
    \end{split} \, .
\end{equation}
Applied to configurations where $M_{L+1} = M_1$, this unitary transformation amounts to fixing $M_{L+1} = \mathbb 1$, leaving an effective total Hilbert space with $L$ degrees of freedom. It follows that the boundary terms associated with $\mathbb 1$ are given by
\begin{align}
    \nn
    {\rm c}X_{L+1,1}^\dag \, \mathbb b^{\Vect_{\mathbb Z_2},\mathbb 1}_1 \, {\rm c}X_{L+1,1} 
    &= |\mathbb 1\ra \la \mathbb 1 |_{L+1} S^x_{1}
    \\[-3pt]
    & \! \stackrel{\rm eff.}{=} S^x_{1} \, ,
    \\
    \nn
    {\rm c}X_{L+1,1}^\dag \, \mathbb b^{\Vect_{\mathbb Z_2},\mathbb 1}_2 \, {\rm c}X_{L+1,1} 
    &= S^z_{L} (S^z |\mathbb 1\ra \la \mathbb 1 |)_{L+1} S^z_{1}
    \\[-3pt]
    & \! \stackrel{\rm eff.}{=} S^z_{L} S^z_{1} \, ,
\end{align}
leading to the Hamiltonian
\begin{equation*}
    {\rm c}X_{L+1,1}^\dag \, \mathbb H^{\Vect_{\mathbb Z_2},\mathbb 1}\,  {\rm c}X_{L+1,1} \stackrel{\rm eff.}{=} -J \sum_{\msf i = 1}^L (S^x_\msf i +g S^z_\msf i S^z_{\msf i+1}) \, ,
\end{equation*}
which is the original untwisted Hamiltonian on periodic boundary conditions. This model enjoys a $\mathbb Z_2$-symmetry generated by $\prod_{\msf i = 1}^L S_\msf i^x$ and as such decomposes into two charge sectors which are even and odd with respect to this symmetry operator, respectively. Let us notate via $([\mathbb 1],\ub 0)$ and $([\mathbb 1],\ub 1)$ the corresponding topological sectors.

Considering now the Hamiltonian twisted by the simple object $m$ in $\Vect_{\mathbb Z_2}$, the Hilbert space consists of $L+1$ degrees of freedom subject to the constraint that $m \otimes M_{L+1} = M_1$. Applying the unitary \eqref{eq:cX} to such configurations amounts to fixing $M_{L+1} = m$, again leaving an effective total Hilbert space with $L$ degrees of freedom. Importantly however, the $\mathbb 1$ and $m$-twisted Hilbert spaces are only effectively the same, but are in fact orthogonal. The boundary terms associated with $m$ are given by
\begin{align}
    \nn
    {\rm c}X_{L+1,1}^\dag \, \mathbb b^{\Vect_{\mathbb Z_2},m}_1 \, {\rm c}X_{L+1,1} 
    &= |\mathbb 1\ra \la \mathbb 1 |_{L+1} S^x_{1} 
    \\
    & \! \stackrel{\rm eff.}{=} S^x_{1} \, ,
    \\
    \nn
    {\rm c}X_{L+1,1}^\dag \, \mathbb b^{\Vect_{\mathbb Z_2},m}_2\,  {\rm c}X_{L+1,1} 
    &=  S^z_{L} (S^z |m\rangle \langle m|)_{L+1} S^z_{1}
    \\
    & \! \stackrel{\rm eff.}{=} - S^z_{L} S^z_{1} \, ,
\end{align}
leading to the Hamiltonian
\begin{align}
    &{\rm c}X_{L+1,1}^\dag \, \mathbb H^{\Vect_{\mathbb Z_2},m} \, {\rm c}X_{L+1,1} 
    \\[-0.6em] \nn
    &\q =  
    - J (S^x_1 - g S^z_{L} S^z_{1}) -
    J\sum_{\msf i = 2}^L (S^x_{\msf i} + g S^z_{\msf i - 1} S^z_{\msf i}) 
\end{align}
referred to as the \emph{antiperiodic} transverse-field Ising model. The model retains the same $\mathbb Z_2$-symmetry as in the periodic case and as such also decomposes into even and odd charge sectors. We notate via $([m],\ub 0)$ and $([m], \ub 1)$ the corresponding topological sectors.

Taking everything together, we find that the transverse-field Ising model possesses four topological sectors, namely $([\mathbb 1], \ub 0)$, $([\mathbb 1], \ub 1)$, $([m], \ub 0)$ and $([m], \ub 1)$. These are labelled by simple objects in $\mc Z(\Vect_{\mathbb Z_2})$ and are in one-to-one correspondence with elementary anyonic excitations of the (2+1)d toric code \cite{KITAEV20032}.

Let us now consider the case $\mc M= \Vect$. As per our construction, boundary conditions for this dual model are labelled by simple objects in $\Fun_{\Vect_{\mathbb Z_2}}(\Vect,\Vect) \cong \Rep(\mathbb Z_2)$ (see app.~\ref{sec:app_Morita}) and by convention the site $\frac{1}{2}$ is defined to be the site $L+\frac{1}{2}$. The Hamiltonian boundary terms defined as per eq.~\eqref{eq:bdryOp} requires MPO tensors evaluating to $\F{\caact}$-symbols of $\Vect$ as a $(\Rep(\mathbb Z_2),\Vect_{\mathbb Z_2})$-bimodule category:
\begin{equation}
    \includeTikz{0}{MPO-dot-dot-1-1-1-1-V-mathbb1-5}{\MPO{dot}{dot}{1}{1}{1}{1}{}{}{}{}{V}{\mathbb 1}{5}} = 1 =  \rho(m)  \includeTikz{0}{MPO-dot-dot-1-1-1-1-V-m-5}{\MPO{dot}{dot}{1}{1}{1}{1}{}{}{}{}{V}{m}{5}}
\end{equation}
where $V \equiv (V,\rho : \mathbb Z_2 \to {\rm End}(V))$ is an irreducible representation of $\mathbb Z_2$. Choosing $V$ to be the trivial representation $\ub 0$ immediately yields 
\begin{equation}
    \mathbb H^{\Vect,\ub 0} = -J \sum_{\msf i=1}^L(S_{\msf i-\frac{1}{2}}^x S_{\msf i+\frac{1}{2}}^x + gS^x_{\msf i+\frac{1}{2}}) \, ,
\end{equation}
i.e. the Kramers-Wannier dual of the transverse-field Ising model with periodic boundary conditions. The $\mathbb Z_2$-symmetry is now generated by $\prod_{\msf i=1}^L S_{\msf i+\frac{1}{2}}^z$ and the topological sectors associated with the even and odd charge sectors are notated via $([\mathbb 1],\ub 0)$ and $([\mathbb 1],\ub 1)$, respectively. Similarly, choosing $V$ to be the sign representation $\ub 1$ yields
\begin{equation*}
    \mathbb H^{\Vect,\ub 1} = 
    - J (S^x_{L+\frac{1}{2}}S^x_{\frac{3}{2}} - g S^z_\frac{3}{2}) - \!
    J\sum_{\msf i = 2}^L (S_{\msf i-\frac{1}{2}}^x S_{\msf i+\frac{1}{2}}^x + gS^z_{\msf i+\frac{1}{2}}) 
\end{equation*}
and the corresponding topological sectors are denoted by $([m],\ub 0)$ and $([m],\ub 1)$.

Let us now establish the mapping of topological sectors under the duality associated with the unique simple object in $\Fun_{\Vect_{\mathbb Z_2}}(\Vect_{\mathbb Z_2},\Vect) \cong \Vect$. We already presented in sec.~\ref{sec:illustration} the MPO intertwiner that performs the mapping of the local symmetric operators. In order to account for boundary conditions it needs to be promoted to intertwining tubes of the form eq.~\eqref{eq:tubeIntertwiner}, which in turn requires the introduction of the following tensors:
\begin{align}
        \label{eq:interVecZ2Vec}
        \includeTikz{0}{spePEPS-obj-dot-1-1-1-1-mathbb1-V-3}{\spePEPS{obj}{dot}{1}{1}{1}{1}{}{}{\mathbb 1}{}{}{V}{3}}
        &= 1 = \rho(m) \includeTikz{0}{spePEPS-obj-dot-1-1-1-1-m-V-3}{\spePEPS{obj}{dot}{1}{1}{1}{1}{}{}{m}{}{}{V}{3}} \, ,
        \\[-2em]
        \includeTikz{0}{spePEPS-dot-obj-1-1-1-1-mathbb1m-mmathbb1-m-4}{\spePEPS{dot}{obj}{1}{1}{1}{1}{\mathbb 1/m\; }{\; m/ \mathbb 1}{}{}{}{m}{4}}
        &= 1 =
        \includeTikz{0}{spePEPS-dot-obj-1-1-1-1-mathbb1m-mathbb1m-mathbb1-4}{\spePEPS{dot}{obj}{1}{1}{1}{1}{\mathbb 1/m\; }{\; \mathbb 1/m}{}{}{}{\mathbb 1}{4}}
        \, ,
\end{align}
for any $(V,\rho) \in \mc I_{\Rep(\mathbb Z_2)}$. Let us consider for instance the topological sector $([\mathbb 1],\ub 1)$, i.e. the odd sector of the periodic transverse-field Ising model $\mathbb H^{\Vect_{\mathbb Z_2},\ub 1}$. Selecting this topological sector amounts to acting with the projector $e^{([\mathbb 1],\ub 1),\mathbb 1,\mathbb 1}_{\Vect_{\mathbb Z_2}| \Vect_{\mathbb Z_2}} = \frac{1}{2}(\fr T_{\Vect_{\mathbb Z_2}|\Vect_{\mathbb Z_2}}^{\mathbb 1, \mathbb 1,\mathbb 1,\mathbb 1,1,1} - \fr T_{\Vect_{\mathbb Z_2}| \Vect_{\mathbb Z_2}}^{\mathbb 1,\mathbb 1,m,m,1,1})$. It follows from eq.~\eqref{eq:interVecZ2Vec} that when pulling this projector through the intertwining tube $\fr T_{\Vect_{\mathbb Z_2}|\Vect}^{\mathbb 1, \ub 0,\mathbb 1, \mathbb 1,1,1}$, it becomes $e^{([m],\ub 0),\mathbb 1,\mathbb 1}_{\Vect| \Vect} = \frac{1}{2}(\fr T_{\Vect|\Vect}^{\ub 1, \ub 1,\mathbb 1,\mathbb 1,1,1} + \fr T_{\Vect| \Vect}^{\ub 1,\ub 1,m,m,1,1})$. Performing analogous computations for the other sectors confirms that under this duality, the topological sectors $([\mathbb 1],\ub 1)$ and $([m],\ub 0)$ are swapped, while the topological sectors $([\mathbb 1],\ub 0)$ and $([m],\ub 1)$ remain unchanged.

\section{Examples with $\Rep(\mc S_3)$ symmetry\label{sec:Examples}}

\noindent
\emph{We present in this section a series of examples that are mathematically non-trivial and physically relevant, thereby showcasing the potential and merits of our approach. The examples we consider all have symmetry operators organized into fusion categories that are in the Morita class of $\Rep(\mc S_3)$, namely the category of (finite-dimensional) representations of the symmetric group $\mc S_3$.}

\subsection{Foreword}

\noindent
Before presenting the examples, let us mention a couple of different ways the constructions presented in this manuscript can be used in practice. 

On the one hand, we can pick a module category $\mc M$ over a given fusion category $\mc D$, make a choice of boundary condition, and construct a lattice Hamiltonian by considering any linear combination of local operators as defined in eq.~\eqref{eq:localOp}. The resulting Hamiltonian is guaranteed to be symmetric with respect to operators organized into $\mc D^\star_\mc M$, which is in the Morita class of $\mc D$ by definition. The same linear combinations of local operators but for a different choice of module category yields a dual lattice Hamiltonian. The corresponding duality operator is then provided by the intertwining tubes given in eq.~\eqref{eq:tubeIntertwiner}. This is the recipe we would typically follow when wishing to define a new family of lattice Hamiltonians satisfying certain symmetry conditions. 

On the other hand, our method can be employed in order to investigate new properties and duality relations of a known lattice model. This first requires to rewrite the Hamiltonian of interest within our framework. To do so, we must identify a suitable choice of input fusion category $\mc D$. Such an input fusion category can be chosen to be any Morita dual of a subcategory of symmetry operators of the starting Hamiltonian. We then need to pick the appropriate $\mc D$-module category so we can find a linear combination of local operators \eqref{eq:localOp} so as to recover the Hamiltonian. Concretely, let us consider for instance a lattice Hamiltonian with a $G$-symmetry so that the category of symmetry operators is the category $\Vect_G$ of $G$-graded vector spaces. For any subgroup $H$ of $G$, any Morita dual of the fusion category $\Vect_H$ is a valid choice of input fusion category $\mc D$, even though $\Vect_H$ does not capture the whole symmetry of the Hamiltonian. There will then be a choice of module category over $\mc D$ such that there exists a linear combination of local operators \eqref{eq:localOp} producing the desired Hamiltonian. Module categories over $\mc D$ then classify duals of this Hamiltonian with respect to the (sub)symmetry encoded into $\Vect_H$. Indeed, any duality relation between two models is always with respect to a given symmetry. This implies in particular that choosing a small input fusion category reduces the number of dual models that can be considered. The same reasoning applies for any one-dimensional lattice model satisfying some categorical symmetry.

This latter scenario is the one we explore in this section. Throughout, the input fusion category is chosen to be the category $\mc D=\Rep(\mc S_3)$ of finite-dimensional representations of the symmetric group $\mc S_3$. For a specific choice of $\Rep(\mc S_3)$-module category, we show how to recover the Heisenberg XXZ model within our framework. After studying its topological sectors, we explore dual models associated with various choices of module categories over $\Rep(\mc S_3)$, and construct the explicit lattice operators performing the corresponding duality transformations.

\subsection{Local operators\label{sec:RepS3localOp}}

\noindent
The input fusion category $\mc D$ of the models we consider is the category $\Rep(\mc S_3)$ of finite-dimensional representations of the symmetric group $\mc S_3 = \la r,s \, | \, r^2 = \mathbb 1 = s^3, rsr=s^2 \ra $. Recall that this fusion category has three simple objects provided by the irreducible representations of the group. We denote these simple objects as $\ub 0,\ub 1,\ub 2$ and refer to them as the trivial, sign and two-dimensional irreducible representations, respectively. The fusion rules read $\ub 0 \otimes Y \simeq Y$, with $Y$ any simple object in $\Rep(\mc S_3)$, $\ub 1 \otimes \ub 1 \simeq \ub 0$, $\ub 1 \otimes \ub 2 \simeq \ub 2$ and $\ub 2 \otimes \ub 2 \simeq \ub 0 \oplus \ub 1 \oplus \ub 2$ (see app.~\ref{sec:app_S3} for a brief review of the element structure and representation theory of $\mc S_3$). 

For every $A \subseteq \mc S_3$ subgroup, the category $\Rep(A)$ is a module category over $\Rep(\mc S_3)$ via the restriction functor ${\rm Res}^{\mc S_3}_A: \Rep(\mc S_3) \to \Rep(A)$ such that $M \cat Y :=  M \otimes {\rm Res}^{\mc S_3}_A(Y) $ for every $M \in \Rep(A)$ and $Y \in \Rep(\mc S_3)$. Every $\Rep(\mc S_3)$-module category is of this form. Given a $\Rep(\mc S_3)$-module category $\mc M$, we consider the Hamiltonian
\begin{equation}
    \label{eq:RepS3Ham}
    \mathbb H^\mc M = J_{\ub 2} \sum_{\msf i} \mathbb b_{\msf i,\ub 2}^{\mc M} - J_{\ub 1} \sum_{\msf i} \mathbb b_{\msf i,\ub 1}^{\mc M}   
\end{equation}
with local operators
\begin{align}
    \label{eq:localOpXXZ}
    \mathbb b_{\msf i,Y}^{\mc M} &= 
    \includeTikz{0}{ladder-1-1-ub2-ub2-ub2-ub2-Y-2}{\ladder{1}{1}{\ub 2}{\ub 2}{\ub 2}{\ub 2}{Y}{2}} \, ,
\end{align}
which are obtained by contracting symmetric tensors evaluating to $\F{\cat}$-symbols of $\mc M$. 
By construction, the Hamiltonian $\mathbb H^\mc M$ is left invariant by the action of MPOs \eqref{eq:symMPO} labelled by objects in $\mc D^\star_\mc M$. In order to write down the Hamiltonian and the symmetry operators more explicitly, we now need to consider the different module categories $\mc M$ separately. 

\subsection{$\mc M = \Vect$}

\noindent
The first $\Rep(\mc S_3)$-module category we consider is $\mc M = \Vect$, which amounts to choosing the trivial subgroup of $\mc S_3$. We notate the unique simple object in $\Vect$ via $\mathbb 1 \simeq \mathbb C$ such that $\mathbb 1 \cat Y \simeq \mathbb 1$ for any $Y \in \mc I_{\mc D}$. For this example, the $\F{\cat}$-symbols reduce to \emph{Clebsch-Gordan} coefficients of $\mc S_3$, i.e. basis vectors of hom-spaces of the form $\Hom_{\Rep(\mc S_3)}(Y_1 \otimes Y_2,Y_3)$:
\begin{equation}
    \big(\F{\cat}^{\mathbb 1 Y_1 Y_2}_{\mathbb 1} \big)^{Y_3,1k}_{\mathbb 1,ij}
    = \CC{Y_1}{Y_2}{Y_3}{i}{j}{k} \, ,
\end{equation}
where $i,j,k$ label basis vectors in the vectors spaces $Y_1$, $Y_2$ and $Y_3$ underlying the corresponding representations, respectively.
We review the definition of these Clebsch-Gordan coefficients in app.~\ref{sec:app_S3}. Given the definition of the local operators \eqref{eq:localOpXXZ}, we only require the following non-vanishing $\F{\cat}$-symbols:
\begin{align*}
    \big(\F{\cat}^{\mathbb 1 \ub 2 \ub 2}_{\mathbb 1}\big)^{\ub 2,12}_{\mathbb 1,11} = 
    \CC{\ub 2}{\ub 2}{\ub 2}{1}{1}{2}  &= 1 
    =
    \CC{\ub 2}{\ub 2}{\ub 2}{2}{2}{1} =
    \big(\F{\cat}^{\mathbb 1 \ub 2 \ub 2}_{\mathbb 1}\big)^{\ub 2,11}_{\mathbb 1,22} \, ,
    \\
    \big(\F{\cat}^{\mathbb 1 \ub 1 \ub 2}_{\mathbb 1}\big)^{\ub 2,11}_{\mathbb 1,11} =
    \CC{\ub 1}{\ub 2}{\ub 2}{1}{1}{1} &= 1
    = - \CC{\ub 1}{\ub 2}{\ub 2}{1}{2}{2} = -
    \big(\F{\cat}^{\mathbb 1 \ub 1 \ub 2}_{\mathbb 1}\big)^{\ub 2,12}_{\mathbb 1,12} \, ,
    \\
    \big(\F{\cat}^{\mathbb 1 \ub 2 \ub 1}_{\mathbb 1}\big)^{\ub 2,12}_{\mathbb 1,21} =
    \CC{\ub 2}{\ub 1}{\ub 2}{2}{1}{2} &= 1
    = - \CC{\ub 2}{\ub 1}{\ub 2}{1}{1}{1} = -
    \big(\F{\cat}^{\mathbb 1 \ub 2 \ub 1}_{\mathbb 1}\big)^{\ub 2,11}_{\mathbb 1,11} \, .
\end{align*}
Let us consider the Hilbert space \eqref{eq:Hilbert} with no regard for the boundary conditions at the moment. Since the module category $\Vect$ has a unique simple object and the local operators $\mathbb b_{\msf i,Y}^{\mc M}$ constrain the $\mc D$-strings to be labelled by $\ub 2$, the only physical degrees of freedom are basis vectors in hom-spaces $\Hom_{\mc M}(\mathbb 1 \cat \ub 2,\mathbb 1) \simeq \mathbb C^2 \simeq \mathbb C[|1 \ra , |2 \ra]$. Therefore, the effective microscopic Hilbert space is isomorphic to $\bigotimes_\msf i \mathbb C^2$. In other words, we are dealing with a Hamiltonian acting on spin-$\frac{1}{2}$ particles located at half-integer sites of the lattice. How does the local operator $\mathbb b^\mc M_{\msf i, \ub 2}$ act on this effective Hilbert space? Notice first that in virtue of the contraction of the symmetric tensors, the local operator $\mathbb b^\mc M_{\msf i, \ub 2}$ is a sum of two terms. It then follows from the definition of the $\F{\cat}$-symbols that it acts by projecting out states for which the hom-space basis vectors at sites $\msf i-\frac{1}{2}$ and $\msf i+\frac{1}{2}$ agree, and acts as $|1,2\ra \mapsto |2,1\ra$ or $|2,1\ra \mapsto |1,2\ra$ otherwise. Similarly, the operator $\mathbb b^\mc M_{\msf i, \ub 1}$ acts as the identity operator when the hom-space basis vectors at sites $\msf i-\frac{1}{2}$ and $\msf i+\frac{1}{2}$ agree, and minus the identity operator otherwise. Putting everything together, we find that the Hamiltonian \eqref{eq:RepS3Ham} in the infinite chain case boils down to
\begin{align*}
    \nn
    \label{eq:XXZ}
    \mathbb H^{\Vect} &= \frac{J_{\ub 2}}{2} \sum_{\msf i} (S^x_{\msf i-\frac{1}{2}} S^x_{\msf i+\frac{1}{2}} + S^y_{\msf i-\frac{1}{2}} S^y_{\msf i+\frac{1}{2}}) + J_{\ub 1} \sum_{\msf i} S^z_{\msf i-\frac{1}{2}} S^z_{\msf i+\frac{1}{2}} 
    \\
    & = J_{\ub 2} \sum_{\msf i} (S^+_{\msf i-\frac{1}{2}} S^-_{\msf i+\frac{1}{2}} + S^-_{\msf i-\frac{1}{2}} S^+_{\msf i+\frac{1}{2}}) + J_{\ub 1} \sum_{\msf i} S^z_{\msf i-\frac{1}{2}} S^z_{\msf i+\frac{1}{2}} 
    \, ,
\end{align*}
where we introduced the notation $S^{\pm} := \frac{1}{2}(S^x \pm iS^y)$. This is the spin-$\frac{1}{2}$ Heisenberg XXZ model.

What is the symmetry of the Hamiltonian $\mathbb H^{\Vect}$? On the one hand, there is a $\mathbb Z_2$ symmetry generated by $\prod_{\msf i} S^x_{\msf i+\frac{1}{2}}$, which acts in particular as $S^{\pm} \mapsto S^{\mp}$. On the other hand, there is a $\mathbb Z_3$ symmetry generated by $\prod_{\msf i} \begin{psmallmatrix} \omega & 0 \\ 0 & \bar \omega \end{psmallmatrix}_{\msf i+\frac{1}{2}}$ with $\omega = e^{2i\pi/3}$, which acts in particular as $S^{\pm} \mapsto \omega^{\pm 1}S^{\pm}$. Crucially, these symmetry operators do not commute so that we have an overall $\mathbb Z_2 \ltimes \mathbb Z_3 \simeq \mc S_3$ symmetry, whereby $\mathbb Z_2$ acts on $\mathbb Z_3$ by complex conjugation. Let us confirm that this is indeed what our general theory predicts. By construction, we know that the Hamiltonian $\mathbb H^{\Vect}$ is left invariant by symmetry operators organized into the fusion category $(\Rep(\mc S_3))^\star_{\Vect}$. But as a fusion category $(\Rep(\mc S_3))^\star_{\Vect}$ is equivalent to $\Vect_{\mc S_3}$ (see app.~\ref{sec:app_Morita}), which means that $\mathbb H^\Vect$ is left invariant by symmetry operators labelled by group variables $g \in \mc S_3$. Our construction further provides an explicit expression for these symmetry operators in the form of MPOs via 
\begin{equation}
    \label{eq:symMPOS3Vec}
    \includeTikz{0}{MPO-dot-dot-1-1-j-i-ub2-g-1}{\MPO{dot}{dot}{1}{1}{j}{i}{}{}{}{}{\ub 2}{g}{1}} = \rho(g)^i_j \, ,
\end{equation}
where $\rho : \mc S_3 \to {\rm End}(\ub 2)$ is the representation matrix of $\ub 2$. It is immediate to confirm that the $\mathbb Z_2$ symmetry is generated by the MPO labelled by the order 2 group element $r \in \mc S_3$, whereas the $\mathbb Z_3$ symmetry is generated by the MPO labelled by the order 3 group element $s \in \mc S_3$.

Let us now examine the topological sectors. Consider a spin chain of length $L+1$. Generally speaking, boundary conditions of the Hamiltonian $\mathbb H^\Vect$ are of the form
\begin{equation}
    S_{L+\frac{3}{2}} := K^{-1}S_\frac{3}{2}K \, ,
\end{equation}
with $S=S^x,S^y,S^z$ and $K \in {\rm End}(\mathbb C^2)$ a unitary matrix. In particular, we can choose $K=\rho(g)$ with $\rho: \mc S_3 \to {\rm End}(\ub 2)$ for any $g \in \mc S_3$. These boundary conditions, which are labelled by simple objects in $(\Rep(\mc S_3))^\star_\Vect \cong \Vect_{\mc S_3}$, do correspond to those predicted by our approach.\footnote{The same way picking a fusion category $\mc D$ amounts to focusing on a specific subsymmetry of the total symmetry of the Hamiltonian, it prescribes a certain subset of boundary conditions compatible with the subsymmetry.} Recall that in general, given a simple object in $\Vect_{\mc S_3}$, the corresponding boundary condition is provided by the local operators defined in eq.~$\eqref{eq:bdryOp}$, where the MPO tensors appearing in these local operators are rotated versions of the symmetry MPO tensor \eqref{eq:symMPOS3Vec}. It follows that boundary conditions are provided by the local operators
\begin{align}
        \nn
        &\mathbb b^{\Vect,g}_{\ub 2} = 
        \rho(g)^1_1 \, S^+_{L+\frac{1}{2}} S^-_{\frac{3}{2}}
        +
        \rho(g)^1_2 \, S^+_{L+\frac{1}{2}} S^+_{\frac{3}{2}} 
        +
        {\rm h.c.} \, ,
        \\
        \label{eq:S3VecBdry}
        &\mathbb b^{\Vect,g}_{\ub 1} = \sigma(g) \, S^z_{L+\frac{1}{2}} S^z_\frac{3}{2} \, ,
\end{align}
where $\sigma: \mc S_3 \to \End(\ub 1)$ is the character of the sign representation $\ub 1$. We can now explicitly check that this definition agrees with that proposed above. For instance, for $\rho(g)=\rho(r)=\begin{psmallmatrix}0 & 1 \\ 1 & 0\end{psmallmatrix}$ one has
\begin{align}
    \nn
    \mathbb b^{\Vect,r}_{\ub 2} &= S^+_{L+\frac{1}{2}}S^+_\frac{3}{2} + S^-_{L+\frac{1}{2}}S^-_\frac{3}{2}\\
    &= S^+_{L+\frac{1}{2}}\rho(r)S^-_\frac{3}{2}\rho(r) + S^-_{L+\frac{1}{2}}\rho(r)S^+_\frac{3}{2}\rho(r)\\
    \mathbb b^{\Vect,r}_{\ub 1} &= - S^z_{L+\frac{1}{2}} S^z_{\frac{3}{2}} = S^z_{L+\frac{1}{2}}\rho(r) S^z_\frac{3}{2}\rho(r)  
    \label{eq:S3VecBdryR}\, ,
\end{align}
which readily agrees with eq.~\eqref{eq:S3VecBdry}. 

Now that we have confirmed that the boundary conditions prescribed by our tensor network approach agree with the usual definitions, let us analyze these boundary conditions in more detail. Naturally, the boundary condition labelled by $g = \mathbb 1$ corresponds to a periodic spin chain. In this case, the Hamiltonian has the same $\mc S_3$ symmetry as in the infinite chain case, and therefore the Hilbert space decomposes into charge sectors labelled by the three irreducible representations of $\mc S_3$. The boundary condition labelled by $g=r$ corresponds to anti-periodic spin chain. Interestingly, it is clear from eq.~\eqref{eq:S3VecBdryR} that the resulting Hamiltonian is no longer $\mc S_3$ symmetric but merely $\mathbb Z_2$ symmetric, and as such decomposes into charge sectors labelled by irreducible representations of $\mathbb Z_2$. This is true for any (twisted) anti-periodic boundary condition labelled by representatives of the conjugacy class $[r] = \{r,rs,rs^2\}$. Finally, the boundary condition labelled by $g=s$ corresponds to a twisted periodic spin chain. We can readily check that the resulting Hamiltonian retains the  $\mathbb Z_3$ symmetry of the periodic case, whereas the $\mathbb Z_2$ symmetry is lost. Therefore, the model decomposes into charge sectors labelled by irreducible representations of $\mathbb Z_3$. Akin to the previous scenario, these statements are valid for any twisted periodic boundary condition labelled by representatives of the conjugacy class $[s] = \{s,s^2\}$. Overall, we find eight topological sectors that are in one-to-one correspondence with simple objects in $\Mod(\mc D(\mc S_3))$, i.e. simple modules over the Drinfel'd double of $\mc S_3$. This is in agreement with the general results presented in the previous section. Indeed, our approach predicts that topological sectors are organized into the Drinfel'd center $\mc Z(\Vect_{\mc S_3})$, which is equivalent to $\Mod(\mc D(\mc S_3))$ as we review in app.~\ref{sec:app_VecG} in the case of an arbitrary finite group $G$.

\subsection{$\mc M = \Rep(\mathbb Z_2)$\label{sec:ExRepZ2}}

\noindent
The second $\Rep(\mc S_3)$-module category we consider is $\mc M = \Rep(\mathbb Z_2)$, whose simple objects are denoted by $\ub 0_{\mathbb Z_2}$ and $\ub 1_{\mathbb Z_2}$. As mentioned in sec.~\ref{sec:RepS3localOp}, the module action is given by $M \cat Y = M \otimes {\rm Res}^{\mc S_3}_{\mathbb Z_2}(Y)$, for every $M \in \Rep(\mathbb Z_2)$ and $Y \in \Rep(\mc S_3)$, with the restriction functor fully specified by its action on the simple objects of $\Rep(\mc S_3)$, namely ${\rm Res}^{\mc S_3}_{\mathbb Z_2}(\ub 0) \simeq \ub 0_{\mathbb Z_2}$, ${\rm Res}^{\mc S_3}_{\mathbb Z_2}(\ub 1) \simeq \ub 1_{\mathbb Z_2}$ and ${\rm Res}^{\mc S_3}_{\mathbb Z_2}(\ub 2) \simeq \ub 0_{\mathbb Z_2} \oplus \ub 1_{\mathbb Z_2}$. The non-vanishing $\F{\cat}$-symbols that are relevant for our construction are given by 
\begin{align*}
    \big(\F{\cat}^{\ub 0_{\mathbb Z_2}\ub 2 \ub 2}_{\ub 1_{\mathbb Z_2}}\big)^{\ub 2,11}_{\ub 1_{\mathbb Z_2},11}
    &=
    \big(\F{\cat}^{\ub 1_{\mathbb Z_2}\ub 2 \ub 2}_{\ub 0_{\mathbb Z_2}}\big)^{\ub 2,11}_{\ub 1_{\mathbb Z_2},11}    
    =
    \big(\F{\cat}^{\ub 1_{\mathbb Z_2}\ub 2 \ub 2}_{\ub 1_{\mathbb Z_2}}\big)^{\ub 2,11}_{\ub 0_{\mathbb Z_2},11}
    \\
    &= \big(\F{\cat}^{\ub 0_{\mathbb Z_2}\ub 2 \ub 2}_{\ub 0_{\mathbb Z_2}}\big)^{\ub 2,11}_{\ub 0_{\mathbb Z_2},11} 
    = \frac{1}{\sqrt{2}} \, ,
    \\
    \big(\F{\cat}^{\ub 1_{\mathbb Z_2}\ub 2 \ub 2}_{\ub 0_{\mathbb Z_2}}\big)^{\ub 2,11}_{\ub 0_{\mathbb Z_2},11} 
    &=
    \big(\F{\cat}^{\ub 0_{\mathbb Z_2}\ub 2 \ub 2}_{\ub 1_{\mathbb Z_2}}\big)^{\ub 2,11}_{\ub 0_{\mathbb Z_2},11}    
    =
    \big(\F{\cat}^{\ub 0_{\mathbb Z_2}\ub 2 \ub 2}_{\ub 0_{\mathbb Z_2}}\big)^{\ub 2,11}_{\ub 1_{\mathbb Z_2},11}
    \\
    &= \big(\F{\cat}^{\ub 1_{\mathbb Z_2}\ub 2 \ub 2}_{\ub 1_{\mathbb Z_2}}\big)^{\ub 2,11}_{\ub 1_{\mathbb Z_2},11} 
    = -\frac{1}{\sqrt{2}} \, ,
\end{align*}
and 
\begin{align*}
    \big(\F{\cat}^{M_1 \ub 1 \ub 2}_{M_2} \big)^{\ub 2,11}_{M_1 \otimes \ub 1_{\mathbb Z_2},11} = 1 =
    \big(\F{\cat}^{M_1 \ub 2 \ub 1}_{M_2} \big)^{\ub 2,11}_{M_1 \otimes \ub 1_{\mathbb Z_2},11} \, ,
\end{align*}
for any $M_1,M_2 \in \mc I_{\mc M}$.

As for the previous choice of module category, let us begin by writing down the Hamiltonian in the infinite spin chain case. The parametrization of the microscopic Hilbert space significantly differs from that of the previous scenario. The local operators $\mathbb b_{\msf i,Y}^{\mc M}$ always constrain the $\mc D$-strings to be labelled by $\ub 2$. Moreover, it follows from the definition of the module action that hom-spaces in $\mc M$ are all one-dimensional. The only fluctuating physical degrees of freedom thus correspond to the $\mc M$-strings. We thus find an effective microscopic Hilbert space that is still isomorphic to $\bigotimes_\msf i \mathbb C^2$, where the spin-$\frac{1}{2}$ particles are now located at integer sites of the lattice. How does the local operator $\mathbb b_{\msf i,\ub 2}^\mc M$ act on this effective Hilbert space? Identifying $|\ub 0_{\mathbb Z_2}\ra$ and $| \ub 1_{\mathbb Z_2}\ra$ with the $+1$ and $-1$ eigenvectors of the Pauli $S^z$ operator, respectively, it follows from the definitions of the $\F{\cat}$-symbols that $\mathbb b_{\msf i, \ub 2}^\mc M$ acts as a projector onto the state 
$|+\ra$ at site $\msf i$ if the degrees of freedom at sites $\msf i-1$ and $\msf i+1$ agree, and minus this projector otherwise. The local $\mathbb b_{\msf i, \ub 1}^\mc M$ simply acts as the Pauli $S^x$ operator at site $\msf i$. Putting everything together, we find that the Hamiltonian \eqref{eq:RepS3Ham} in the infinite chain case boils down to
\begin{align*}
    \mathbb H^{\Rep(\mathbb Z_2)} = \frac{J_{\ub 2}}{2} \sum_{\msf i} (S^z_{\msf i-1} \mathbb 1_\msf i S^z_{\msf i+1} + S^z_{\msf i-1} S^x_{\msf i} S^z_{\msf i+1})
    - J_{\ub 1} \sum_{\msf i} S^x_{\msf i} \, .
\end{align*}
What is the symmetry of this Hamiltonian? Unlike the Hamiltonian $\mathbb H^{\Vect}$, it is very difficult to identify the symmetry without relying on the fact that we obtained this Hamiltonian using tensor networks satisfying certain pulling-through conditions. Specifically, we know by construction that the Hamiltonian $\mathbb H^{\Rep(\mathbb Z_2)}$ is left invariant by symmetry operators organized into the fusion category $(\Rep(\mc S_3))^\star_{\Rep(\mathbb Z_2)}$, which happens to be equivalent to $\Rep(\mc S_3)$. Given the  representative $X$ of an isomorphism class of simple objects in $\Rep(\mc S_3)$, the corresponding symmetry operator is provided by an MPO generated by tensors that evaluate to $\F{\caact}$-symbols of $\Rep(\mathbb Z_2)$ as a $(\Rep(\mc S_3),\Rep(\mc S_3))$-bimodule category. Naturally, the symmetry MPO associated with the trivial representation $\ub 0$ acts trivially, whereas the symmetry MPO labelled by $\ub 1$ can readily be checked to act as $\prod_\msf i S^x_\msf i$, which does commute the Hamiltonian $\mathbb H^{\Rep(\mathbb Z_2)}$. What about the symmetry MPO labelled by $\ub 2$? Let us try to write it down as explicitly as possible. The non-vanishing $\F{\caact}$-symbols relevant to this operator are
\begin{align*}
    \big(\F{\caact}^{\ub 2 \ub 1_{\mathbb Z_2} \ub 2}_{\ub 0_{\mathbb Z_2}} \big)^{\ub 0_{\mathbb Z_2},11}_{\ub 0_{\mathbb Z_2},11} 
    =
    \big(\F{\caact}^{\ub 2 \ub 0_{\mathbb Z_2} \ub 2}_{\ub 1_{\mathbb Z_2}} \big)^{\ub 0_{\mathbb Z_2},11}_{\ub 0_{\mathbb Z_2},11} 
    &= \frac{-\sqrt{3}}{2} \, ,
    \\
    \big(\F{\caact}^{\ub 2 \ub 0_{\mathbb Z_2} \ub 2}_{\ub 0_{\mathbb Z_2}} \big)^{\ub 1_{\mathbb Z_2},11}_{\ub 0_{\mathbb Z_2},11} 
    =
    \big(\F{\caact}^{\ub 2 \ub 0_{\mathbb Z_2} \ub 2}_{\ub 0_{\mathbb Z_2}} \big)^{\ub 0_{\mathbb Z_2},11}_{\ub 1_{\mathbb Z_2},11}
    &= \frac{-\sqrt{3}}{2} \, ,
    \\
    \big(\F{\caact}^{\ub 2 \ub 0_{\mathbb Z_2} \ub 2}_{\ub 1_{\mathbb Z_2}} \big)^{\ub 1_{\mathbb Z_2},11}_{\ub 1_{\mathbb Z_2},11}
    =
    \big(\F{\caact}^{\ub 2 \ub 1_{\mathbb Z_2} \ub 2}_{\ub 0_{\mathbb Z_2}} \big)^{\ub 1_{\mathbb Z_2},11}_{\ub 1_{\mathbb Z_2},11} 
    &= \frac{\sqrt{3}}{2} \, ,
    \\
    \big(\F{\caact}^{\ub 2 \ub 1_{\mathbb Z_2} \ub 2}_{\ub 1_{\mathbb Z_2}} \big)^{\ub 0_{\mathbb Z_2},11}_{\ub 1_{\mathbb Z_2},11} 
    =
    \big(\F{\caact}^{\ub 2 \ub 1_{\mathbb Z_2} \ub 2}_{\ub 1_{\mathbb Z_2}} \big)^{\ub 1_{\mathbb Z_2},11}_{\ub 0_{\mathbb Z_2},11}
    &= \frac{\sqrt{3}}{2} \, ,
\end{align*}
\begin{align*}
    \big(\F{\caact}^{\ub 2 \ub 0_{\mathbb Z_2} \ub 2}_{\ub 0_{\mathbb Z_2}} \big)^{\ub 0_{\mathbb Z_2},11}_{\ub 0_{\mathbb Z_2},11} 
    =
    \big(\F{\caact}^{\ub 2 \ub 1_{\mathbb Z_2} \ub 2}_{\ub 1_{\mathbb Z_2}} \big)^{\ub 1_{\mathbb Z_2},11}_{\ub 1_{\mathbb Z_2},11} 
    &= -\frac{1}{2} \, ,
    \\
    \big(\F{\caact}^{\ub 2 \ub 0_{\mathbb Z_2} \ub 2}_{\ub 0_{\mathbb Z_2}} \big)^{\ub 1_{\mathbb Z_2},11}_{\ub 1_{\mathbb Z_2},11} 
    =
    \big(\F{\caact}^{\ub 2 \ub 1_{\mathbb Z_2} \ub 2}_{\ub 1_{\mathbb Z_2}} \big)^{\ub 0_{\mathbb Z_2},11}_{\ub 0_{\mathbb Z_2},11} 
    &= \frac{1}{2} \, ,
    \\
    \big(\F{\caact}^{\ub 2 \ub 1_{\mathbb Z_2} \ub 2}_{\ub 0_{\mathbb Z_2}} \big)^{\ub 1_{\mathbb Z_2},11}_{\ub 0_{\mathbb Z_2},11} 
    =
    \big(\F{\caact}^{\ub 2 \ub 0_{\mathbb Z_2} \ub 2}_{\ub 1_{\mathbb Z_2}} \big)^{\ub 0_{\mathbb Z_2},11}_{\ub 1_{\mathbb Z_2},11} 
    &= \frac{1}{2} \, ,
    \\
    \big(\F{\caact}^{\ub 2 \ub 1_{\mathbb Z_2} \ub 2}_{\ub 0_{\mathbb Z_2}} \big)^{\ub 0_{\mathbb Z_2},11}_{\ub 1_{\mathbb Z_2},11}
    =
    \big(\F{\caact}^{\ub 2 \ub 0_{\mathbb Z_2} \ub 2}_{\ub 1_{\mathbb Z_2}} \big)^{\ub 1_{\mathbb Z_2},11}_{\ub 0_{\mathbb Z_2},11} 
    &= \frac{1}{2} 
\end{align*}
and
\begin{align*}
    \big(\F{\caact}^{\ub 2 \ub 0_{\mathbb Z_2} \ub 1}_{\ub 0_{\mathbb Z_2}} \big)^{\ub 1_{\mathbb Z_2},11}_{\ub 1_{\mathbb Z_2},11}
    =
    \big(\F{\caact}^{\ub 2 \ub 1_{\mathbb Z_2} \ub 1}_{\ub 1_{\mathbb Z_2}} \big)^{\ub 0_{\mathbb Z_2},11}_{\ub 0_{\mathbb Z_2},11} &= 1 \, ,
    \\
    \big(\F{\caact}^{\ub 2 \ub 0_{\mathbb Z_2} \ub 1}_{\ub 1_{\mathbb Z_2}} \big)^{\ub 0_{\mathbb Z_2},11}_{\ub 1_{\mathbb Z_2},11}
    =
    \big(\F{\caact}^{\ub 2 \ub 1_{\mathbb Z_2} \ub 1}_{\ub 0_{\mathbb Z_2}} \big)^{\ub 1_{\mathbb Z_2},11}_{\ub 0_{\mathbb Z_2},11} &= -1 \, .
\end{align*}
We notice immediately that the first set of $\F{\caact}$-symbols satisfy a symmetry condition whereby the entries only depend on the number of times the representation $\ub 1_{\mathbb Z_2}$ occurs in the symbol. This symmetry condition can be exploited in order to rewrite the symmetry MPO associated with the above symbols as an MPO of the form
\begin{equation}
    \includeTikz{0}{simpMPO}{\simpMPO}
\end{equation}
with building blocks
\begin{equation}
    \includeTikz{0}{simpMPOTensor-a-b}{\simpMPOTensor{a}{b}{}{}{}{}} \equiv 
    \frac{1}{2}
    \begin{pmatrix}
    -\mathbb 1 & -\sqrt{3}S^x \\ -\sqrt{3}\mathbb 1 & S^x
    \end{pmatrix}_{\! ab} \, .
\end{equation}
This MPO should be interpreted as an operator acting on the integer sites of the effective Hilbert space. Let us now use this alternative form of the symmetry MPO labelled by $\ub 2$ to explicitly check its commutation relation with  $\mathbb H^{\Rep(\mathbb Z_2)}$. The building block defined above verifies the following symmetry conditions:
\begin{align*}
    \!\!\! \includeTikz{0}{simpMPOTensor-Sz-Sz-Sz}{\simpMPOTensor{}{}{}{S^z}{S^z}{S^z}} \!\!\! = \!\!\!     \includeTikz{0}{simpMPOTensor-O-Sz-Sz}{\simpMPOTensor{}{}{O}{S^z}{}{S^z}} \!\!\! = \!\!\!
    \includeTikz{0}{simpMPOTensor-Sz-mathbb1+Sx-O-mathbb1+Sx}{\simpMPOTensor{}{}{S^z}{\mathbb 1 + S^x}{O}{\mathbb 1 + S^x}} \!\!\! = \!\!\!
    \includeTikz{0}{simpMPOTensor}{\simpMPOTensor{}{}{}{}{}{}} \!\! ,
\end{align*}
where $O := \frac{1}{2}(\sqrt{3}S^x-S^z)$. Let us briefly derive the last equality. Since $(\mathbb 1+S^x)|0/1\ra = \sqrt{2}|+\ra = \sqrt{2} H |0\ra$ with $H$ the Hadamard matrix, and
\begin{align}
    \includeTikz{0}{simpMPOTensor-a-b-H-H}{\simpMPOTensor{a}{b}{}{H}{}{H}} &= \frac{1}{2}
    \begin{pmatrix}
        -1 & -\sqrt{3} \\ -\sqrt{3} & 1  
    \end{pmatrix}_{\! ab}
    \\[-2em] \nn
    &= \frac{1}{2}(-\sqrt{3}S^x-S^z)_{ab} \, ,
\end{align}
it is implied by $S^z \frac{1}{2}(-\sqrt{3}S^x-S^z)O = \frac{1}{2}(-\sqrt{3}S^x-S^x)$. It then follows from the three symmetry conditions above that the MPO commutes with $S^z_{\msf i-1}(\mathbb 1+S^x)_\msf i S^z_{\msf i+1}$, for any $\msf i$, thus it commutes with $\mathbb H^{\Rep(\mathbb Z_2)}$, thereby confirming that the MPO labelled by $\ub 2$ is indeed a symmetry operator. Putting everything together, this confirms the $\Rep(\mc S_3)$ symmetry of $\mathbb H^{\Rep(\mathbb Z_2)}$. The exercise we just carried out exemplifies how non-trivial it may be to explicitly confirm certain properties of a given Hamiltonian, although these properties are immediate once the model has been recast within our framework.

Let us now examine the topological sectors. Consider a spin chain of length $L+1$. We know from our general construction that topological sectors are in one-to-one correspondence with simple objects of $\mc Z(\Rep(\mc S_3))$, which is equivalent to $\mc Z(\Vect_{\mc S_3})$ considered in the previous scenario in virtue of the Morita equivalence between $\Rep(\mc S_3)$ and $\Vect_{\mc S_3}$ (see app.~\ref{sec:app_Morita}). Therefore, we must find eight topological sectors corresponding to the simple objects $\Mod(\mc D(\mc S_3))$ as for $\mc M=\Vect$. The Hamiltonian $\mathbb H^{\Rep(\mathbb Z_2)}$ admits three types of boundary conditions labelled by simple objects in $\Rep(\mc S_3)$. Naturally, choosing $\ub 0$ identifies the degrees of freedom at sites $L+1$ and $1$ and thus corresponds to a periodic chain, in which case the Hamiltonian has the same $\Rep(\mc S_3)$ symmetry as in the infinite case. We can then check that the Hilbert space decomposes into charge sectors indexed by conjugacy classes of the group, providing the simple objects labelled by $([\mathbb 1],\ub 0)$, $([r],\ub 0_{\mathbb Z_2})$ and $([s],\ub 0_{\mathbb Z_3})$ of $\mc Z(\Rep(\mc S_3))$ in the notation of app.~\ref{sec:app_Hopf}. More generally, given a boundary condition labelled by a simple object $A$ in $\Rep(\mc S_3)$, the corresponding Hilbert space decomposes in charge sectors indexed by the simple objects of $\mc Z(\Rep(S_3))$, which, when treated as (typically not simple) objects in $\Rep(\mc S_3)$, contain $A$ as subrepresentation (see app.~\ref{sec:app_S3}). We can obtain the explicit boundary conditions associated with the other simple objects in $\Rep(\mc S_3)$ applying our general construction. Recall that a boundary condition is provided by the local operators defined in eq.~$\eqref{eq:bdryOp}$, where the MPO tensors appearing in these local operators are rotated versions of the symmetry MPOs tensor evaluating to the $\F{\caact}$-symbols defined previously. Concretely, the boundary condition labelled by $\ub 1$ identifies the degree of freedom at site $L+1$ and the image of that at site $1$ under the Pauli $S^x$ operator, so we still have an effective Hilbert space with $L$ degrees of freedom, and the local operator reads
\begin{align}
    \mathbb b^{\Rep(\mathbb Z_2),\ub 1}_{\ub 2} &= \frac{1}{2}(S^z_L \mathbb 1_1 S^z_{2} - S^z_L S^x_1 S^z_2)\, ,\\
    \mathbb b^{\Rep(\mathbb Z_2),\ub 1}_{\ub 1} &= S^x_1 \, .
\end{align}
The corresponding charge sectors are then given by $([\mathbb 1],\ub 1)$, $([r],\ub 1_{\mathbb Z_2})$ and $([s],\ub 0_{\mathbb Z_3})$. Finally, our construction stipulates that the boundary conditions labelled by $\ub 2$ is provided by 
\begin{align}
    \label{eq:ExRepZ2NonTrivBdry}    
    \mathbb b^{\Rep(\mathbb Z_2),\ub 2}_{\ub 2} &= \frac{1}{4}(-S^z_L S^z_{L+1}S^z_1S^z_2
    +\sqrt{3}S^z_L S^x_{L+1}S^z_1S^z_2 
    \\
    \nn
    &+\sqrt{3} S^z_L S^z_{L+1}S^x_1S^z_2 
    +S^z_L S^x_{L+1}S^x_1S^z_2) \, ,
    \\
    \mathbb b^{\Rep(\mathbb Z_2),\ub 2}_{\ub 1} &= - S^x_{L+1}S_1^x
\end{align}
and the corresponding charge sectors are labelled by $([\mathbb 1],\ub 2)$, $([r], \ub 0_{\mathbb Z_2})$, $([r], \ub 1_{\mathbb Z_2})$, $([s],\ub 1_{\mathbb Z_3})$ and $([s],\ub 1_{\mathbb Z_3}^*)$.
Notice the doubling of physical degrees of freedom taking place at the boundary in contrast to the other cases, i.e. this model is effectively defined on a chain of $L+1$ spins as opposed to $L$. This is an explicit example of a \emph{non-abelian} boundary condition. In sec.~\ref{sec:dualityRepS3}, we shall provide the lattice operator performing the duality relation between $\mathbb H^{\Rep(\mathbb Z_2)}$ and $\mathbb H^\Vect$, as well as the mapping of all the topological sectors under this duality relation.

\subsection{$\mc M=\Rep(\mathbb Z_3)$}

\noindent
We pursue the exploration of this Morita class of models with $\mc M=\Rep(\mathbb Z_3)$, whose simple objects are denoted by $\ub 0_{\mathbb Z_3}$, $\ub 1_{\mathbb Z_3}$ and $\ub 1^*_{\mathbb Z_3}$. As earlier, the module action is given by $M \cat Y = M \otimes {\rm Res}^{\mc S_3}_{\mathbb Z_3}(Y)$, for every $M \in \Rep(\mathbb Z_3)$ and $Y \in \Rep(\mc S_3)$, with the restriction fully specified by its action on the simple objects of $\Rep(\mc S_3)$, namely ${\rm Res}^{\mc S_3}_{\mathbb Z_3}(\ub 0) \simeq \ub 0_{\mathbb Z_3}$, ${\rm Res}^{\mc S_3}_{\mathbb Z_3}(\ub 1) \simeq \ub 0_{\mathbb Z_3}$ and ${\rm Res}^{\mc S_3}_{\mathbb Z_3}(\ub 2) \simeq \ub 1_{\mathbb Z_3} \oplus \ub 1^*_{\mathbb Z_3}$. The non-vanishing $\F{\cat}$-symbols relevant for our construction are 
\begin{align*}
    \big( \F{\cat}^{\ub 0_{\mathbb Z_3} \ub 1 \ub 2}_{\ub 1_{\mathbb Z_3}} \big)^{\ub 2,11}_{\ub 0_{\mathbb Z_3},11}
    = - \big( \F{\cat}^{\ub 0_{\mathbb Z_3} \ub 1 \ub 2}_{\ub 1^*_{\mathbb Z_3}} \big)^{\ub 2,11}_{\ub 0_{\mathbb Z_3},11} = 1 \, ,
    \\
    \big( \F{\cat}^{\ub 0_{\mathbb Z_3} \ub 2 \ub 1}_{\ub 1_{\mathbb Z_3}} \big)^{\ub 2,11}_{\ub 0_{\mathbb Z_3},11} 
    =
    \big( \F{\cat}^{\ub 0_{\mathbb Z_3} \ub 2 \ub 1}_{\ub 1^*_{\mathbb Z_3}} \big)^{\ub 2,11}_{\ub 0_{\mathbb Z_3},11} = 1 \, ,
    \\
    \big( \F{\cat}^{\ub 1_{\mathbb Z_3} \ub 1 \ub 2}_{\ub 0_{\mathbb Z_3}} \big)^{\ub 2,11}_{\ub 1_{\mathbb Z_3},11} 
    = - 
    \big( \F{\cat}^{\ub 1_{\mathbb Z_3} \ub 1 \ub 2}_{\ub 1^*_{\mathbb Z_3}} \big)^{\ub 2,11}_{\ub 1_{\mathbb Z_3},11} = 1 \, ,
    \\
   \big( \F{\cat}^{\ub 1_{\mathbb Z_3} \ub 2 \ub 1}_{\ub 0_{\mathbb Z_3}} \big)^{\ub 2,11}_{\ub 1_{\mathbb Z_3},11} 
   = -
    \big( \F{\cat}^{\ub 1_{\mathbb Z_3} \ub 2 \ub 1}_{\ub 1^*_{\mathbb Z_3}} \big)^{\ub 2,11}_{\ub 1_{\mathbb Z_3},11} = 1 \, ,
    \\
    \big( \F{\cat}^{\ub 1^*_{\mathbb Z_3} \ub 1 \ub 2}_{\ub 0_{\mathbb Z_3}} \big)^{\ub 2,11}_{\ub 1^*_{\mathbb Z_3},11} 
    = -
    \big( \F{\cat}^{\ub 1^*_{\mathbb Z_3} \ub 1 \ub 2}_{\ub 1_{\mathbb Z_3}} \big)^{\ub 2,11}_{\ub 1^*_{\mathbb Z_3},11} = 1 \, ,
    \\
    \big( \F{\cat}^{\ub 1^*_{\mathbb Z_3} \ub 2 \ub 1}_{\ub 0_{\mathbb Z_3}} \big)^{\ub 2,11}_{\ub 1^*_{\mathbb Z_3},11}
    =
    \big( \F{\cat}^{\ub 1^*_{\mathbb Z_3} \ub 2 \ub 1}_{\ub 1_{\mathbb Z_3}} \big)^{\ub 2,11}_{\ub 1^*_{\mathbb Z_3},11} = -1 
\end{align*}
and
\begin{equation*}
    \big( \F{\cat}^{Y_1 \ub 2 \ub 2}_{Y_2}\big)^{\ub 2,11}_{Y_3,11} = 1 \, ,
\end{equation*}
where $Y_1,Y_2,Y_3 \in \Rep(\mc S_3)$ are all required to be distinct from one another, i.e. each irreducible representation of $\mc S_3$ can only appear once in $\{Y_1,Y_2,Y_3\}$.

As usual, we begin by writing down the Hamiltonian in the infinite chain case. The microscopic Hilbert space underlying this model differs from that of the previous scenarios. First of all, the local operators still constrain the $\mc D$-strings to be labelled by $\ub 2$ and the hom-spaces in $\mc M$ are all one-dimensional. The only fluctuating physical degrees of freedom thus correspond to the $\mc M$-strings. Since the category $\mc M$ counts three isomorphism classes of simple objects, we are now dealing with a system of spin-1 particles, in sharp contrast with the previous cases. Moreover, given an object $M$ in $\Rep(\mathbb Z_3)$, it follows from the fusion rules in $\Rep(\mathbb Z_3)$ that $M$ does not appear in the decomposition of $M \cat \ub 2 \simeq M \otimes (\ub 1_{\mathbb Z_3} \oplus \ub 1_{\mathbb Z_3}^\star)$. As such, strands $\msf i$ and $\msf i+1$ cannot be labelled by the same object $M$---this is confirmed by the definition of the $\F{\cat}$-symbols above. We thus find an effective microscopic Hilbert space of spin-1 particles located at integer sites of the lattice, which is not a tensor product of local Hilbert spaces given the kinematical constraint we just described. We notate this microscopic Hilbert space via $\mc H^{\Rep(\mathbb Z_3)}$. Within this microscopic Hilbert space, it follows from the definition of the $\F{\cat}$-symbols that $\mathbb b_{\msf i, \ub 2}^\mc M$ modifies the label of the strand $\msf i$ to whichever other label---if any---is allowed by the kinematical constraint, and acts as the zero operator if this is not possible. Similarly, $\mathbb b_{\msf i, \ub 1}^\mc M$ acts on the strands $\msf i-1$, $\msf i$ and $\msf i+1$ as the identity operator whenever the labels of the strands $\msf i-1$ and $\msf i+1$ are identical and minus the identity operator otherwise. Putting everything together, the Hamiltonian is given by 
\begin{align}
        \mathbb H^{\Rep(\mathbb Z_3)} = \;&J_{\ub 2} \sum_{\msf i} (\ms S^x_{\msf i} + ({\ms S}^x_{\msf i})^\dagger) 
        \\[-.3em] \nn
        - &\frac{J_{\ub 1}}{3} \sum_{\msf i}  \big( \ms S^z_{\msf i-1}\mathbb 1_{\msf i} ({\ms S}^z_{\msf i+1})^\dagger - \ms S^z_{\msf i-1} \ms S^z_{\msf i} \ms S^z_{\msf i+1} 
        + \text{h.c.} \big)
\end{align}
with
\begin{equation}
    \ms S^x := \begin{pmatrix}
        0 & 1 & 0\\
        0 & 0 & 1\\
        1 & 0 & 0
    \end{pmatrix} ,\q
    \ms S^z := \begin{pmatrix}
        1 & 0 & 0\\
        0 & \omega & 0\\
        0 & 0 & \omega^2
    \end{pmatrix} ,
\end{equation}
so that $\ms S^x \ms S^z = \omega \ms S^z \ms S^x$. Notice that given certain configurations of spin-1 variables, acting with $\ms S_{\msf i}^x + ({\ms S}^x_{\msf i})^\dagger$ may bring the corresponding state outside of $\mc H^{\Rep(\mathbb Z_2)}$, in which case it is projected out by definition of the $\F{\cat}$-symbols.

Our construction predicts that the Hamiltonian $\mathbb H^{\Rep(\mathbb Z_3)}$ is left invariant by operators organized into the fusion category $(\Rep(\mc S_3))^\star_{\Rep(\mathbb Z_3)}\cong \Vect_{\mc S_3}$, i.e. $\mathbb H^{\Rep(\mathbb Z_3)}$ has an $\mc S_3$ symmetry. This symmetry works as follows: Firstly, notice that $\Rep(\mathbb Z_3)$ and $\Vect_{\mathbb Z_3}$ are equivalent as fusion categories. Secondly, identifying the simple objects $\mathbb C_g$ in $\Vect_{\mathbb Z_3}$ with the corresponding left cosets $M = g \mathbb Z_2$, the left $\Vect_{\mc S_3}$-module structure of $\Vect_{\mathbb Z_3}$ is given by $\mathbb C_g \act M := (g \msf r(M))\mathbb Z_2$ for any $g \in \mc S_3$ and $M \in \mc S_3/\mathbb Z_2$, where $\msf r(M)$ denotes the representative of $M$. Therefore, the symmetry operator labelled by $g \in \mc S_3$ simply acts on the local Hilbert space $\mathbb C^3$ associated with every spin-1 particle by permutation of the coordinates. The fact that $\mathbb H^{\Rep(\mathbb Z_3)}$ commutes with these symmetry operators finally follows from $\ms S^z \ms S^z = \omega \ms S^z \ms S^x$, $\ms C (\ms S^x + (\ms S^x)^\dagger) = (\ms S^x + (\ms S^x)^\dagger) \ms C$ and $\ms C \ms S^z = (\ms S^z)^\dagger \ms C$, where
\begin{align}
    \ms C := 
    \begin{pmatrix}
        1 & 0 & 0\\
        0 & 0 & 1\\
        0 & 1 & 0
    \end{pmatrix}  .
\end{align}
This can be readily confirmed using the symmetry MPOs whose building blocks evaluate to $\F{\caact}$-symbols of $\Rep(\mathbb Z_3)$ as a $(\Vect_{\mc S_3},\Rep(\mc S_3))$-bimodule category.

Boundary conditions are labelled by simple objects in $\Vect_{\mc S_3}$ and can be implemented using our general recipe. The analysis of the topological sectors then parallel that of the model $\mathbb H^{\Vect}$, that is, conjugacy classes of $\mc S_3$ define equivalence classes of boundary conditions and the corresponding charge sectors are labelled by irreducible representations of the centralizer of the conjugacy class. Topological sectors are found to correspond to simple objects of $\mc Z(\Vect_{\mc S_3})$ as expected.

\subsection{$\mc M=\Rep(\mc S_3)$}

\noindent 
The remaining $\Rep(\mc S_3)$-module category is $\Rep(\mc S_3)$ itself. This model is both the simplest model to define---since all the data we need is provided by the monoidal structure of $\Rep(\mc S_3)$---and the most difficult model to analyse explicitly. Indeed, 
contrary to the previous scenarios, we do not know how to rewrite the resulting Hamiltonian in terms of spin operators so we can hardly make the local operators and the symmetry operators as explicit. On the bright side, this illustrates the need for a systematic category theoretical approach in general. 

By definition, the $\F{\cat}$-symbols of $\Rep(\mc S_3)$ as a $\Rep(\mc S_3)$-module category coinciding with the $\F{}$-symbols of $\Rep(\mc S_3)$. The latter are typically referred to as $6j$-symbols and are obtained by contracting Clebsch-Gordan coefficients. It was shown in ref.~\cite{Lootens:2021tet} that in the case where we choose module categories over themselves, the lattice models constructed following our approach boils down to so-called \emph{anyonic chains} \cite{PhysRevLett.98.160409,PhysRevB.87.235120,PhysRevLett.101.050401,PhysRevLett.103.070401,ardonneAnyonChain,Buican:2017rxc}. In particular, the $\Rep(\mc S_3)$ anyonic chain associated with our choice of local operator \eqref{eq:localOpXXZ} was studied in \cite{Finch_2013,PhysRevB.94.085138}. As such, we shall not review this model explicitly here and encourage the reader to consult the above references for detail. Let us merely stress the fact that since $\Fun_{\Rep(\mc S_3)}(\Rep(\mc S_3),\Rep(\mc S_3)) \cong \Rep(\mc S_3)$ the model is left invariant by symmetry operators encoded into $\Rep(\mc S_3)$ as for $\mathbb H^{\Rep(\mathbb Z_2)}$.

\subsection{Duality $\Rep(\mathbb Z_2) \to \Vect$\label{sec:dualityRepS3}}

\noindent
By definition, the Hamiltonians $\mathbb H^{\Vect}$, $\mathbb H^{\Rep(\mathbb Z_2)}$, $\mathbb H^{\Rep(\mathbb Z_3)}$ and $\mathbb H^{\Rep(\mc S_3)}$ constructed above are all dual to one another. Duality relies upon the fact that they only differ by a choice of module category over $\Rep(\mc S_3)$. Crucially, symmetry operators of these various models are encoded into categories that are Morita equivalent, although they are not necessarily equivalent as fusion categories. Morita equivalence ensures that the center of the categories of symmetry operators are equivalent as braided fusion categories, which is crucial to being able to map the topological sectors of one model onto those of another. We explained in sec.~\ref{sec:tubeInt} how to explicitly perform such a mapping via intertwining tubes. We illustrate in this section our construction with the duality between the models $\mathbb H^{\Vect}$ and $\mathbb H^{\Rep(\mathbb Z_2)}$.

Let us first compute the mapping of local operators. Note that this mapping immediately provides the duality relation in the infinite chain case. Inspecting the local operators, we notice that $\mathbb H^{\Vect}$ can be obtained from $\mathbb H^{\mathbb Z_2}$ via the mapping
\begin{equation}
    \label{eq:ExDualityMappings}
    \begin{split}
        S^x_\msf i &\mapsto -S^z_{\msf i-\frac{1}{2}} S^z_{\msf i+\frac{1}{2}} 
        \\
        S^z_{\msf i} S^z_{\msf i+1}  & \mapsto S^x_{\msf i+\frac{1}{2}}
    \end{split} \, .
\end{equation}
Indeed, due to $S^y = i S^xS^z$, we have for instance
\begin{align}
    \mathbb b_{\msf i, \ub 2}^{\Rep(\mathbb Z_2)} = \; &S^z_{\msf i-1}\mathbb 1_\msf i S^z_{\msf i+1} + S^z_{\msf i-1}S^x_\msf i S^z_{\msf i+1} 
    \\ \nn 
    &\mapsto
    S^x_{\msf i-\frac{1}{2}}S^x_{\msf i+\frac{1}{2}} + S^y_{\msf i-\frac{1}{2}}S^y_{\msf i+\frac{1}{2}} = \mathbb b_{\msf i, \ub 2}^{\Vect} 
    \, .
\end{align}
Let us now prove that this mapping is provided by a $\Rep(\mc S_3)$-module functor in $\Fun_{\Rep(\mc S_3)}(\Rep(\mathbb Z_2),\Vect)$, as predicted by the general construction. First of all, we can show that $\Fun_{\Rep(\mc S_3)}(\Rep(\mathbb Z_2),\Vect) \cong \Rep(\mathbb Z_3)$, so that we distinguish three duality maps between $\mathbb H^\Vect$ and $\mathbb H^{\Rep(\mathbb Z_2)}$ labelled by simple objects in $\Rep(\mathbb Z_3)$. These duality maps cannot be distinguished locally in the sense that they all perform the same transformations of local operators. Locally then, without loss of generality, we can focus on the duality operator labelled by the simple object $\ub 0_{\mathbb Z_3}$ in $\Rep(\mathbb Z_3)$. Globally however, they may perform different mappings of the states within topological sectors, and one needs to consider all the duality operators.

Constructing the MPO intertwiners that perform the transformation of the local operators only requires one type of tensors, namely those that evaluate to the $\omF{\ub 0_{\mathbb Z_3}\,}$-symbols of the $\Rep(\mc S_3)$-module functor in $\Fun_{\Rep(\mc S_3)}(\Rep(\mathbb Z_2),\Vect)$ associated with $\ub 0_{\mathbb Z_3} \in \Rep(\mathbb Z_3)$. However, directly computing these tensors may prove challenging. Conveniently, these can be obtained via the following composition of module functors:  
\begin{align}
    \nn
    &\Fun_{\Rep(
    \mc S_3)}(\Rep(\mc S_3),\Vect) \times \Fun_{\Rep(\mc S_3)}(\Rep(\mathbb Z_2),\Rep(\mc S_3)) 
    \\ 
    & \q \to \Fun_{\Rep(\mc S_3)}(\Rep(\mathbb Z_2),\Vect) \, .
\end{align}
On the one hand, the MPO intertwiner labelled by the unique object in $\Fun_{\Rep(\mc S_3)}(\Rep(\mc S_3),\Vect) \cong \Vect$ evaluates to Clebsch-Gordan coefficients of $\Rep(\mc S_3)$, i.e.
\begin{equation}
    \includeTikz{0}{MPO-dot-i-j-l-1-Y_3-Y_1-ub2-2}{\MPO{}{dot}{i}{j}{l}{1}{}{}{Y_3}{Y_1}{\ub 2}{}{2}} =\CC{Y_1}{\ub 2}{Y_3}{i}{l}{j} \, .
\end{equation}
On the other hand, simple objects $X$ in $\Fun_{\Rep(\mc S_3)}(\Rep(\mathbb Z_2),\Rep(\mc S_3)) \cong \Rep(\mathbb Z_2)^{\rm op}$ label MPO intertwiners whose building blocks read
\begin{equation}
    \includeTikz{0}{MPO-mod-1-1-1-1-Y_1-Y_3-M_2-M_1-ub2-X-4}{\MPO{mod}{}{1}{1}{1}{1}{Y_1}{Y_3}{M_2}{M_1}{\ub 2}{X}{4}} = 
    \big(\Fbar{\cat}^{X Y_1 \ub 2}_{M_2}\big)^{Y_3,11}_{M_1,11} \, ,
\end{equation}
where $\F{\cat}$ here refers to the $\Rep(\mc S_3)$-module associator of $\Rep(\mathbb Z_2)$. We provided the relevant entries for these tensors in sec.~\ref{sec:ExRepZ2}. Importantly, composing module functors labelled by simple objects in the relevant categories do not yield simple objects in $\Fun_{\Rep(\mc S_3)}(\Rep(\mathbb Z_2),\Vect)$. This means that after contracting the MPO intertwiners presented above, we must decompose the resulting MPO into simple blocks. Doing so, we are able to find the MPO intertwiner labelled by $\ub 0_{\mathbb Z_3}$ whose building block is of the form
\begin{equation}
    \label{eq:ExMPOInt}
    \includeTikz{0}{MPO-mod-dot-1-1-i-1-M_2-M_1-ub2-ub0_-mathbbZ_3-1}{\MPO{mod}{dot}{1}{1}{i}{1}{}{}{M_2}{M_1}{\ub 2}{\ub 0_{\mathbb Z_3}}{1}} =
    \big( \omF{\ub 0_{\mathbb Z_3} \, }^{M_1 \ub 2}_{\mathbb 1}\big)^{M_2,11}_{\mathbb 1,1i} 
\end{equation}
for any $M_1,M_2 \in \mc I_{\Rep(\mathbb Z_2)}$ and basis vector $|i=1,2 \ra$ in $\Hom_{\Vect}(\mathbb C \cat \ub 2,\mathbb C) \simeq \mathbb C^2$.
The non-vanishing $\omF{\ub 0_{\mathbb Z_3}\,}$-symbols that are relevant to our computations are found to be
\begin{align*}
    \big( \omF{\ub 0_{\mathbb Z_3}\,}^{\ub 0_{\mathbb Z_2} \ub 2}_{\mathbb 1} \big)^{\ub 0_{\mathbb Z_2},11}_{\mathbb 1,11}
    =
    \big( \omF{\ub 0_{\mathbb Z_3}\,}^{\ub 0_{\mathbb Z_2} \ub 2}_{\mathbb 1} \big)^{\ub 0_{\mathbb Z_2},11}_{\mathbb 1,12}
    &= \frac{1}{\sqrt{2}} \, ,
    \\
    \big( \omF{\ub 0_{\mathbb Z_3}\,}^{\ub 0_{\mathbb Z_2} \ub 2}_{\mathbb 1} \big)^{\ub 1_{\mathbb Z_2},11}_{\mathbb 1,11}
    =
    -\big( \omF{\ub 0_{\mathbb Z_3}\,}^{\ub 0_{\mathbb Z_2} \ub 2}_{\mathbb 1} \big)^{\ub 1_{\mathbb Z_2},11}_{\mathbb 1,12}
    &= \frac{1}{\sqrt{2}} \, ,
    \\
    \big( \omF{\ub 0_{\mathbb Z_3}\,}^{\ub 1_{\mathbb Z_2} \ub 2}_{\mathbb 1} \big)^{\ub 0_{\mathbb Z_2},11}_{\mathbb 1,11}
    =
    -\big( \omF{\ub 0_{\mathbb Z_3}\,}^{\ub 1_{\mathbb Z_2} \ub 2}_{\mathbb 1} \big)^{\ub 0_{\mathbb Z_2},11}_{\mathbb 1,12}
    &= -\frac{1}{\sqrt{2}} \, ,
    \\
    \big( \omF{\ub 0_{\mathbb Z_3}\,}^{\ub 1_{\mathbb Z_2} \ub 2}_{\mathbb 1} \big)^{\ub 1_{\mathbb Z_2},11}_{\mathbb 1,11}
    =
    \big( \omF{\ub 0_{\mathbb Z_3}\,}^{\ub 1_{\mathbb Z_2} \ub 2}_{\mathbb 1} \big)^{\ub 1_{\mathbb Z_2},11}_{\mathbb 1,12}
    &= -\frac{1}{\sqrt{2}} \, .
\end{align*}
Let us check that this MPO intertwiner indeed performs the mappings given in eq.~\eqref{eq:ExDualityMappings}. Given the specific form of \eqref{eq:ExMPOInt} and the values of the $\omF{\ub 0_{\mathbb Z_3} \,}$-symbols, one can rewrite the resulting MPO intertwiner in a more traditional form as 
\begin{equation}
    \includeTikz{0}{simpMPOInt}{\simpMPOInt}
\end{equation}
with building blocks
\begin{equation}
    \begin{split}
        \delta(a,j)\delta(b,j) &\equiv \includeTikz{-14}{deltaTensor-a-b-j}{\deltaTensor{a}{b}{j}} 
        \\[-2.5em]
        \raisebox{2pt}{\includeTikz{12}{simpMPOIntTensor-a-b-i}{\simpMPOIntTensor{a}{b}{i}{}{}{}}} &\equiv \frac{1}{\sqrt{2}}(-1)^{b|i-1|+a|i-2|}
    \end{split}\, .
\end{equation}
It is now immediate that the MPO tensor verifies the following symmetry conditions
\begin{equation*}
    \!\!\!\!  \includeTikz{12}{simpMPOIntTensor-Sz-Sx}{\simpMPOIntTensor{}{}{}{}{S^z}{S^x}}
    \!\!\!  = \!  -  \!\!\!   \includeTikz{12}{simpMPOIntTensor-Sx-Sz}{\simpMPOIntTensor{}{}{}{S^x}{S^z}{}}
    \!\!\! = \!\!\! \includeTikz{12}{simpMPOIntTensor-Sz-Sx-Sz}{\simpMPOIntTensor{}{}{}{S^z}{S^x}{S^z}}
    \!\!\!  = \!\!\! \includeTikz{12}{simpMPOIntTensor}{\simpMPOIntTensor{}{}{}{}{}{}} \!\!\! ,
\end{equation*}
which in turn provide the mappings given in eq.~\eqref{eq:ExDualityMappings}, as expected.

We checked how the MPO intertwiners labelled by simple objects in $\Fun_{\Rep(\mc S_3)}(\Rep(\mathbb Z_2),\Vect)$ transform the local symmetric operators entering the definition of $\mathbb H^{\Rep(\mathbb Z_2)}$ into those entering the definition of $\mathbb H^{\Vect}$. In order to fully establish the duality relations between these models, we must further explain how topological sectors are mapped under these transformations. We know from the general construction that topological sectors of $\mathbb H^{\Vect}$ and $\mathbb H^{\Rep(\mathbb Z_2)}$ are associated with isomorphism classes of simple objects in $\mc Z(\Vect_{\mc S_3})$ and $\mc Z(\Rep(\mc S_3))$, respectively. Naturally, since $\Vect_{\mc S_3}$ and $\Rep(\mc S_3)$ are Morita equivalent, we have $\mc Z(\Vect_{\mc S_3}) \cong \mc Z(\Rep(\mc S_3))$, which in turns guarantees the correspondence of topological sectors. However, it does not mean that under duality the topological sectors are mapped identically. It is indeed possible for permutations of sectors to take place, as already illustrated for the Kramers-Wannier duality. Let us investigate in some detail such a scenario. 

Let us consider the boundary condition of $\mathbb H^{\Rep(\mathbb Z_2)}$ labelled by the simple object $\ub 2$ in $\Rep(\mc S_3)$. Concretely, we found that the local operators associated with this boundary condition are given by eq.~\eqref{eq:ExRepZ2NonTrivBdry}. We further explained in sec.~\ref{sec:ExRepZ2} that, given this boundary condition, the model decomposes into five charge sectors. In order to consider a specific topological sector of the model, it is thus required to project the model onto the corresponding charge sector. This can be performed via the projectors described in app.~\ref{sec:app_RepG}, which are constructed from the half-braiding tensors associated with the simple objects in $\mc Z(\Rep(\mc S_3))$ labelling the topological sectors of interest. Concretely, let us consider the topological sector labelled by the simple object $([\mathbb 1], \ub 2)$. As an object of $\mc Z(\Rep(\mc S_3))$, $([\mathbb 1], \ub 2)$ is also an object of $\Rep(\mc S_3)$, namely $\ub 2$ itself. We wish to compute what this topological operator is mapped to under the duality transformation labelled by $\ub 0_{\mathbb Z_3}$.

Carrying out this computation amounts to deriving the intertwining tubes for all boundary conditions defined in sec.~\ref{sec:tubeInt}, applying the projectors associated with all topological sectors on both side of the intertwining tube, and identifying which topological sectors that have non-vanishing overlap with $([\mathbb 1],\ub 2)$. The MPO intertwiners provided by eq.~\eqref{eq:ExMPOInt} are only one component of the intertwining tubes \eqref{eq:tubeIntertwiner}. Indeed, we further require three-valent tensors that evaluate to $\F{\fr F}$-symbols capturing the composition $ \Fun_{\Rep(\mc S_3)}(\Vect,\Rep(\mathbb Z_2)) \times \Fun_{\Rep(\mc S_3)}(\Rep(\mathbb Z_2),\Vect) \to \Fun_{\Rep(\mc S_3)}(\Rep(\mathbb Z_2). \Rep(\mathbb Z_2))$ of module functors:
\begin{equation}
    \begin{split}
        \includeTikz{0}{spePEPS-mod-dot-1-1-1-1-M-X-X-B-1}{\spePEPS{mod}{dot}{1}{1}{1}{1}{}{}{M}{X'}{X}{B}{1}}
        &\equiv
        \big( \Fbar{\fr F}^{B X M}_{\mathbb 1}\big)^{\mathbb 1,11}_{X',11}
        \\[-2em]
        \includeTikz{0}{spePEPS-dot-mod-1-1-1-1-M_1-M_2-X-X-A-2}{\spePEPS{dot}{mod}{1}{1}{1}{1}{M_1}{M_2}{}{X}{X'}{A}{2}}
        &\equiv
        \big( \F{\fr F}^{X A M_1}_{\mathbb 1}\big)^{M_2,11}_{X',11}
    \end{split}\, ,
\end{equation}
for any $X,X' \in \mc I_{\Rep(\mathbb Z_3)}$, $A \in \mc I_{\Rep(\mc S_3)}$, $B \in \mc I_{\Vect_{\mc S_3}}$ and $M,M_1,M_2 \in \mc I_{\Rep(\mathbb Z_2)}$. Since we are interested in the fate of the topological sector $([\mathbb 1],\ub 2)$ of $\mathbb H^{\Rep(\mathbb Z_2)}$ under the duality operator labelled by $\ub 0_{\mathbb Z_3}$, we fix $A = \ub 2$ and $X = \ub 0_{\mathbb Z_3}$, which in turn constrains $X'$ to be $\ub 1_{\mathbb Z_3} \oplus \ub 1^*_{\mathbb Z_3}$. We then apply apply the projector associated with the simple object $([\mathbb 1],\ub 2)$ of $\mc Z(\Rep(\mc S_3))$ so as to select the corresponding topological sector amongst all possible charge sectors compatible with the boundary condition labelled by $\ub 2$. The question is then, for which $B \in \mc I_{\Vect_{\mc S_3}}$ and charge sectors of $\mathbb H^{\Vect}$ associated with the boundary condition labelled by $B$ acting with the corresponding projector on the intertwining tube provides a non-vanishing operator? We find that $B$ must be equal to $\mathbb C_{s} \oplus \mathbb C_{s^2}$, where we should think of $\mathbb C$ as the vector space underlying $\ub 0_{\mathbb Z_3}$, yielding the topological sector associated with the simple object $([s],\ub 0_{\mathbb Z_3})$ of $\mc Z(\Vect_{\mc S_3})$. The $\F{\fr F}$-symbols entering the definition of the intertwining tube that are relevant to this mapping are given by
\begin{align*}
    \big( \F{\fr F}^{\ub 0_{\mathbb Z_3} \ub 0 \ub 0_{\mathbb Z_2}}_{\mathbb 1}\big)^{0_{\mathbb Z_2},11}_{\ub 0_{\mathbb Z_3},11} 
    = 
    \big( \F{\fr F}^{\ub 0_{\mathbb Z_3} \ub 0 \ub 1_{\mathbb Z_2}}_{\mathbb 1}\big)^{1_{\mathbb Z_2},11}_{\ub 0_{\mathbb Z_3},11} 
    &= 1 \, ,
    \\
    \big( \F{\fr F}^{\ub 0_{\mathbb Z_3} \ub 2 \ub 0_{\mathbb Z_2}}_{\mathbb 1}\big)^{0_{\mathbb Z_2},11}_{\ub 1_{\mathbb Z_3},11}
    =
    -i \big( \F{\fr F}^{\ub 0_{\mathbb Z_3} \ub 2 \ub 0_{\mathbb Z_2}}_{\mathbb 1}\big)^{1_{\mathbb Z_2},11}_{\ub 1_{\mathbb Z_3},11}
    &= \frac{1}{\sqrt{2}} \, ,
    \\
    \big( \F{\fr F}^{\ub 0_{\mathbb Z_3} \ub 2 \ub 1_{\mathbb Z_2}}_{\mathbb 1}\big)^{1_{\mathbb Z_2},11}_{\ub 1_{\mathbb Z_3},11}
    =
   -i \big( \F{\fr F}^{\ub 0_{\mathbb Z_3} \ub 2 \ub 1_{\mathbb Z_2}}_{\mathbb 1}\big)^{0_{\mathbb Z_2},11}_{\ub 1_{\mathbb Z_3},11}
   &= \frac{-1}{\sqrt{2}} \, ,
\end{align*}
and
\begin{align*}
    \big( \F{\fr F}^{\mathbb 1 \ub 0_{\mathbb Z_3} \ub 0_{\mathbb Z_2}}_{\mathbb 1}\big)^{\mathbb 1,11}_{\ub 0_{\mathbb Z_3},11} = 
    \big( \F{\fr F}^{\mathbb 1 \ub 0_{\mathbb Z_3} \ub 1_{\mathbb Z_2}}_{\mathbb 1}\big)^{\mathbb 1,11}_{\ub 0_{\mathbb Z_3},11} &= 1 \, ,
    \\
    \big( \F{\fr F}^{s \ub 0_{\mathbb Z_3} \ub 0_{\mathbb Z_2}}_{\mathbb 1}\big)^{\mathbb 1,11}_{\ub 1_{\mathbb Z_3},11} = 
    \big( \F{\fr F}^{s \ub 0_{\mathbb Z_3} \ub 1_{\mathbb Z_2}}_{\mathbb 1}\big)^{\mathbb 1,11}_{\ub 1_{\mathbb Z_3},11} &= 1 \, ,
    \\
    \big( \F{\fr F}^{s^2 \ub 0_{\mathbb Z_3} \ub 0_{\mathbb Z_2}}_{\mathbb 1}\big)^{\mathbb 1,11}_{\ub 1^*_{\mathbb Z_3},11} = 
    \big( \F{\fr F}^{s^2 \ub 0_{\mathbb Z_3} \ub 1_{\mathbb Z_2}}_{\mathbb 1}\big)^{\mathbb 1,11}_{\ub 1^*_{\mathbb Z_3},11} &= 1 \, .
\end{align*}
In short, this shows that under the duality operator labelled by the $\Rep(\mc S_3)$-module functor in $\Fun_{\Rep(\mc S_3)}(\Rep(\mathbb Z_2),\Vect)\cong \Rep(\mathbb Z_3)$ identified with the simple object $\ub 0_{\mathbb Z_3}$, the topological sector $([\mathbb 1],\ub 2)$ of $\mathbb H^{\Rep(\mathbb Z_2)}$ is mapped to the topological sector $([s],\ub 0_{\mathbb Z_3})$ of $\mathbb H^{\Vect}$. Conversely, we can show that the topological sector $([s],\ub 0_{\mathbb Z_3})$ is mapped to $([\mathbb 1],\ub 2)$. We can similarly show that the remaining topological sectors are mapped identically. This permutation of topological sector corresponds to the non-trivial braided autoequivalence of $\mc Z(\Rep(\mc S_3))$ in $\msf{BrEq}(\mc Z(\Rep(\mc S_3))) \simeq \mathbb Z_2$ (see app.~\ref{sec:app_Morita}).

\section{Discussion}

\noindent
\emph{We conclude our manuscript with a discussion of concrete applications of some of the results presented in this manuscript and comments on possible generalizations and extensions.}

\subsection{Application: symmetric tensor networks}

\noindent 
Our study of the interplay between duality transformations and closed boundary conditions can be directly applied to the numerical diagonalization of symmetric Hamiltonians within the variational class of symmetry-preserving tensor networks. One standard approach for this problem involves decomposing the Hamiltonian into symmetric tensors and working in a fusion basis for the Hilbert space \cite{PhysRevA.82.050301,WEICHSELBAUM20122972,SCHMOLL2020168232}. In this basis, the matrix elements of the Hamiltonian can be obtained by invoking the recoupling theory of the symmetric tensors associated with the fusion category $\mc D$, and the setting effectively becomes equivalent to that of one-dimensional models covered in this manuscript. For closed boundary conditions, the sectors are organized into $\mc Z(\mc D)$, and one requires the half-braiding tensors as discussed for instance in app.~\ref{sec:app_double} to thread the corresponding flux through the closed loop \cite{Vanhove:2021zop}. For \emph{modular} tensor categories, there is an equivalence $\mc Z(\mc D) \cong \mc D \boxtimes \mc D^{\text op}$ so that half-braidings can be obtained from the braiding of $\mc D$; this case was studied in detail in \cite{PhysRevB.82.115126,PhysRevB.82.125118,PhysRevB.86.155111}. 

In the setup described above, the original Hamiltonian is given in terms of symmetric tensors associated with a choice of right $\mc D$-module category $\mc M$. However, when working in the usual fusion basis, we are effectively replacing the Hamiltonian by a dual Hamiltonian obtained by choosing $\mc M = \mc D$. This means that, when performing symmetric tensor network computations, one is generically simulating a dual model that has the same spectrum but typically much smaller degeneracies for given sectors. This last property is responsible for the computational advantage gained by working with symmetry preserving tensor networks. In order to simulate the full model with all possible boundary conditions, it is clear that a detailed understanding of how sectors are mapped into one another under duality is required, which we obtained in this manuscript.

\subsection{Open boundary conditions}

\noindent
The results presented in this manuscript focus on \emph{closed} boundary conditions, raising the question, what about \emph{open} boundary conditions? 
Accommodating open boundary conditions requires an extension of the framework employed in this manuscript.

We showed in this manuscript that given a fusion category $\mc D$ and a choice of $\mc D$-module category $\mc M$ we can construct local operators that commute with symmetric operators organized into the fusion category $\mc D^{\star}_\mc M := \Fun_\mc D(\mc M,\mc M)$. In this context, we loosely define open boundary conditions as equivalence classes of extensions of the one-dimensional system to its boundary components in a way that is compatible with the `bulk' $\mc D^\star_\mc M$-symmetry. More precisely, we require open boundary conditions to be organized into categories that are equipped with a $\mc D^\star_\mc M$-action, i.e. $\mc D^\star_\mc M$-module categories. Given a pair $(\mc P,\mc Q)$ of $\mc D^\star_\mc M$-module categories, we would then consider microscopic Hilbert spaces of the form
\begin{equation*}
    \mathbb C \bigg[ \!\!  \includeTikz{0}{openBdry}{\openBdry} \!\! \bigg] ,
\end{equation*}
over objects $\{M \in \mc I_\mc M\}$, $\{Y \in \mc I_\mc D\}$, $P \in \mc I_{\mc P}$, $Q \in \mc I_{\mc Q}$, $X_0 \in \mc I_{\Fun_{\mc D^\star_\mc M}(\mc M,\mc P)}$, $X_{L+1} \in \mc I_{\Fun_{\mc D^\star_\mc M}(\mc Q,\mc M)}$, and basis vectors $\{i\}$ in the appropriate hom-spaces. The resulting model would have symmetry operators organized into $\Fun_{\mc D^\star_\mc M}(\mc P, \mc Q)$. The same way topological sectors correspond to simple objects in the center $\mc Z(\mc D^\star_\mc M)$ in the closed case, we would find that topological sectors in the open case define the bicategory $\Mod(\mc D^\star_\mc M)$ of $\mc D^\star_\mc M$-module categories, $\mc D^\star_\mc M$-module functors and $\mc D^\star_\mc M$-module natural transformations.  

As discussed in the introduction, mathematically, spherical fusion categories serve as input data for topological quantum field theories via the Turaev-Viro-Barrett-Westbury construction \cite{Turaev:1992hq,Barrett:1993ab}. Crucially, these topological quantum field theories are \emph{fully-extended} in the sense that they capture locality all the way down to the point. In this context, the Drinfel'd center $\mc Z(\mc C)$ of $\mc C$ corresponds to the quantum invariant assigned to the circle by the state-sum \cite{Freed:1995fn}, whereas $\Mod(\mc C)$ is identified with the quantum invariant assigned to the point. Crucially, these invariants are related via a so-called `crossing with the circle' condition stipulating that $\msf{Dim} \, \Mod(\mc C) \cong \mc Z(\mc C)$, where $\msf{Dim}$ is an appropriate categorification of the notion of dimension of a vector space suited to bicategories (see ref.~\cite{bartlett,Bullivant:2021pkd} for the case of $\mc C = \Vect_G$). This relation formalizes the process whereby identifying the endpoints of a model with open boundary conditions yields a model with closed boundary conditions. We postpone a systematic study of dualities in one-dimensional quantum lattice models with open boundary conditions to another manuscript.

\subsection{Higher dimensions}

\noindent
The results presented in this manuscript can be largely extended to two-dimensional quantum lattice models following the ethos of categorification. Loosely speaking, it amounts to replacing fusion (1-)categories and module (1-)categories in our exposition by fusion 2-categories and module 2-categories. More specifically, given an input fusion 2-category and a choice of module 2-category over it, local symmetric operators akin to those considered in this manuscript can be constructed \cite{CDMod2}. These in turn define Hamiltonians that commute with operators organized into the (higher) Morita dual of the input fusion 2-category with respect to the chosen module 2-category. Similarly, duality relations are also encoded into module 2-functors between distinct module 2-categories. Boundary conditions and topological sectors will then be related to representations of higher-dimensional tube algebras as considered in ref.~\cite{Delcamp:2017pcw,Bullivant:2019fmk,Bullivant:2019tbp,Bullivant:2020xhy} and to the higher-categorical center of the symmetry fusion 2-category. 

As an example, given a two-dimensional model with a $G$-symmetry, we can construct a tensor network operator performing the gauging of the symmetry and show that the resulting dual model commutes with \emph{projected entangled pair operators} forming the fusion 2-category $\msf{2Rep}(G)$ of `2-representations' of the group $G$ \cite{CDMod2,Bartsch:2022mpm,Bhardwaj:2022lsg}, which is Morita equivalent to the fusion 2-category $\msf{2Vec}_G$ of $G$-graded 2-vector spaces \cite{CDMod2}.

Recently, a closely related point of view has been embraced by the high-energy community under the name of `sandwich construction' or `symmetry topological field theory' \cite{Freed:2022qnc}. So far, this approach has been mostly employed in the continuum, although lattice versions of these ideas have appeared in the past \cite{PhysRevLett.112.247202,PhysRevLett.121.177203,Aasen:2020jwb,Vanhove_2022,Chatterjee:2022kxb,Chatterjee:2022jll}. Succinctly, this approach amounts to realising the partition function of a $d$-dimensional theory as the interval compactification of a ($d$+1)-dimensional topological field theory with two types of boundaries: one hosting a gapped boundary condition and another hosting a `physical' typically non-topological boundary condition. In two spacetime dimensions on the lattice, the choice of three-dimensional topological field theory and its gapped boundary conditions amounts to our choices of input fusion category and module category over it, whereas the data of the physical boundary condition is encodes our choice of abstract algebra of operators. Explicitly performing the interval compactification, this would immediately recover our construction, or its completely analogue formulation in terms of anyonic chains presented in ref.~\cite{Lootens:2021tet}. We note that in principle the sandwich construction produces a partition function, but this can be related to our Hamiltonian construction via the transfer matrix and the standard quantum to classical mapping.

\bigskip\bigskip\noindent
{\bf Acknowledgements:} We thank Jacob Bridgeman, Paul Fendley and Jutho Haegeman for inspiring discussions and insightful comments, and in particular Gerardo Ortiz for collaboration on the prequel \cite{Lootens:2021tet}. This work has received funding from the Research Foundation Flanders (FWO) through postdoctoral fellowship No.~1228522N awarded to CD and doctoral fellowship No.~1184722N awarded to LL.

\titleformat{name=\section}
{\normalfont}
{\centering {\textsc{App.} \thesection \; \raisebox{1pt}{\textbar} \;}}
{0pt}
{\normalsize\bfseries\centering}

\newpage
\appendix
\section{Morita equivalence\label{sec:app_Morita}}

\noindent
\emph{In this appendix, we collect a few results about (categorical) Morita equivalence of fusion categories.}

\subsection{Motivating examples}

\noindent
Consider the fusion category $\Vect_G$ of $G$-graded vector spaces. Recall that it is the $\mathbb C$-linear category with simple objects the one-dimensional vectors spaces $\mathbb C_g$, $g \in G$, such that $\mathbb C_g \otimes \mathbb C_h \simeq \mathbb C_{gh}$ and $\Hom(\mathbb C_g, \mathbb C_h) = \delta_{g,h} \mathbb C$ for every $g,h \in G$. (Indecomposable) module categories over $\Vect_G$ are labelled by pairs $(A,\psi)$ with $A \subseteq G$ a subgroup and $\psi$ a representative of a cohomology class in $H^2(A,{\rm U}(1))$, such that the collection of simple objects is provided by $G/A$ \cite{2002math......2130O}. Choosing $A = G$, we find that $\Vect$ is a (right) $\Vect_G$-module category via the forgetful functors $\Vect_G \to \Vect$. It is a well-known result that the category $(\Vect_G)^\star_\Vect := \Fun_{\Vect_G}(\Vect,\Vect)$ is equivalent to $\Rep(G)$ \cite{etingof2016tensor}. Let us briefly review this derivation. A functor $F:\Vect \to \Vect$ is fully specified by the vector space $V:= F(\mathbb C)$ it assigns to the unique simple object $\mathbb C$ in $\Vect$. The $\Vect_G$-module functor structure in turn provides natural isomorphisms prescribed by
\begin{equation*}
    \omega_g \in \Hom_{\Vect}(F(\mathbb C \otimes \mathbb C_g), F(\mathbb C) \otimes \mathbb C_g) = {\rm End}_{\mathbb C}(V) \, ,
\end{equation*}
for every $g \in G$.
It follows from the defining coherence relation of module functors that $(V,\omega : g \mapsto \omega_g)$ defines a representation of $G$. Similarly, we can show that module natural transformations between $\Vect_G$-module endofunctors of $\Vect$ correspond to intertwiners of $G$. Putting everything together, we have $(\Vect_G)^\star_\Vect \cong \Rep(G)$.

As we shall see below, this derivation demonstrates  that $\Vect_G$ and $\Rep(G)$ are Morita equivalent. A consequence of this result is that indecomposable module categories over $\Rep(G)$ are also parametrized by pairs $(A,\psi)$. Indeed, for any $(A,\psi)$, the category $\Rep^\psi(A)$ of projective representations of $A$ can be endowed with the structure of a (right) $\Rep(G)$-module category via the restriction functor ${\rm Res}^G_A: \Rep(G) \to \Rep(A)$. As one would expect, we have $(\Rep(G))^\star_\Vect := \Fun_{\Rep(G)}(\Vect,\Vect) \cong \Vect_G$. The derivation of this result is more subtle than the previous one, as such we shall merely sketch here and refer the reader to ref.~\cite{etingof2016tensor} for details. First, remember that $\Rep(G)$ can be equivalently defined as the category $\Mod(\mathbb C[G])$ of modules over the group algebra, whereas $\Vect_G$ is equivalent to the category $\Mod(\mathbb C^G)$ of module over the algebra of functions on $G$. As already mentioned, a functor $F: \Vect \to \Vect$ is fully specified by a vector space $V := F(\mathbb C)$. A $\Rep(G)$-module structure on $F$ then provides natural isomorphisms prescribed by
\begin{equation*}
    \omega_{U} \in \Hom_{\Vect}(F(\mathbb C \cat U),F(\mathbb C) \cat U) = {\rm End}_{\mathbb C}(V \otimes {\rm Res}(U)) \, ,
\end{equation*}
for every $U \in \Rep(G)$, satisfying the defining coherence relation. Equivalently, the module structure of $F$ prescribes a homomorphism $V \to V \otimes {\rm Nat}({\rm Res},{\rm Res})$, where ${\rm Nat}({\rm Res})$ denotes here the vector space of natural endotransformations of ${\rm Res}: \Rep(G) \to \Vect$. By definition, a vector in ${\rm Nat}({\rm Res},{\rm Res})$ assigns to every object $U$ in $\Rep(G)$ a morphism ${\rm Res}(U)\to {\rm Res}(U)$ in $\Vect$ satisfying a naturality condition with respect to any intertwiner of $G$. It turns out that this vector space ${\rm Nat}({\rm Res}, {\rm Res})$ can be equipped with a canonical Hopf algebraic structure. It follows from the defining axioms of module functors that the map $V \to V \otimes {\rm Nat}({\rm Res},{\rm Res})$ endows $V$ with a right \emph{comodule} structure over ${\rm Nat}({\rm Res},{\rm Res})$. We can show that a natural transformation ${\rm Res} \to {\rm Res}$ is fully specified by its component on the regular representation $\mathbb C[G]$. 
But endomorphic intertwiners of the regular representation are given by right multiplication by an element in $\mathbb C[G]$, and thus the component of the natural transformation on the regular representation must amount to left multiplication by some element in $\mathbb C[G]$.\footnote{Alternatively, we can invoke the fact that the forgetful functor ${\rm Res}$ is naturally isomorphic to $\Hom_{\Rep(G)}(\mathbb C[G],-)$ where $\mathbb C[G]$ is thought as the regular representation of $G$. It then follows from the \emph{Yoneda lemma} that ${\rm Nat}({\rm Res},{\rm Res})$ is equivalent to $\mathbb C[G]$ as a vector space and then as a Hopf algebra.} So we have a right comodule structure over ${\rm Nat}({\rm Res},{\rm Res}) \simeq \mathbb C[G]$, which is the same thing as a left module structure over the dual of the Hopf algebra $\mathbb C[G]$, namely $\mathbb C^G$. Putting everything together we find $(\Rep(G))^\star_{\Vect} \cong \Mod(\mathbb C^G) \cong \Vect_G$, as expected. We shall now put these results in the context of Morita equivalence.

\subsection{Definition\label{sec:app_defMorita}}

\noindent
Given two fusion categories $\mc C$ and $\mc D$, they are said to be (categorically) Morita equivalent if there exists a left $\mc C$-module category $\mc M$ such that $\mc D^\star_\mc M \cong \mc C$, or equivalently if there exists a right $\mc D$-module category such that $\mc C^\star_\mc M \cong \mc D^{\rm op}$, or still equivalently if there exists an invertible $(\mc C, \mc D)$-bimodule category $\mc M$ \cite{etingof2016tensor,etingof2010fusion}.

Consider for instance a fusion category $\mc C$ as a (left) module category over itself. The functor $X \mapsto (- \otimes X): \mc C^{\rm op} \to \mc C^\star_\mc C$ is a monoidal equivalence, whereby the module structure is provided by
\begin{align*}
    (Y \otimes -) \otimes X \xrightarrow{\alpha_{Y,-,X}} Y \otimes (- \otimes X) \, ,
\end{align*}
for every $Y \in \Ob(\mc C)$. This is the statement that $\mc C$ and $\mc C^{\rm op}$ are Morita equivalent. Notice that the monoidal structure in $\mc C^\star_\mc C$ is provided by the composition of $\mc C$-module functors, which explains why the monoidal equivalence is with $\mc C^{\rm op}$ instead of $\mc C$.

Let us consider an interesting application of the derivation above. Given a fusion category $\mc C$, we can construct another fusion category $\mc Z(\mc C)$ known as the \emph{center} of $\mc C$. Objects in $\mc Z(\mc C)$ consists of pairs $(X,R_{-,X})$, where $X$ is an object of $\mc C$ and $R_{-,X}: - \otimes X \xrightarrow{\sim} X \otimes -$ is a collection of natural isomorphisms known as \emph{half-braidings}. These half-braidings are required to satisfy an `hexagon axiom' involving the monoidal associator of $\mc C$. It turns out that the center $\mc Z(\mc C)$ can be identified with the category $\Fun_{\mc C|\mc C}(\mc C,\mc C)$ of $(\mc C,\mc C)$-bimodule functors. Indeed, any functor in $\Fun_{\mc C| \mc C}(\mc C, \mc C)$ is in particular a functor in $\Fun_{\mc C}(\mc C,\mc C) \cong \mc C^{\rm op}$ and hence is of the form $- \otimes X$ with $X \in \Ob(\mc C)$. The right $\mc C$-module structure then imposes the existence of natural isomorphisms
\begin{align*}
    (- \otimes Y) \otimes X \xrightarrow{\sim} (- \otimes X) \otimes Y
\end{align*}
for every $Y \in \Ob(\mc C)$, yielding half-braiding isomorphisms $R_{-,X}:- \otimes X \xrightarrow{\sim} X \otimes -$ in virtue of $\mc C^\star_\mc C \cong \mc C^{\rm op}$. It follows from the defining coherence relation of module functors that these half-braiding isomorphisms fulfill the expected `hexagon axioms'. Monoidal structure in $\mc Z(\mc C)$ finally corresponds to the composition of $(\mc C,\mc C)$-bimodule functors. Notably, it follows immediately from $\Fun_{\mc C| \mc C}(\mc C,\mc C) \cong (\mc C \boxtimes \mc C^{\rm op})^\star_\mc C$ that $\mc Z(\mc C^\star_\mc C) \cong \mc Z(\mc C)$ as expected. More generally, for any $\mc C$-module category $\mc M$, we have $\mc Z(\mc C^\star_\mc M) \cong \mc Z(\mc C)$. We review this important result below.

\subsection{Invariant}

\noindent
Let $\mc C$ and $\mc D$ be two fusion categories and $\mc M \equiv (\mc M,\act, \cat, \alpha^{\act},\alpha^{\cat},\alpha^{\caact})$ an invertible $(\mc C,\mc D)$-bimodule category. There is a well-known braided monoidal equivalence $\mc Z(\mc C) \cong \mc Z(\mc D)$, making the Drinfel'd center an invariant of Morita equivalence \cite{MUGER200381,etingof2016tensor}. Let us examine this property. Consider an object $(X,R_{-,X})$ in $\mc Z( \mc C)$. The endofunctor $X \act -$ of $\mc M$ is equipped with a left $\mc C$- and right $\mc D$-module structures via
\begin{align*}
    X \act (X' \act -)
    &\xrightarrow{(\alpha^{\act})^{-1}_{X,X',-}}
    (X \otimes X') \act - 
    \\[-.5em]
    &\xrightarrow{R_{X',X}^{-1} \act {\rm id}_{-}}
    (X' \otimes X) \act -
    \\[-.5em]
    &\xrightarrow{\alpha^{\act}_{X' ,X, -}}
    X' \act (X \act -) \,
\end{align*}
and
\begin{align*}
    X \act (- \cat Y) \xrightarrow{(\alpha^{\caact})^{-1}_{X,-,Y}} (X \act -) \cat Y \, ,
\end{align*}
respectively,
for any $X' \in \Ob(\mc C)$ and $Y \in \Ob(\mc D)$. The `hexagon axioms' satisfied by the half-braiding isomorphisms $R_{-,X}$ as well as the various `pentagon axioms' fulfilled by the module associators ensure that the defining coherence relations of right and left module functors are satisfied. 

Conversely, let us consider an object $\Fun_{\mc C|\mc D}(\mc M,\mc M)$. By definition, it is in particular a right $\mc D$-module functor in $\Fun_{\mc D^{\rm op}}(\mc M, \mc M)$. In virtue of $\Fun_{\mc D^{\rm op}}(\mc M,\mc M) \cong \mc C$, it is of the form $X \act -$ with $X \in \Ob(\mc C)$. The $(\mc C, \mc D)$-bimodule structure is provided by natural isomorphisms
\begin{align*}
    X \act ((X' \act -) \cat Y) \xrightarrow{\sim} (X' \act (X \act -)) \cat Y \, ,
\end{align*}
for any $X' \in \Ob(\mc C)$, $Y \in \Ob(\mc D)$. These yield the following collection of isomorphisms of right $\mc D$-module endofunctors:
\begin{align*}
    (X' \otimes X) \act (- \cat Y) 
    &\xrightarrow{(\alpha^{\caact})^{-1}_{X'\otimes X,- , Y} }
    ((X' \otimes X) \act -) \cat Y
    \\[-.5em]
    &\xrightarrow{\alpha^{\act}_{X',X,-} \cat {\rm id}_Y}
    (X' \act (X \act -)) \cat Y
    \\
    &\xrightarrow{\sim}
    X \act ((X' \act -) \cat Y)
    \\[-.4em]
    &\xrightarrow{{\rm id}_{X} \act \alpha^{\caact}_{X',-,Y}}
    X \act (X' \act (- \cat Y))
    \\[-.5em]
    &\xrightarrow{(\alpha^{\act})^{-1}_{X,X',-} \cat {\rm id}_Y}
    (X \otimes X') \act (- \cat Y) \, ,
\end{align*}
which invoking $\Fun_{\mc D^{\rm op}}(\mc M, \mc M) \cong \mc C$ in turn yield a collection of natural isomorphisms $R_{-,X} : - \otimes X  \xrightarrow{\sim} X \otimes -$ in $\mc C$. It follows from the definition of module functors that the natural isomorphisms $R_{-,X}$ in $\mc C$ satisfy the defining `hexagon axioms' of half-braidings. Keeping in mind that the monoidal structure in $\Fun_{\mc C|\mc D}(\mc M,\mc M)$ is provided by the composition of module functors, this establishes the monoidal equivalence $\mc Z(\mc C) \cong \Fun_{\mc C|\mc D}(\mc M,\mc M)$, which can be lifted to a braided monoidal equivalence. By symmetry, we similarly find $\mc Z(\mc D) \cong \Fun_{\mc C|\mc D}(\mc M,\mc M)$ whereby objects $(Y,R_{Y,-})$ in $\mc Z(\mc D)$ are identified with $(\mc C,\mc D)$-module functors $- \cat Y$, and thus
\begin{align*}
    \mc Z(\mc C) \xrightarrow{\sim} \Fun_{\mc C|\mc D}(\mc M,\mc M) \xleftarrow{\sim} \mc Z(\mc D) \, .
\end{align*}
Finally, the induced braided monoidal equivalence $\mc Z(\mc C) \xrightarrow{\sim} \mc Z(\mc D)$ is provided by a functor $F:\mc Z(\mc C) \to \mc Z(\mc D)$ such that there is an isomorphism $X \act - \simeq - \cat F(X)$ of $(\mc C, \mc D)$-bimodule functors for every $X \in \Ob(\mc C)$. This in turn motivates the definition of our tube intertwiners as maps between the tubes associated with $\mc Z(\mc C)$ and $\mc Z(\mc D)$ that `commute' with $\mc M$ (see eq.~\eqref{eq:tubeIntertwiner}).

A particularly compact way of stating the result presented in this section is that the centre construction defines a 2-functor $\mc Z: \underline{\msf{BrPic}} \to \underline{\msf{EqBr}}$, where $\underline{\msf{BrPic}}$ refer to the \emph{Brauer-Picard} 2-groupoid of fusion categories and invertible bimodule categories, whereas $\underline{\msf{EqBr}}$ denotes the 2-groupoid of braided fusion categories and braided equivalences \cite{etingof2010fusion}. This implies in particular the group isomorphism $\msf{BrPic}(\mc C) \cong \msf{EqBr}(\mc Z(\mc C))$ between the Brauer-Picard group of invertible $(\mc C,\mc C)$-bimodules and braided auto-equivalences of $\mc Z(\mc C)$.

\section{Quantum double\label{sec:app_double}}

\noindent
\emph{In this appendix, we illustrate with a detailed example the equivalence between the category of representations of the tube category and the Drinfel'd center of the underlying fusion category.}

\subsection{Hopf algebra\label{sec:app_Hopf}}
\noindent
Let $G$ be a finite group with identity element $\mathbb 1$. We denote by $\mathbb C[G]$ the group ring of $G$ and $\mathbb C^G$ the space of functions on $G$. The quantum double $\mc D(G) \equiv (\mc D(G),\star,\eta,\Delta,\epsilon,S,R)$ of $G$ is a quasi-triangular Hopf algebra whose underlying vector space is isomorphic to $\mathbb C[G] \otimes \mathbb C^G$. Denoting by $a \otimes \delta_g \equiv \qd{a}{g}$, with $a,g \in G$, basis elements of $\mc D(G)$, the Hopf algebraic structure is defined by
\begin{alignat*}{3}
    (\qd{a}{g})\star (\qd{b}{h}) 
    &:= \delta_{a^{-1}ga,h} (\qd{ab}{g})
    && \text{(multiplication)}
    \\
    \eta(z) 
    &:= z \sum_{g \in G}\qd{\mathbb 1}{g}
    && \text{(unit)}
    \\
    \Delta(\qd{a}{g}) 
    &:= \sum_{h \in G} \qd{a}{h} \otimes \qd{a}{h^{-1}g}
    &\;\;\; & \text{(comultiplication)}
    \\
    \epsilon(\qd{a}{g}) 
    &:= \delta_{g,\mathbb 1}
    && \text{(counit)}
    \\
    S(\qd{a}{g}) 
    &:= \qd{a^{-1}}{a^{-1}g^{-1}a}
    && \text{(antipode)}
\end{alignat*}
for all $a,b,g,h \in G$ and $z \in \mathbb C$. The definitions above imply that both $\mathbb C[G]$ and $\mathbb C^G$ are equipped with the structures of Hopf subalgebras of $\mc D(G)$ and we denote by $\iota_{\mathbb C[G]}$ and $\iota_{\mathbb C^G}$ the corresponding embedding maps. The quasi-triangularity is then provided by the invertible element
\begin{equation}
    \label{eq:Rmatrix}
    R \equiv (\iota_{\mathbb C[G]} \otimes \iota_{\mathbb C^G})(\tilde R)
    \q {\rm with} \q \tilde R := \sum_{g \in G}\qd{g}{g} \, .
\end{equation}
Let us consider the braided monoidal category $\Mod(\mc D(G))$ of left modules over the quantum double $\mc D(G)$. It follows from $\mathbb C[G]$ and $\mathbb C^G$ forming subalgebras of $\mc D(G)$ as well as the relation
\begin{equation*}
    \qd{a}{g} = (\qd{\mathbb 1}{g}) \star \Big(\sum_{h \in G}\qd{a}{h}\Big) 
    = \Big(\sum_{h \in G}\qd{a}{h}\Big) \star (\qd{\mathbb 1}{a^{-1}ga})
\end{equation*}
that an object in $\Mod(\mc D(G))$ is a vector space $\mc V$ equipped with left $\mathbb C[G]$- and $\mathbb C^G$-module structures satisfying the straightening formula
\begin{equation*}
    \delta_g \act (a \act |\nu \ra) = a \act (\delta_{a^{-1}ga} \act |\nu \ra) \, ,
\end{equation*} 
for any $a,g \in G$ and $|\nu \ra \in \mc V$. In particular, simple modules of $\mc D(G)$ are labeled by pairs $([c_1],\hat V)$ with $[c_1]$ a conjugacy class of $G$ with representative $c_1$ and $\hat V$ a simple left module over the group ring $\mathbb C[Z_{[c_1]}]$ of the centralizer $Z_{[c_1]}=\{g \in G \, | \, gc_1 = c_1g\}$ of $[c_1]$ in $G$. Denoting the constituents of $[c_1]$ by $\{c_i\}_{i=1,\ldots,|[c_1]|}$, we have
\begin{equation*}
    \mc V = {\rm Span}_\mathbb C\{\, |c_i,v \ra \}_{\substack{ \hspace{-10pt} \forall \, i=1,\ldots,|[c_1]| \\ \forall \, v = 1 , \ldots, {\rm dim}(\hat V)}} 
\end{equation*}
so that ${\rm dim}(\mc V) = |[c_1]|\cdot {\rm dim}(\hat V)$.
Introducing the set $Q_{[c_1]} = \{q_{c_i} \in G\}_{i=1,\ldots,|[c_1]|}$ such that $c_1 = q_{c_i}^{-1}c_iq_{c_i}$ and $q_{c_1} = \mathbb 1$, the action of $\mc D(G)$ on $\mc V$ is provided by
\begin{equation*}
    (\qd{a}{g})\act | c_j,v \ra = \delta_{g,ac_ja^{-1}} \big| ac_ja^{-1}\ra \otimes  (q^{-1}_{ac_ja^{-1}}aq_j) \act | v \big\ra \, ,
\end{equation*}
for any $a,g \in G$. We shall often implicitly make use of the equivalence between simple objects $\mc V$ labeled by $([c_1],\hat V)$ in $\Mod(\mc D(G))$ and irreducible representations $(\mc V, \rho: \mc D(G) \to \End(\mc V))$ such that $(\qd{a}{g}) \act |c_i,v \ra = \rho(\qd{a}{g})|c_i,v \ra$ and
\begin{align*}
    \rho(\qd{a}{g})^{i\tilde v}_{j v} &\equiv \la c_i,\tilde v | \rho(\qd{a}{g}) | c_j,v \ra 
    \\
    & = \delta_{g,c_i} \, \delta_{g,ac_ja^{-1}} \, \hat\rho(q^{-1}_{c_i}a q_{c_j})_v^{\tilde v} \, ,
\end{align*}
where $(\hat V,\hat \rho :  Z_{[c_1]} \to \End(\hat V))$ is an irreducible representation of $Z_{[c_1]} \subset G$. The monoidal structure in $\Mod(\mc D(G))$ can then be conveniently defined in terms of the tensor product $(\mc V \otimes \mc W, (\rho \otimes \sigma) \circ \Delta)$ of representations $(\mc V,\rho)$ and $(\mc W,\sigma)$, whereas the braiding is provided by the isomorphism
\begin{align*}
    \mc R &= (\sigma \otimes \rho)(R) \circ {\rm swap}
    \\
    &: \mc V \otimes \mc W \xrightarrow{\sim} \mc W \otimes \mc V \, ,
\end{align*}
where `swap' simply permutes the order of vector spaces in the tensor product.

\bigskip \noindent
In the context of our work, the quantum double $\mc D(G)$ of a finite group $G$ arises---up to a normalization factor---as the \emph{groupoid algebra} of the tube category $\msf{Tube}(\Vect_G)$. By definition $\msf{Tube}(\Vect_G)$ is the category with object-set $G$ and morphisms of the form $a : g \to a^{-1}ga$. Its groupoid algebra is then defined as the algebra with underlying vector space $\mathbb C[ \, | g \! \xrightarrow{a} \! a^{-1}ga \ra \,]$, over $a,g \in G$, and algebra product \begin{equation*}
    |g \! \xrightarrow{a} \! a^{-1}ga \ra \star 
    | h \! \xrightarrow{b} \! b^{-1}hb \ra = \delta_{a^{-1}ga,h} \, | g \xrightarrow{ab} (ab)^{-1}g (ab)\ra \, .
\end{equation*}
Representations of the tube category defined as objects in $\Fun(\msf{Tube}(\Vect_G),\Vect)$ then correspond to the modules over $\mc D(G)$ as defined in this section.

\subsection{Drinfel'd center $\mc Z(\Vect_G)$\label{sec:app_VecG}}

\noindent
There is a famous braided monoidal equivalence between $\Mod(\mc D(G))$ and the Drinfel'd center $\mc Z(\Vect_G)$ of the category of $G$-graded vector spaces. Given an object $(\mc V,\rho)$ in $\Mod(\mc D(G))$, we shall now review how to obtain the corresponding object in $\mc Z(\Vect_G)$ (see \cite{2018arXiv180805060G} for more details). Since $\qd{\mathbb 1}{g} \star \qd{\mathbb 1}{g} = \qd{\mathbb 1}{g}$, the matrix $\rho(\qd{\mathbb 1}{g})$ is a projector and a fortiori it is diagonalizable. Furthermore, since $\qd{\mathbb 1}{g} \star \qd{\mathbb 1}{h} = \qd{\mathbb 1}{h} \star \qd{\mathbb 1}{g}$, the matrices $\rho(\qd{\mathbb 1}{g})$ and $\rho(\qd{\mathbb 1}{h})$ commute for every $g,h \in G$, and thus the set $\{\rho(\qd{\mathbb 1}{g})\}_{g \in G}$ is simultaneously diagonalizable w.r.t. a basis notated via $\{|\nu \ra \}_{\nu=1,\ldots,{\rm dim}(\mc V)}$. We denote by $\mc V_g$ the subspace of $\mc V$ given by ${\rm Im}\, \rho(\qd{\mathbb 1}{g})$. Due to the unit element of $\mc D(G)$ being provided by $\sum_{g \in G}\qd{\mathbb 1}{g}$, given $\nu \in \{1,\ldots,{\rm dim}(\mc V)\}$, there is a unique $g \in G$ such that $|\nu \ra \in {\rm Im}\, \rho(\qd{\mathbb 1}{g})$. Consequently, $\mc V$ decomposes as
\begin{align*}
    \mc V = \bigoplus_{g \in G} \mc V_g \, .
\end{align*}
We deduce that the left $\mc D(G)$-module $\mc V$ has the structure of a $G$-graded vector space. Given two objects $(\mc V,\rho)$ and $(\mc W,\sigma)$ of $\Mod(\mc D(G))$, we further recover from
\begin{equation*}
    [(\rho \otimes \sigma) \circ \Delta](\qd{\mathbb 1}{g}) = \sum_{h \in G} \rho (\qd{\mathbb 1}{h}) \otimes \sigma (\qd{\mathbb 1}{h^{-1}g})
\end{equation*}
the monoidal structure of $\Vect_G$, i.e.
\begin{equation*}
    (\mc V \otimes \mc W)_g = \bigoplus_{h \in G}\mc V_h \otimes \mc W_{h^{-1}g} \, .
\end{equation*}
Recall that we denote by $\mathbb C_a$ the one-dimensional $G$-graded vector space such that $(\mathbb C_a)_b = \delta_{a,b} \mathbb C$. For any $a \in G$, we consider the map 
\begin{align*}
    R_a := \Big( \rho \big( \sum_{g \in G}a\delta_g \big) \otimes {\rm id } \Big) \circ {\rm swap} : \mathbb C_a \otimes \mc V \to \mc V \otimes \mathbb C_a \, .
\end{align*}
Since $\rho(\qd{\mathbb 1}{g})$ is an isomorphism $\mc V_g \to \mc V_g$ that acts as the zero map on $\mc V_{h \neq g}$, the relations
\begin{align*}
    \rho(a\delta_g) \circ \rho(\mathbb 1\delta_{a^{-1}ga}) = \rho(a\delta_g) = \rho(\mathbb 1\delta_g) \circ \rho(a\delta_g) 
\end{align*}
induces that $\rho(\qd{a}{g})$ must be a linear map $\mc V_{a^{-1}ga} \to \mc V_g$ that acts as the zero map on $\mc V_{h \neq a^{-1}ga}$. Therefore, the restriction of $R_a$ to $\mc V_{a^{-1}g}$ is a map $\mc V_{a^{-1}g} \to \mc V_{ga^{-1}}$. It follows from 
\begin{equation*}
    \mathbb C_a \otimes \mc V = \bigoplus_{g \in G} \mc V_{a^{-1}g}
    \q {\rm and} \q
    \mc V \otimes \mathbb C_a = \bigoplus_{g \in G} \mc V_{ga^{-1}}
\end{equation*}
that $R_a$ preserves the grading. Furthermore,  $R_a$ is invertible with 
\begin{align*}
    R_a^{-1} := \Big( {\rm id} \otimes \rho \big( \sum_{g \in G} \qd{a^{-1}}{g} \big) \Big) \circ {\rm swap} 
\end{align*}
and thus $R_a$ is a (natural) $G$-grading preserving isomorphism $\mathbb C_a \otimes \mc V \xrightarrow{\sim} \mc V \otimes \mathbb C_a$. Due to $\rho: \mc D(G) \to \End(\mc V)$ being a homomorphism, we can readily check that
\begin{align*}
    R_{ab} = (R_a \otimes {\rm id}_{\mathbb C_b}) \circ ({\rm id}_{\mathbb C_a} \otimes R_{b}) \, ,
\end{align*}
for any $a,b \in G$.
Putting everything together, we obtain that $(\mc V,R_{-,\mc V})$ with $R_{\mathbb C_a,\mc V} \equiv R_a$ defines an object of $\mc Z(\Vect_G)$. Given a simple object $(\mc V,\rho)$ in $\Mod(\mc D(G))$ labeled by $([c_1],\hat V)$, the corresponding graded object in $\mc Z(\Vect_G)$ is the object $\mc V = \bigoplus_{g \in G}\mc V_g$ such that $\mc V_g \simeq \hat V$ for every $g \in [c_1]$ and zero otherwise. The corresponding half-braiding isomorphisms are conveniently encoded into the so-called \emph{half-braiding} tensor expressed graphically as:
\begin{align*}
    \includeTikz{0}{halfBraiding-mcV-mathbbC_a}{\halfBraiding{\mc V}{\mathbb C_a}} \!\!\!\!\!\!\!
    := \sum_{g \in G}\! \sum_{i=1}^{|[c_1]|}
    \delta_{g,c_i} \,
    \hat \rho (q_{g}^{-1}aq_{a^{-1}ga})
    \!\!\! \includeTikz{0}{resolution-mathbbC_-g-mathbbC_-ga-mathbbC_-a--1ga-mathbbC_a-1-1}{\resolution{\mathbb C_{g}}{\;\;\mathbb C_{ga}}{\mathbb C_{a^{-1}ga}}{\mathbb C_a}{1}{1}} \, .
\end{align*}
This half-braiding tensor can in turn be used to explicitly compute tube projectors onto the topological sector labeled by the simple object $(\mc V, \rho) \in \mc Z(\Vect_G)$ in the case where $\mc D^\star_\mc M \cong \Vect_G$:
\begin{align*}
    \frac{1}{|G|}\sum_{a \in G} \;\;\; \includeTikz{0}{idempotent-mcV-mathbbC_a}{\idempotent{\mc V}{\mathbb C_a}} \;\;\;\, 
    \equiv \frac{1}{|G|} \sum_{a,g \in G} \rho(\qd{a}{g}) \, ,
\end{align*}
where the dotted lines indicate periodic boundary conditions as in the main text. A representation of these projectors on the Hilbert spaces considered in the main text in the basis of matrix units is given by
\begin{equation*}
    \frac{1}{\rm dim \mc V}\sum_{c_i,c_j,v,\tilde v} e^{(\mc V, \rho),(\mathbb C_{c_i})_v,(\mathbb C_{c_j})_{\tilde v}}_{\mc M | \mc M}
\end{equation*}
from which we can infer the coefficients defining the matrix units in terms of tubes as
\begin{equation}
    \Omega^{(\mc V, \rho),(\mathbb C_{c_i})_v,(\mathbb C_{c_j})_{\tilde v}}_{\mathbb C_a,\mathbb C_{c_i a},1,1} := \hat \rho(q^{-1}_{c_i}a q_{c_j})^{\tilde v}_{v}.
\end{equation}

\subsection{Drinfel'd center $\mc Z(\Rep(G))$\label{sec:app_RepG}}

\noindent
In virtue of the Morita equivalence between $\Rep(G)$ and $\Vect_G$, we know the Drinfel'd centers are equivalent, and thus there is also a braided monoidal equivalence between $\Mod(\mc D(G))$  and $\mc Z(\Rep(G))$. As before, we are only interested in computing the simple objects in $\mc Z(\Rep(G))$ corresponding to those in $\Mod(\mc D(G))$. Let $\mc V$ be a simple left module over $\mc D(G)$ labeled by $([c_1],\hat V)$. We mentioned earlier that $\mc V$ is in particular a left module over both $\mathbb C[G]$ and $\mathbb C^G$ with 
\begin{equation}
    \label{eq:actionsSubAlg}
    \begin{split}
    a &\act |c_j,v \ra = |ac_ja^{-1}\ra \otimes (q^{-1}_{ac_ja^{-1}}aq_{c_j}) \act | v \ra
    \\
    \delta_g &\act |c_j,v\ra = \delta_{g,c_j}|c_j, v \ra \, ,
    \end{split}
\end{equation}
for every $a,g \in G$, $j \in \{1,\ldots,|[c_1]|\}$ and $v \in \{1,\ldots,{\rm dim} \, \hat V\}$. Given any left $\mathbb C[G]$-module $W$, we define 
\begin{equation*}
    \arraycolsep=1.4pt
    \begin{array}{ccccl}
        R_{W,\mc V} & : & W \otimes \mc V & \to & \mc V \otimes W
        \\
        & : & \,| w \ra \otimes | \nu \ra & \mapsto & {\rm swap} \circ \tilde R( |w \ra \otimes | \nu\ra )
    \end{array} \, ,
\end{equation*}
where $\tilde R$ was introduced in eq.~\eqref{eq:Rmatrix}.
By definition, we have $\tilde R(| w \ra \otimes | \nu \ra) = \sum_{g \in G}(g \act |w \ra) \otimes (\delta_g \act | \nu \ra)$, where $\mc V$ is treated here as a left $\mathbb C^G$-module. 
It follows from $(\Delta \otimes {\rm id}_{\mc D(G)})(\tilde R) = \tilde R_{13} \star \tilde R_{23}$ that $R_{-,\mc V}$ is a half-braiding so that $(\mc V,R_{-,\mc V})$ is an object of $\mc Z(\Rep(G))$. Choosing $| \nu \ra \equiv | c_j,v\ra$, we have in particular
\begin{equation*}
    R_{W,\mc V}( |w \ra \otimes | \nu \ra) = |c_j,v \ra \otimes c_j \act |w\ra \, .
\end{equation*}
Ultimately, we are interested in the half-braiding tensors $\Omega^\mc V$ defined as follows: Given two objects $V,W$ in $\Rep(G)$, the hom-space $\Hom_{\Rep(G)}(W \otimes V , V \otimes W)$ has a basis provided by morphisms of the form $(f^{VW}_{U})_j \circ (f^U_{WV})_i$ where $U$ is a simple object in $\Rep(G)$ and $1 \leq i,j \leq {\dim} \, \Hom_{\Rep(G)}(W \otimes V,U)$. We wish to express the half-braiding isomorphisms as linear combinations of such basis vectors. In order to choose a basis for hom-spaces $\Hom_{\Rep(G)}(W \otimes \mc V ,X)$, it is convenient to first decompose the simple object $\mc V$ in $\Mod(\mc D(G))$ into a direct sum of simple objects in $\Rep(G)$. This is typically a tedious exercise but it can be facilitated by remembering that, as a left module over $\mathbb C[G]$, the simple left module over $\mc D(G)$ labeled by $([c_1],\hat V)$ corresponds to the induced $\mathbb C[G]$-module of the simple $\mathbb C[Z_{[c_1]}]$-module $\hat V$.
The decomposition of this induced module can then be obtained by invoking Frobenius reciprocity, which states that the multiplicity of the $\mathbb C[G]$-module $W$ in the induced $\mathbb C[H]$-module ${\rm Ind}^G_H(V)$ of $V$, where $H \subset G$ is a subgroup, equals the multiplicity of the $\mathbb C[H]$-module $V$ in the restriction ${\rm Res}^G_H(W)$ of the $\mathbb C[G]$-module $W$. In symbols,
\begin{equation*}
    {\rm mult}_W({\rm Ind}^G_H(V)) = {\rm mult}_V({\rm Res}^G_H(W)) \,.
\end{equation*}
Equipped with a decomposition of $\mc V$, a basis for the hom-spaces mentioned above can be readily derived. The coefficients of the decomposition of the half-braiding into such a basis define the non-vanishing components of the half-braiding tensor $\Omega^\mc V$. 
Graphically, we represent this decomposition as the resolution of a crossing:
\begin{equation*}
    \includeTikz{0}{halfBraiding-mcV-W}{\halfBraiding{\mc V}{W}} 
    \!\!\! := \sum_{V_i,V'_j,U}\sum_{k,\tilde k} \, \Omega^{\mc V,V_{i},V'_j}_{W,U,k,\tilde k}
    \includeTikz{0}{resolution-V-U-V-W-k-tildek}{\resolution{V'}{U}{V}{W}{k}{\tilde k}} \, ,
\end{equation*}
where the first sum on the r.h.s. is over representatives of isomorphism classes in $\mc I_{\Rep(G)}$ of simple objects $U,V,V'$ in $\Rep(G)$ with multiplicities $i,j$. We provide the explicit definition of $\Omega^\mc V$ for a specific example below.

As before, the half-braiding tensor can be used to explicitly compute tube projectors onto the topological sector labeled by the simple object $\mc V \in \mc Z(\Rep(G))$ whenever $\mc D^\star_\mc M \cong \Rep(G)$:
\begin{align*}
    \frac{1}{|G|}\sum_{W \in \mc I_{\Rep(G)}} \!\!\!\!\!  {\rm dim} \, W \;\;\;\; \includeTikz{0}{idempotent-mcV-W}{\idempotent{\mc V}{W}} \;\;\;\;\,  .
\end{align*}
A representation of these projectors on the Hilbert spaces considered in the main text in the basis of matrix units is given by
\begin{equation*}
    \frac{1}{\rm dim \mc V}\sum_{V_i,V'_j} e^{\mc V,V_i,V'_j}_{\mc M | \mc M} \, ,
\end{equation*}
from which we obtain that the coefficients defining the matrix units in terms of tubes are the matrix elements of the half-braiding tensor $\Omega^{\mc V}$.

\subsection{Example: $\mc D(\mc S_3)$\label{sec:app_S3}}

\noindent
Let us illustrate the procedure described above by computing the half-braiding tensor for the Drinfel'd double of the permutation group on the set of three elements. Although the non-vanishing components of this tensor have previously appeared (see e.g. ref~\cite{PhysRevB.97.195154}), we have not found in the literature any systematic analytical derivation. We provide here such a derivation for pedagogical purposes.

Let $\mc S_3$ be the group with presentation $\la r,s \, | \, r^2 = s^3 = (rs)^2 = \mathbb 1 \ra$. The set $\{\mathbb 1,r ,rs ,rs^2,s,s^2\}$ of group elements is the union of the conjugacy classes $[\mathbb 1] = \{\mathbb 1\}$, $[r] = \{r,rs,rs^2\}$ and $[s] = \{s,s^2\}$. The associated centralizer subgroups are $Z_{[\mathbb 1]} \simeq \mc S_3$, $Z_{[r]} = \{\mathbb 1,r\} \simeq \mathbb Z_2$ and $Z_{[s]} = \{\mathbb 1,s,s^2\} \simeq \mathbb Z_3$. Furthermore, we have $\mc Q_{[s]} = \{\mathbb 1,r\}$ and $\mc Q_{[r]} = \{\mathbb 1,s,s^2\}$. There are three irreducible representations, namely $\{\ub 0, \ub 1, \ub 2\}$, referred to as the trivial, sign and two-dimensional representation, respectively:
\begin{align*}
	\setlength{\tabcolsep}{0.33em}
	\begin{tabularx}{0.86\columnwidth}{c|cccccc}
        $\rho(g)$ & $\mathbb 1$ & $r$ & $rs$ & $rs^2$ & $s$ & $s^2$
        \\
        \midrule
        $V = \ub 0$ & ${\sss 1}$ & ${\sss 1}$ & ${\sss 1}$ & ${\sss 1}$ & ${\sss 1}$ & ${\sss 1}$
        \\[0.2em]
        $V = \underbar 1$ & ${\sss 1}$ & ${\sss -1}$ & ${\sss -1}$ & ${\sss -1}$ & ${\sss 1}$ & ${\sss 1}$
        \\[0.2em]
        $V = \underbar 2$ 
        & $\repMat{1}{0}{0}{1}$
        & $\repMat{0}{1}{1}{0}$
        & $\repMat{0}{\bar \omega}{\omega}{0}$
        & $\repMat{0}{\omega}{\bar \omega}{0}$
        & $\repMat{\omega}{0}{0}{\bar \omega}$
        & $\repMat{\bar \omega}{0}{0}{\omega}$
	\end{tabularx}
    \, ,
\end{align*}
where $\omega = e^{2i\pi/3}$.
The monoidal structure in $\Rep(\mc S_3)$ is provided by $\ub 0 \otimes V \simeq V$, with $V$ any simple object in $\Rep(\mc S_3)$, $\ub 1 \otimes \ub 1 \simeq \ub 0$, $\ub 1 \otimes \ub 2 \simeq \ub 2$ and $\ub 2 \otimes \ub 2 \simeq \ub 0 \oplus \ub 1 \oplus \ub 2$. Given three simple objects $V$, $W$ and $U$ in $\Rep(\mc S_3)$, we denote our choice of basis vector for the one-dimensional hom-space $\Hom_{\Rep(\mc S_3)}(W \otimes V, U)$ by
\begin{align*}
    \arraycolsep=1.4pt
    \begin{array}{ccccl}
        \CC{W}{V}{U}{-}{-}{-}  & : & W \otimes V & \to & U
        \\
        & : & | w \ra \otimes | v \ra & \mapsto & \sum_{u}
        \CC{W}{V}{U}{w}{v}{u} | u \ra 
    \end{array} \, ,
\end{align*}
where the sum is over basis vector labels in the vector space $U$. The non-vanishing matrix components of these intertwiners are provided by
\begin{align*}
    \CC{\ub 0}{\ub 0}{\ub 0}{1}{1}{1}    
    = \CC{\ub 0}{\ub 1}{\ub 1}{1}{1}{1}
    = \CC{\ub 1}{\ub 0}{\ub 1}{1}{1}{1}
    = -\CC{\ub 1}{\ub 1}{\ub 0}{1}{1}{1} 
    &= 1 \, ,
    \\
    \CC{\ub 2}{\ub 0}{\ub 2}{1}{1}{1}
    = \CC{\ub 2}{\ub 0}{\ub 2}{2}{1}{2}
    = \CC{\ub 0}{\ub 2}{\ub 2}{1}{1}{1}
    = \CC{\ub 0}{\ub 2}{\ub 2}{1}{2}{2} 
    &= 1 \, ,
    \\
    \CC{\ub 1}{\ub 2}{\ub 2}{1}{1}{1}
    = \CC{\ub 2}{\ub 1}{\ub 2}{2}{1}{2}
    = - \CC{\ub 2}{\ub 1}{\ub 2}{1}{1}{1}
    = - \CC{\ub 1}{\ub 2}{\ub 2}{1}{2}{2} 
    &= 1 \, ,
    \\
    \CC{\ub 2}{\ub 2}{\ub 0}{1}{2}{1}
    = \CC{\ub 2}{\ub 2}{\ub 0}{2}{1}{1} 
    = -\CC{\ub 2}{\ub 2}{\ub 1}{1}{2}{1}
    = \CC{\ub 2}{\ub 2}{\ub 1}{2}{1}{1} &= \frac{1}{\sqrt{2}} \, ,
    \\
    \CC{\ub 2}{\ub 2}{\ub 2}{1}{1}{2} 
    = \CC{\ub 2}{\ub 2}{\ub 2}{2}{2}{1} &= 1 \, .
\end{align*}
Henceforth, we refer to these entries as the Clebsch-Gordan coefficients. Matrix components of the basis vectors in the hom-space $\Hom_{\Rep(\mc S_3)}(W \otimes V, V \otimes W)$ are depicted as
\begin{equation*}
    \includeTikz{0}{resolution-V-U-V-W-1-1}{\resolution{V}{U}{V}{W}{1}{1}} \!\!\!\!\!\!
    \equiv 
    \sum_{\substack{u,v,w \\ \tilde v, \tilde w}} \CC{W}{V}{U}{w}{v}{u} \CC{V}{W}{U}{\tilde v}{\tilde w}{u}^* \! (|\tilde v \ra \otimes |\tilde w \ra)(\la  w | \otimes \la v |) \, .
\end{equation*}
We denote by $\{\ub 0_{\mathbb Z_2}, \ub 1_{\mathbb Z_2}\}$ the irreducible representations of the centralizer subgroup $\mathbb Z_2$ such that the character of $\ub 1_{\mathbb Z_2}$ maps $r$ to $-1$, and by $\{\ub 0_{\mathbb Z_3},\ub 1_{\mathbb Z_3}, \ub 1^*_{\mathbb Z_3}\}$ the irreducible representations of the centralizer subgroup $\mathbb Z_3$ such that the characters of $\ub 1_{\mathbb Z_3}$ and $\ub 1^*_{\mathbb Z_3}$ map $s$ to $\omega$ and $\bar \omega$, respectively. The eight simple objects in $\Mod(\mc D(\mc S_3))$ are then labeled by $([\mathbb 1], \ub 0)$, $([\mathbb 1], \ub 1)$, $([\mathbb 1], \ub 2)$, $([r],\ub 0_{\mathbb Z_2})$, $([r], \ub 1_{\mathbb Z_2})$, $([s],\ub 0_{\mathbb Z_3})$, $([s], \ub 1_{\mathbb Z_3})$ and $([s],\ub 1^*_{\mathbb Z_3})$. Let us now compute the half-braiding tensors associated with each simple object. Since simple objects in $\Mod(\mc D(\mc S_3))$ labeled by $[\mathbb 1]$ correspond to irreducible representations of $\mc S_3$, it follows immediately from the definition of the Clebsch-Gordan coefficients that for every simple object $(W,\sigma)$ in $\Rep(\mc S_3)$.
\begin{align*}
    \includeTikz{0}{halfBraiding-mathbb1ub0-W}{\halfBraiding{([\mathbb 1],\ub 0)}{W}} 
    \hspace{-1.3em} 
    =   
    \includeTikz{0}{resolution-ub0-W-ub0-W-1-1}{\resolution{\ub 0}{W}{\ub 0}{W}{1}{1}}  ,
\end{align*}
\begin{align*}
    \includeTikz{0}{halfBraiding-mathbb1ub1-W}{\halfBraiding{([\mathbb 1],\ub 1)}{W}} 
    \hspace{-1.3em}  
    = \,
    {\rm tr} \, \sigma(s) \!\!
    \includeTikz{0}{resolution-ub1-Wotimesub1-ub1-W-1-1}{\resolution{\ub 1}{\;\;\;\;\,W  \otimes \ub 1}{\ub 1}{W}{1}{1}} ,
\end{align*}
and
\begin{align*}
    \includeTikz{0}{halfBraiding-mathbb1ub2-W}{\halfBraiding{([\mathbb 1],\ub 2)}{W}}
    \hspace{-1.3em} 
    = \, \sum_{U}
    \begin{pmatrix}
        0 & 0 & 1
        \\
        0 & 0 & -1
        \\
        1 & -1 & 1
    \end{pmatrix}_{ \!\! WU}
    \hspace{-1em}
    \includeTikz{0}{resolution-ub2-U-ub2-W-1-1}{\resolution{\ub 2}{U}{\ub 2}{W}{1}{1}} .
\end{align*}
The other simple objects in $\Mod(\mc D(\mc S_3))$ require to work out the decomposition of the corresponding induced representations in $\Rep(\mc S_3)$. Let us consider the simple object $\mc V$ labeled by $([s],\ub 1_{\mathbb Z_3})$. The restriction of $\ub 2$ to $\mathbb Z_3$ boils down to a sum of two characters. More precisely, for $(\rho,  V = \ub 2)$, we have ${\rm tr} \, \rho(s) = -1 = \omega + \bar \omega$. It follows that ${\rm Res}^{\mc S_3}_{\mathbb Z_3}(\ub 2) \simeq \ub 1_{\mathbb Z_3} \oplus \ub 1_{\mathbb Z_3}^*$ and thus $ {\rm Ind}^{\mc S_3}_{\mathbb Z_3}(\ub 1_{\mathbb Z_3}) \simeq \ub 2$ such that the basis vectors are provided by $|s,1 \ra$ and $|s^2,1\ra$. Although this case is similar to the previous one, the half-braiding isomorphism is not trivial anymore with $R_{W,\mc V}(|w \ra \otimes |s/s^2,1 \ra) = |s/s^2,1 \ra \otimes \hat \rho(s/s^2)|w \ra$. When $W \simeq \ub 2$, we have for instance $R_{\ub 2,\mc V}(|1 \ra_{\ub 2} \otimes |s^2,1 \ra) = |s ,1 \ra \otimes \bar \omega | 1 \ra_{\ub 2}$ and $R_{\ub 2,\mc V}(|1 \ra_{\ub 2} \otimes | s,1 \ra) = |s,1\ra \otimes \omega |1 \ra_{\ub 2}$. It then follows from the definition of the Clebsch-Gordan coefficients that
\begin{align*}
    \includeTikz{0}{halfBraiding-sub1_-mathbbZ_3}{\halfBraiding{([s],\ub 1_{\mathbb Z_3})}{W}}
    \hspace{-1.3em} 
    = \, \sum_{U}
    \begin{pmatrix}
        0 & 0 & 1
        \\
        0 & 0 & -1
        \\
        \bar \omega & - \bar \omega & \omega
    \end{pmatrix}_{ \!\! WU}
    \hspace{-1em}
    \includeTikz{0}{resolution-ub2-U-ub2-W-1-1}{\resolution{\ub 2}{U}{\ub 2}{W}{1}{1}} .
\end{align*}
Similarly, we find ${\rm Ind}^{\mc S_3}_{\mathbb Z_3}(\ub 1^*_{\mathbb Z_3}) \simeq \ub 2$ and 
\begin{align*}
    \includeTikz{0}{halfBraiding-sub1conj_-mathbbZ_3}{\halfBraiding{([s],\ub 1^*_{\mathbb Z_3})}{W}}
    \hspace{-1.3em} 
    = \, \sum_{U}
    \begin{pmatrix}
        0 & 0 & 1
        \\
        0 & 0 & -1
        \\
        \omega & - \omega & \bar \omega
    \end{pmatrix}_{ \!\! WU}
    \hspace{-1em}
    \includeTikz{0}{resolution-ub2-U-ub2-W-1-1}{\resolution{\ub 2}{U}{\ub 2}{W}{1}{1}} .
\end{align*}
Let us now consider the simple object $([s],\ub 0_{\mathbb Z_3})$. 
As the induced representation of a one-dimensional trivial representation, we have ${\rm Ind}^{\mc S_3}_{\mathbb Z_3}(\ub 0_{\mathbb Z_3}) \simeq \mathbb C[\mc S_3 / \mathbb Z_3] \simeq \ub 0 \oplus \ub 1$. This isomorphism is such that the one-dimensional vector spaces $\ub 0$ and $\ub 1$ are spanned by $| 1 \ra_{\ub 0} \equiv |s,1 \ra + |s^2,1 \ra$ and $|1 \ra_{\ub 1} \equiv |s,1 \ra - | s^2,1 \ra$, respectively. This can be confirmed using the explicit action of $\mc S_3$ on $([s],\ub 0_{\mathbb Z_3})$ provided by eq.~\eqref{eq:actionsSubAlg}. But
\begin{align*}
    R_{\ub 2,\mc V}(|1\ra_{\ub 2} \otimes |1 \ra_{\ub 0}) 
    &= \omega |s,1\ra \otimes |1\ra_{\ub 2} + \bar \omega |s^2,1\ra \otimes |1 \ra_{\ub 2}
    \\
    &= \Big( -\frac{1}{2} |1\ra_{\ub 0} - i \frac{\sqrt{3}}{2} |1\ra_{\ub 1} \Big) \otimes | 1 \ra_{\ub 2} \, ,
\end{align*}
and similarly
\begin{equation*}
    R_{\ub 2,\mc V}(|1\ra_{\ub 2} \otimes |1\ra_{\ub 1})
    = \Big(-\frac{1}{2}| 1 \ra_{\ub 1} +i\frac{\sqrt{3}}{2}|1 \ra_{\ub 0}\Big) \otimes |1\ra_{\ub 2} \, .
\end{equation*}
It follows that
\begin{align*}
    \includeTikz{0}{halfBraiding-sub0_-mathbbZ_3}{\halfBraiding{([s],\ub 0_{\mathbb Z_3})}{W}}
    \hspace{-1.3em} 
    &= 
    \frac{{\rm tr}\, \sigma(s)}{{\rm dim} \, W} \!\!
    \includeTikz{0}{resolution-ub0-W-ub0-W-1-1}{\resolution{\ub 0}{W}{\ub 0}{W}{1}{1}}
    \!\!\! + \frac{1}{{\rm dim} \, W} \!\!
    \includeTikz{0}{resolution-ub1-Wotimesub1-ub1-W-1-1}{\resolution{\ub 1}{\;\;\;\;\, W \otimes \ub 1}{\ub 1}{W}{1}{1}} 
    \\
    & -
    i\delta_{W,\ub 2} \frac{\sqrt{3}}{2} \!\! \includeTikz{0}{resolution-ub0-ub2-ub1-ub2-1-1}{\resolution{\ub 0}{\ub 2}{\ub 1}{\ub 2}{1}{1}}
    \!\!\! +
    i\delta_{W,\ub 2} \frac{\sqrt{3}}{2} \!\! \includeTikz{0}{resolution-ub1-ub2-ub0-ub2-1-1}{\resolution{\ub 1}{\ub 2}{\ub 0}{\ub 2}{1}{1}}  \, .  
\end{align*}
The remaining cases are slightly more involved.  We begin with the simple object labeled by $([r],\ub 1_{\mathbb Z_2})$. What is the decomposition of ${\rm Ind}^{\mc S_3}_{\mathbb Z_2}(\ub 1_{\mathbb Z_2})$? Let us consider the restrictions of the irreducible representations of $\mc S_3$ to $\mathbb Z_2$, and check the multiplicity of $\underbar 1_{\mathbb Z_2}$. We know that $\underbar 1_{\mathbb Z_2}$ does not appear in $\ub 0$ since the odd element $a$ acts trivially on it. On the other hand, we have the obvious fact that $\ub 1$ restricts to $\ub 1_{\mathbb Z_2}$. Since ${\rm Ind}^{\mc S_3}_{\mathbb Z_2}(\ub 1_{\mathbb Z_2})$ is three-dimensional as an irreducible representation of $\mc D(\mc S_3)$, we must have ${\rm Ind}^{\mc S_3}_{\mathbb Z_2}(\ub 1_{\mathbb Z_2}) \simeq \ub 1 \oplus \ub 2$. Invoking eq.~\eqref{eq:actionsSubAlg}, the basis vectors are found to be
\begin{align*}
    | 1 \ra_{\ub 1} 
    &\equiv 
    | r ,1\ra + |rs,1 \ra + |rs^2,1 \ra \, ,
    \\
    | 1 \ra_{\ub 2} 
    &\equiv 
    | r ,1\ra + \bar \omega|rs,1 \ra + \omega|rs^2,1 \ra \, ,
    \\
    | 2 \ra_{\ub 2} 
    &\equiv 
    - | r ,1\ra - \omega |rs,1 \ra -  \bar \omega |rs^2,1 \ra \, .
\end{align*}
Noticing in particular that 
\begin{align*}
    R_{\ub 2,\mc V}(|1\ra_{\ub 2} \otimes |1\ra_{\ub 1}) &= \!\!\!\! \sum_{g=r,rs,rs^2}
    \!\!\! |g,1 \ra \otimes \hat \rho(g)|1 \ra_{\ub 2} 
    \\
    &= -|2 \ra_{\ub 2} \otimes |2 \ra_{\ub 2} \, ,    
\end{align*}
it follows from the definition of the Clebsch-Gordan coefficients that
\begin{align*}
    \includeTikz{0}{halfBraiding-rub1_-mathbbZ_2}{\halfBraiding{([r],\ub 1_{\mathbb Z_2})}{W}}
    \hspace{-1.5em} 
    &= 
    {\rm tr}\, \sigma(r) \!\!\!\!\!
    \includeTikz{0}{resolution-ub1-Wotimesub1-ub1-W-1-1}{\resolution{\ub 1}{\;\;\;\;\,W  \otimes \ub 1}{\ub 1}{W}{1}{1}}    
    \\[-1.3em]
    &+ \sum_{U}
    \begin{pmatrix}
        0 & 0 & 1
        \\
        0 & 0 & 1
        \\
        -1 & -1 & 0
    \end{pmatrix}_{\!\!\! WU}
    \hspace{-1em}
    \includeTikz{0}{resolution-ub2-U-ub2-W-1-1}{\resolution{\ub 2}{U}{\ub 2}{W}{1}{1}}
    \\
    & +
    \delta_{W,\ub 2} \!\!
    \includeTikz{0}{resolution-ub1-ub2-ub2-ub2-1-1}{\resolution{\ub 1}{\ub 2}{\ub 2}{\ub 2}{1}{1}}
    \!\!\! - \delta_{W,\ub 2} \!\!
    \includeTikz{0}{resolution-ub2-ub2-ub1-ub2-1-1}{\resolution{\ub 2}{\ub 2}{\ub 1}{\ub 2}{1}{1}}  \, . 
\end{align*}

Similarly, we find ${\rm Ind}^{\mc S_3}_{\mathbb Z_2}(\ub 0_{\mathbb Z_2}) \simeq \ub 0 \oplus \ub 2$ and  
\begin{align*}
    \includeTikz{0}{halfBraiding-rub0_-mathbbZ_2}{\halfBraiding{([r],\ub 0_{\mathbb Z_2})}{W}}
    \hspace{-1.5em} 
    &= 
    {\rm tr}\, \sigma(r) \!\!\!\!\!
    \includeTikz{0}{resolution-ub0-W-ub0-W-1-1}{\resolution{\ub 0}{W}{\ub 0}{W}{1}{1}}    
    \!\!\! + \sum_{U}
    \begin{pmatrix}
        0 & 0 & 1
        \\
        0 & 0 & 1
        \\
        1 & 1 & 0
    \end{pmatrix}_{\!\!\! WU}
    \hspace{-1em}
    \includeTikz{0}{resolution-ub2-U-ub2-W-1-1}{\resolution{\ub 2}{U}{\ub 2}{W}{1}{1}}
    \\
    & +
    \delta_{W,\ub 2} \!\!
    \includeTikz{0}{resolution-ub0-ub2-ub2-ub2-1-1}{\resolution{\ub 0}{\ub 2}{\ub 2}{\ub 2}{1}{1}}
    \!\!\! + \delta_{W,\ub 2} \!\!
    \includeTikz{0}{resolution-ub2-ub2-ub0-ub2-1-1}{\resolution{\ub 2}{\ub 2}{\ub 0}{\ub 2}{1}{1}}  \, . 
\end{align*}
This concludes the computation of the half-braiding tensor for $\mc Z(\Rep(\mc S_3))$.

\newpage
%

\end{document}